\documentclass[prd,aps,twocolumn,showpacs,floatfix,longbibliography,superscriptaddress]{revtex4-2}

\usepackage{mathrsfs}
\usepackage{rotating}
\usepackage{color}
\usepackage{graphicx}           % Include figure files
\usepackage{array,booktabs,amsmath}
\usepackage{dcolumn}            % Align table columns on decimal point
\usepackage{bm}                 % bold math
\usepackage{hyperref}           % hyerlinks in PDF
\usepackage[latin1]{inputenc}   % don't bother me when using special characters
\usepackage{pstricks}           % e.g. colors can be used in the text
\usepackage{amsmath,amssymb}
\usepackage{version}
\usepackage{epstopdf}
\DeclareGraphicsExtensions{.eps,.ps}

\usepackage{ulem}
\usepackage{float}
\usepackage{booktabs}
\usepackage{subfigure}
\usepackage{graphicx}
\usepackage{dcolumn}
\usepackage{bm}
\usepackage{ulem}
\usepackage{setspace}

\begin{document}
%%\preprint{APS/123-QED}
%\title{Semiclassical Analysis of Axion-Like Particle Emission via Nonlinear Compton-Like Scattering in Intense Laser Fields}
\title{Semiclassical analysis of axion-like particle emission via nonlinear Compton-like scattering in intense laser fields}

\author{Pei-Lun He}
\email{peilunhe@sjtu.edu.cn}
\affiliation{State Key Laboratory of Dark Matter Physics, Key Laboratory for Laser Plasmas (Ministry of Education) and School of Physics and Astronomy, Collaborative Innovation Center for IFSA (CICIFSA), Shanghai Jiao Tong University, Shanghai 200240, China}
\author{En-Rui Zhou}
\affiliation{State Key Laboratory of Dark Matter Physics, Key Laboratory for Laser Plasmas (Ministry of Education) and School of Physics and Astronomy, Collaborative Innovation Center for IFSA (CICIFSA), Shanghai Jiao Tong University, Shanghai 200240, China}
%\author{Feng He}
%\email{fhe@sjtu.edu.cn}
%\affiliation{State Key Laboratory of Dark Matter Physics, Key Laboratory for Laser Plasmas (Ministry of Education) and School of Physics and Astronomy, Collaborative Innovation Center for IFSA (CICIFSA), Shanghai Jiao Tong University, Shanghai 200240, China}
\author{Yue-Yue Chen}
\email{yue-yue.chen@shnu.edu.cn}
\affiliation{Department of Physics, Shanghai Normal University, Shanghai 200234, China}

\date{\today}
\begin{abstract}
We investigate the production of axion-like particles through nonlinear Compton-like scattering in intense laser fields using the Baier-Katkov operator method.
By explicitly constructing the eikonal spinor wave function, we utilize the semiclassical nature of relativistic electrons, which simplifies the theoretical derivation and circumvents the need to evaluate certain operator products.
The electron spin-resolved axion emission rate is obtained under the local constant field approximation, with explicit calculations demonstrating that transverse acceleration dominates the radiation yields.
The electron spin dependence of the axion emission rate is found to differ significantly from that of photon emission by analytically examining the asymptotic behavior of the radiation in both the weak- and strong-field limits and by numerically exploring the intermediate regime.
The derived spin-resolved axion emission rate can be directly incorporated into existing semiclassical Monte Carlo algorithms developed for strong-field photon processes, enabling efficient modeling of axion-like particle generation. Our results provide promising avenues for the experimental detection and control of axion-like particles with high-intensity laser facilities.
\end{abstract}

\maketitle
\section{Introduction}
The axion was initially proposed to address the strong CP problem in quantum chromodynamics (QCD), which arises from vacuum tunneling effects induced by instantons \cite{POLYAKOV197582, PhysRevLett.37.172, CALLAN1976334}. One of the key parameters in axion models is the energy scale $f_a$ at which the Peccei-Quinn (PQ) symmetry breaks \cite{PhysRevD.16.1791, PhysRevLett.38.1440}, which is inversely proportional to both the axion mass and its coupling constant. 
The original Peccei-Quinn-Weinberg-Wilczek model \cite{PhysRevLett.40.279, PhysRevLett.40.223} associated $f_a$ with the electroweak scale and was subsequently ruled out due to inconsistencies with experimental observations \cite{kim1987light}. With the introduction of a much higher PQ symmetry-breaking scale in subsequent developments \cite{PhysRevLett.43.103, SHIFMAN1980493, Zhitnitsky:1980tq,dine1981simple, PECCEI1986435}, the axion acquires a significantly lighter mass and smaller coupling constants, thus resolving the earlier conflicts with experiments. The extremely weak coupling with ordinary matter also makes axion a compelling dark matter candidate apart from being a solution to the strong CP problem \cite{marsh2016axion}. 

Moreover, axion-like particles (ALPs) - a term used to distinguish pseudoparticles that have similar properties to axion in QCD but without a prior relationship between mass and coupling constant - are quite universal in string theory \cite{witten1984some,arvanitaki2010string}. In condensed matter physics, axion-like quasiparticles appear in magnetic topological insulators \cite{wu2016quantized,okada2016terahertz}, antiferromagnetic topological insulators \cite{zhang2019topological,liu2020robust}, and Weyl semimetals \cite{PhysRevLett.58.1799,gooth2019axionic}. Theoretical investigations of pseudoscalar particles with light mass and small coupling constants continue to flourish across different branches of physics.

Despite the appealing properties of axion-like particles (ALPs), their extremely weak couplings to ordinary matter make them nearly undetectable, rendering their experimental verification one of the most significant challenges in modern physics. Extensive efforts are underway to search for ALP signatures through cosmological and astronomical observations, as detailed in several comprehensive review articles~\cite{IRASTORZA201889,di2020landscape,RevModPhys.93.015004,carenza2025axion}. In addition, a variety of laboratory-based searches have been developed, including light-shining-through-walls experiments~\cite{pugnat2014search,ballou2015new}, vacuum magnetic birefringence experiments~\cite{della2016pvlas}, beam dump experiments~\cite{konaka1986search,davier1989unambiguous,bross1991search}, and fixed-target experiments~\cite{abrahamyan2011search}. 
However, the sensitivity of these traditional laboratory approaches is generally limited by the maximum attainable static magnetic field strength and beam intensity in current experimental setups.
Owing to rapid advances in modern laser technology, there has been growing interest in studying pseudoscalar interactions with intense laser fields~\cite{mendoncca2007axion,gies2009strong,dobrich2010axion}. The spin-averaged axion emission rate via nonlinear Compton-like scattering has been derived in Refs.~\cite{dillon2018alp,king2019axion,PhysRevD.111.055001} for the case of a plane-wave laser pulse, where the electron is described by the Volkov wave function~\cite{wolkow1935klasse}. 
These laser-based approaches offer a promising alternative for ALP detection~\cite{King:2018qbq,PhysRevD.99.035048,PhysRevD.106.115034,huang2022axion,an2024modeling,an2025situ}, potentially overcoming limitations of conventional magnetic-field-based experiments by exploiting the extreme intensities now achievable with modern lasers~\cite{Yoon:21}.

One of the universal features of quantum electrodynamics (QED) in intense laser fields is that the radiation formation time~\cite{baier2005concept} is much shorter than the period of the external field.
Consequently, the laser field can be treated as a locally constant field (LCF) during each photon emission event, allowing radiation in intense laser fields to be described as an incoherent sum of instantaneous synchrotron-like emissions, with interference between successive events neglected~\cite{ritus1985quantum,di2010quantum}. The Monte Carlo model of strong-field QED incorporates this feature and describes the dynamics in a semiclassical manner~\cite{elkina2011qed,ridgers2014modelling,green2015simla,gonoskov2015extended,zhuang2023laser}. Specifically, at each time step, the electron's classical motion is calculated via the Lorentz equation, and photon emission is then probabilistically determined using random sampling based on the local instantaneous emission spectrum, which is derived under the assumption that the laser field can be approximated as an LCF during the short formation time of the emission process. The Baier-Katkov method ~\cite{schwinger1954quantum,baier1968processes,baier1968quasiclassical,berestetskii2012quantum} provides a powerful approach to derive the quantum radiation probability within the LCF approximation (LCFA), which is a key ingredient of the semiclassical Monte Carlo model, by utilizing the electron's classical trajectory as the basis for calculation while accounting for photon recoil effects through the commutation relations of operators~\cite{baier1998electromagnetic,baier2009recent,bogdanov2019semiclassical}.
%Combined with LCFA, this method allows radiation in intense laser fields to be treated as an incoherent sum of synchrotron emissions at each instant, with neglected corrections being relativistically suppressed.

The semiclassical Monte Carlo model of QED can be extended to include spin degrees of freedom~\cite{chen2022electron,li2023strong}. In the spin-extended model, classical spin precession is governed by the Bargmann--Michel--Telegdi equation~\cite{bargmann1959precession}, while additional random numbers are generated to determine whether a spin-flip occurs, according to the spin-resolved synchrotron radiation probabilities~\cite{chen2022electron,li2023strong}. The spin-resolved semiclassical Monte Carlo model has proven valuable for simulating the generation of polarized photons~\cite{li2020polarized,dai2022photon,PhysRevD.110.012008}, electrons~\cite{li2019ultrarelativistic,li2022helicity}, and positrons~\cite{chen2019polarized,li2020production} in strong electromagnetic fields. 
The simple theoretical structure of the model also facilitates its integration into particle-in-cell simulation~\cite{gong2021retrieving,PhysRevLett.129.035001,gong2023electron,PhysRevLett.131.225101,PhysRevLett.131.175101}. 
Analogous to photon emission processes, a formula describing the electron spin-resolved axion emission rate would be highly beneficial, as it enables the inclusion of axion dynamics in the semiclassical Monte Carlo model~\cite{JointSub}.

%In the context of synchrotron radiation \cite{schwinger1954quantum,baier1968processes,baier1968quasiclassical}, the magnetic field is time-independent, so the electron dynamics can be described by a time-independent Hamiltonian. 
%prevents direct evaluation of how the essentially time-dependent laser field affects the dynamics, although the correction should be relativistically suppressed.
%The quantum radiation theory is thus formulated via a time-independent Hamiltonian, which, however, prevents direct evaluation of how the essentially time-dependent laser field affects the dynamics, although the correction should be relativistically suppressed.

%In the previous treatment, the Hamiltonian is time-independent, which is exact for synchrotron radiation where the magnetic field is constant.

%While the use of a time-independent Hamiltonian approach is physically justified in the previous derivation, the laser field is essentially time-dependent, making the rigorous treatment of longitudinal acceleration not possible, though it is indeed relativistically suppressed as shown by explicit calculation in this work. 

In this work, we apply the Baier-Katkov method to study electron spin-resolved ALP production in intense laser fields. The structure of the paper is as follows. Section~\ref{section_Conventions} summarizes the theoretical conventions used throughout the paper. In Sec.~\ref{section_Semiclassic_spinor}, we construct the explicit form of the semiclassical spinor wave function under the eikonal approximation, justifying the semiclassical description of relativistic electron dynamics in intense electromagnetic fields. Section~\ref{section_Production_Probability} presents the detailed theoretical derivation of the ALP production rate via nonlinear Compton-like scattering and derives a compact, closed-form expression for the electron spin-resolved radiation rate under the LCFA. In Sec.~\ref{SectionRadiation}, we numerically investigate ALP radiation properties and analytically examine the asymptotic features of spin dependence in both the weak- and strong-field limits. Finally, Sec.~\ref{section_Conclusion} summarizes the theoretical assumptions underlying the semiclassical model and their physical implications, and concludes with an outlook for future applications. Two appendices provide detailed integration formulas and explicit calculations demonstrating that contributions from longitudinal acceleration to the ALP yield are relativistically suppressed.

\section{Conventions}\label{section_Conventions}

Throughout this work, natural units are employed with $m_e = \hbar = c = 1$, unless stated otherwise. 
The metric convention for Minkowski spacetime specified as
\begin{equation}
g_{\mu\nu}= \begin{pmatrix}
-1 & 0 & 0 & 0 \\
0 & 1 & 0 & 0 \\
0 & 0 & 1 & 0 \\
0 & 0 & 0 & 1 \\
\end{pmatrix},
\end{equation}
and the Dirac representation for gamma matrices:
\begin{equation}
\gamma^0=\begin{pmatrix}
1 & 0 \\
0 & -1
\end{pmatrix}, \quad
\gamma^i=\begin{pmatrix}
0 & \sigma^i \\
-\sigma^i & 0
\end{pmatrix}, \quad
\gamma^5=\begin{pmatrix}
0 & 1 \\
1 & 0
\end{pmatrix},
\end{equation}
where $\sigma^i$ denotes the Pauli matrices.

The Lagrangian density describing the interaction between axions and electrons in intense laser fields is given by 
\begin{equation} 
\begin{aligned} 
&\mathcal{L} = \bar{\psi} \left[ i\gamma^\mu (\partial_\mu + i g_e A_\mu) - m_e \right] \psi \\
& - \frac{1}{2}\partial_\mu \phi\, \partial^\mu \phi - \frac{1}{2}m_\phi^2 \phi^2 - g_{ea} \phi\, \bar{\psi} \gamma^5 \psi, 
\end{aligned} 
\end{equation} 
where $m_e$ is the electron mass, $m_\phi$ the axion mass, $g_{ea}$ the axion-electron coupling constant, $g_e$ the QED coupling constant, $\psi$ the electron field, $\phi$ the axion field, $\gamma^\mu$ are the Dirac matrices, and repeated indices follow the Einstein summation convention.
We do not assume any prior relation between the axion mass and coupling constants for generality, but we refer to ALPs as axions hereafter for simplicity.
Following the Furry picture~\cite{PhysRev.81.115}, the photon field has a nonzero vacuum expectation value $A_\mu(\boldsymbol{x}, t)$ in the presence of an intense laser field, which is a solution of Maxwell's equations. The laser field is considered intense in the sense that the classical nonlinear parameter $a_0=\frac{g_eA_0}{m_e} \gg 1$~\cite{di2012extremely,fedotov2023advances}, where $A_0$ is the amplitude of the laser vector potential. The field dependencies on spacetime coordinates are implicitly assumed for compactness when there is no confusion.

The following assumptions are made for the nonlinear Compton-like process in intense laser fields:
\begin{itemize}
\item The electron motion is relativistic both before and after axion emission.
\item The emitted axions are also relativistic.
\item The electron dynamics can be treated semiclassically.
\end{itemize}
The criteria and justifications for these assumptions are provided in detail below for typical parameter regimes, with a summary given in Sec.~\ref{section_Conclusion}.

\section{Semiclassical spinor}\label{section_Semiclassic_spinor}

To construct a semiclassical spinor that matches the boundary condition of the incoming electron, we adopt the eikonal approximation. In intense laser fields, the Dirac equation reads
\begin{equation}
\left[i \gamma^\mu D_\mu - m_e \right] \psi = 0, \label{DiracEq}
\end{equation}
where $D_\mu = \partial_\mu + i g_e A_\mu(\boldsymbol{x}, t)$ is the covariant derivative.
To obtain its semiclassical solution, we recast the Dirac equation into its second-order form
\begin{equation}
\left[D_\mu D^\mu - m_e^2 - \frac{g_e}{2} \sigma^{\mu\nu} F_{\mu\nu}\right]\psi = 0, \label{DiracEq2}
\end{equation}
where $F_{\mu\nu}(\boldsymbol{x}, t) = \frac{-i}{g_e} [D_\mu, D_\nu]$ is the electromagnetic field tensor, and $\sigma^{\mu\nu} = \frac{i}{2}[\gamma^\mu, \gamma^\nu]$.
Upon restoring the Planck constant in Eq.~(\ref{DiracEq2}), we can see that the term $\frac{g_e}{2} \sigma^{\mu \nu}F_{\mu\nu}$ is a quantum correction that is linear in $\hbar$. Therefore, this term is negligible under the Wentzel-Kramers-Brillouin (WKB) approximation, and the remaining equation to be solved is an auxiliary scalar equation
\begin{equation}
\left[\left(\partial_\mu+i g_e A_\mu\right)\left(\partial^\mu+i g_e A^\mu\right)-m_e^2\right]\varphi=0.\label{KGeq}
\end{equation}
To solve Eq.~(\ref{KGeq}), we make the ansatz
\begin{equation}
\varphi(\boldsymbol{x}, t) \approx N_0\, e^{i S_0(\boldsymbol{x}, t)}, \label{Ansatzeq}
\end{equation}
where $N_0$ is a normalization constant and $S_0(\boldsymbol{x}, t)$ is the action field determined by the relativistic Hamilton-Jacobi equation
\begin{equation}
\partial_t S_0(\boldsymbol{x}, t) = -\sqrt{\boldsymbol{p}(\boldsymbol{x}, t)^2 + m_e^2}, \label{eqHJ}
\end{equation}
with $\boldsymbol{p}(\boldsymbol{x}, t) \equiv \boldsymbol{\nabla} S_0(\boldsymbol{x}, t) + g_e \boldsymbol{A}(\boldsymbol{x}, t)$ denoting the momentum field.
Taking the time derivative of $\boldsymbol{p}(\boldsymbol{x}, t)$, the equation of motion reads
\begin{equation}
\frac{\mathrm{d}}{\mathrm{d}t} \boldsymbol{p}(\boldsymbol{x}, t) = g_e \partial_t \boldsymbol{A}(\boldsymbol{x}, t) - g_e \boldsymbol{v}(\boldsymbol{x}, t) \times \boldsymbol{B}(\boldsymbol{x}, t), \label{eqHJ2}
\end{equation}
where $\boldsymbol{B}(\boldsymbol{x}, t) = \boldsymbol{\nabla} \times \boldsymbol{A}(\boldsymbol{x}, t)$ is the magnetic field, $\boldsymbol{v}(\boldsymbol{x}, t) \equiv \frac{\boldsymbol{p}(\boldsymbol{x}, t)}{\sqrt{\boldsymbol{p}(\boldsymbol{x}, t)^2 + m_e^2}}$ the velocity field, and $\frac{\mathrm{d}}{\mathrm{d}t} = \frac{\partial}{\partial t} + \boldsymbol{v}(\boldsymbol{x}, t) \cdot \boldsymbol{\nabla}$ denotes the total time derivative.
Equation~(\ref{eqHJ2}) can be solved using the classical trajectory specified by
\begin{equation}
\frac{\mathrm{d}}{\mathrm{d}t}\boldsymbol{x}_c = \boldsymbol{\nabla}_{\boldsymbol{p}_c} H_c , \quad 
\frac{\mathrm{d}}{\mathrm{d}t}\boldsymbol{p}_c = -\boldsymbol{\nabla}_{\boldsymbol{x}_c} H_c , \label{LorentzEquation}
\end{equation}
where $H_c = \sqrt{(\boldsymbol{p}_c + g_e \boldsymbol{A}(\boldsymbol{x}_c, t))^2 + m_e^2}$ is the classical Hamiltonian.
With the boundary condition that the classical trajectory $\boldsymbol{x}_c(t)$ reaches the position $\boldsymbol{x}$ at instant $t$, the solution to Eq.~(\ref{eqHJ2}) is given by
\begin{equation}
\boldsymbol{p}(\boldsymbol{x}, t) = \left. \boldsymbol{p}_c(t) \right|_{\boldsymbol{x}_c(t) = \boldsymbol{x}}, \label{p_solution}
\end{equation}
and the action field $S_0(\boldsymbol{x}, t)$ is obtained by integrating Eq.~(\ref{eqHJ}) along the classical trajectory.
For our purposes, the explicit expression of the WKB wave function is of little relevance, but its qualitative features matter.

The relativistic nature of electron motion provides the physical justification for the semiclassical Monte Carlo description of strong-field QED processes. For typical relativistic electron beams, the incoming electron wave packet is well localized in momentum space, with a divergence angle less than $0.1~\mathrm{rad}$ and an energy spread below $10\%$~\cite{esarey2009physics}. This implies that the expectation value of the momentum operator is $\left\langle \hat{\boldsymbol{p}}\right\rangle = \boldsymbol{p}_c(t)$, with a corresponding standard deviation of order $\sigma \sim \left| \boldsymbol{p}_c(t) \right|$. Similarly, the expectation value of the position operator is $\left\langle \hat{\boldsymbol{x}}\right\rangle = \boldsymbol{x}_c(t)$, which coincides with the classical trajectory described by Eq.~(\ref{LorentzEquation}), and the associated standard deviation is of order $1/\sigma$ according to the uncertainty principle.
For an electron with GeV-scale energy interacting with a laser of wavelength $\lambda_L = 800~\mathrm{nm}$, the ratio $1/(\sigma \lambda_L) \sim 10^{-8}$ indicates that the electron is highly localized in space compared to the characteristic scale of the laser field. Thus, it is justified to substitute the classical trajectory obtained from Eq.~(\ref{LorentzEquation}) for the corresponding operator value when taking expectation values. This approximation is valid provided that operators involving products such as $\hat{\boldsymbol{p}} \cdot \hat{\boldsymbol{x}}$ are absent, which would otherwise give rise to uncertainties of order unity.

With the auxiliary scalar wave function $\langle \boldsymbol{x}| i \rangle = \varphi_i(\boldsymbol{x}, t_0)$, the spinor wave function for the electron can be constructed as
\begin{equation}
\psi(\boldsymbol{x}, t) = \frac{u(\hat{\boldsymbol{p}}(t))}{\sqrt{2 \hat{H}(t)}}\, U(t, t_0)\, \varphi_i(\boldsymbol{x}, t_0), \label{psii}
\end{equation}
where the time evolution is governed by the unitary operator
\begin{equation}
U(t, t_0) = T \left\{ \exp \left[ -i \int_{t_0}^{t} \mathrm{d}\tau\, \hat{H}(\tau) \right] \right\},
\end{equation}
with $T$ denoting the time-ordering operator, $\hat{H}(t) = \sqrt{\hat{\boldsymbol{p}}(t)^2 + m_e^2}$ the Hamiltonian operator, and $\hat{\boldsymbol{p}}(t) \equiv -i \boldsymbol{\nabla} + g_e \boldsymbol{A}(\boldsymbol{x}, t)$ the momentum operator. 
The need to solve the time-dependent trajectory explicitly is circumvented by the use of $U(t, t_0)$. Substituting Eq.~(\ref{psii}) into the Dirac equation and noting that the spacetime derivative acting on the spinor basis yields corrections of order $\hbar$, the spinor basis is determined as
\begin{equation}
u(\hat{\boldsymbol{p}}(t))
= 
\begin{pmatrix}
\sqrt{\hat{H}(t) + m_e}\; w_i \\
\dfrac{\boldsymbol{\sigma} \cdot \hat{\boldsymbol{p}}(t)}{\sqrt{\hat{H}(t) + m_e}}\; w_i
\end{pmatrix}, \label{psiibasis}
\end{equation}
where the corresponding polarization matrix is given by
\begin{equation}
w_i w_i^\dagger = \frac{1}{2}\left(1 + \boldsymbol{\sigma} \cdot \boldsymbol{\zeta}_i\right),
\end{equation}
with $\boldsymbol{\zeta}_i$ denoting the electron spin direction in its rest frame.
Operator ordering within the spinor basis of Eq.~(\ref{psiibasis}) is not important in the semiclassical approximation, since the commutators are linear in $\hbar$.

\section{Axion Production Probability}\label{section_Production_Probability}

\subsection{General formalism}
\begin{figure}
\centering
\includegraphics[width=0.45\textwidth]{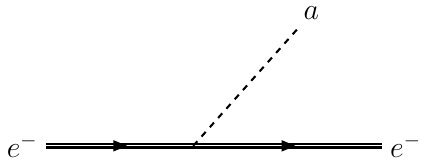}
\caption{Illustration of axion emission in intense laser fields via nonlinear Compton-like scattering. The double lines represent the laser-dressed incoming and outgoing electrons described by semiclassical Dirac spinors within the eikonal approximation, while the dashed line represents the outgoing axion.}
\label{FigFeyn}
\end{figure}

The leading-order contribution to axion emission by an electron in intense laser fields is illustrated in Fig.~\ref{FigFeyn}, which is analogous to nonlinear Compton scattering for photon emission. The incoming and outgoing electrons are treated semiclassically using Dirac spinors under the eikonal approximation [Eq.~(\ref{psii})], and the axion is emitted via a nonlinear Compton-like scattering process.
The transition amplitude for this nonlinear Compton-like axion emission can be expressed as
\begin{equation}
\begin{aligned}
a_{fi} 
= -\frac{i g_{ea}}{\sqrt{2 \omega}} \int \! \mathrm{d}t \, 
   \left\langle f \right| \hat{Q}(t) \left| i \right\rangle \exp(i \omega t), 
\end{aligned}
\label{a_fiX}
\end{equation} 
where the pseudoscalar density operator $\hat{Q}(t)$ is defined as
\begin{equation}
\begin{aligned}
\hat{Q}(t) 
= U^\dagger(t, t_0) 
   \frac{\bar{u}_f(\hat{\boldsymbol{p}}(t))}{\sqrt{2 \hat{H}(t)}}   \gamma^5 
   \exp(-i \boldsymbol{k} \cdot \hat{\boldsymbol{x}}) 
   \frac{u_i(\hat{\boldsymbol{p}}(t))}{\sqrt{2 \hat{H}(t)}}  U(t, t_0).
\end{aligned}
\label{Q}
\end{equation}
Here, $\boldsymbol{k} = k\boldsymbol{n}$ denotes the axion momentum, $\boldsymbol{n}$ the unit vector along the axion propagation direction, and $\omega = \sqrt{\boldsymbol{k}^2 + m_a^2}$ the axion energy.

After summing over all final electron states $\left| f \right\rangle$, the axion production probability is obtained from the transition amplitude [Eq.~(\ref{a_fiX})] as
\begin{equation}
\frac{\mathrm{d}^3 P}{\mathrm{d}^3 k} = \frac{g_{ea}^2}{16} \frac{1}{\pi^3 \omega} \int \mathrm{d} t \int \mathrm{d} t'\, \left\langle i \left| \hat{Q}^\dagger(t') \hat{Q}(t) \right| i \right\rangle\, e^{i \omega (t - t')},
\label{dPexp}
\end{equation}
where the effects of axion emission recoil are incorporated via
\begin{equation}
\begin{aligned}
\hat{Q}^\dagger(t') \hat{Q}(t) =\ & U^\dagger(t', t_0) \frac{u_i^\dagger(\hat{\boldsymbol{p}}(t'))}{\sqrt{2 \hat{H}(t')}} \gamma^5 \gamma^0 \frac{u_f(\hat{\boldsymbol{p}}'(t'))}{\sqrt{2 \hat{H}'(t')}} U'(t', t) \\
& \times \frac{u_f^\dagger(\hat{\boldsymbol{p}}'(t))}{\sqrt{2 \hat{H}'(t)}} \gamma^0 \gamma^5 \frac{u_i(\hat{\boldsymbol{p}}(t))}{\sqrt{2 \hat{H}(t)}} U(t, t_0),
\end{aligned}
\label{prefactor}
\end{equation}
with $\hat{\boldsymbol{p}}' = \hat{\boldsymbol{p}} - \boldsymbol{k}$ being the recoil-corrected momentum operator, $\hat{H}'(t) = \sqrt{\hat{\boldsymbol{p}}'(t)^2 + m_e^2}$ the corresponding Hamiltonian operator, and $U'(t', t) = T \left\{ \exp \left( -i \int_{t}^{t'} \mathrm{d}\tau\, \hat{H}'(\tau) \right) \right\}$ the associated time evolution operator. In the derivation, the identity $e^{i \boldsymbol{k} \cdot \hat{\boldsymbol{x}}}\, \hat{\boldsymbol{p}}\, e^{-i \boldsymbol{k} \cdot \hat{\boldsymbol{x}}} = \hat{\boldsymbol{p}} - \boldsymbol{k}$ is used.

The expression for the prefactor given by Eq.~(\ref{prefactor}) involves only momentum operators. Thus, when acting on the semiclassical wave function, Eq.~(\ref{dPexp}) yields
\begin{equation}
\left\langle i \left| \hat{Q}^\dagger(t') \hat{Q}(t) \right| i \right\rangle  e^{ i \omega (t-t') }
= Q_c^\dagger(t') Q_c(t) \exp\bigl(i S_c\bigr),
\label{PrefactorC}
\end{equation}
where the classical action $S_c$ is defined as
\begin{equation}
S_c 
= \int_{t}^{t'} \! \mathrm{d}\tau \left( 
   \sqrt{\boldsymbol{p}_c(\tau)^2 + m_e^2} 
   - \sqrt{\boldsymbol{p}_c'(\tau)^2 + m_e^2} - \omega \right).
\label{RadiationForm1}
\end{equation}
Here, the pseudoscalar density is given by
\begin{equation}
Q_c(t)
= \frac{u_f^\dagger( \boldsymbol{p}_c'(t))}{\sqrt{2 H'_c}}\, \gamma^0 \gamma^5\, \frac{u_i( \boldsymbol{p}_c(t) )}{\sqrt{2 H_c}},
\end{equation}
where the subscript $c$ indicates that the operators have been replaced with their classical values, corresponding to the trajectory from Eq.~(\ref{LorentzEquation}).

Our approach circumvents the need to evaluate the operator $e^{i\boldsymbol{k} \cdot \hat{\boldsymbol{x}}(t')} e^{-i\boldsymbol{k} \cdot \hat{\boldsymbol{x}}(t)}$, which in previous literature was handled either using the Baker--Campbell--Hausdorff identity~\cite{schwinger1954quantum} or by solving differential equations~\cite{baier1968processes,baier1968quasiclassical}, thereby simplifying the calculation.
The presence of the operator $e^{i\boldsymbol{k} \cdot \hat{\boldsymbol{x}}(t')} e^{-i\boldsymbol{k} \cdot \hat{\boldsymbol{x}}(t)}$ prevents the direct substitution of classical trajectory values, since the magnitude of the emitted axion momentum $\boldsymbol{k}$ is of the same order as $|\boldsymbol{p}_c(t)|$ [see Eq.~(\ref{omegac})], while the uncertainty in $\boldsymbol{x}$ is of order $1/|\boldsymbol{p}_c(t)|$. As a result, the uncertainty in the product is non-negligible and must be properly accounted for to include the recoil effect of axion emission.

Up to this point, the main approximation employed for the nonlinear Compton-like axion emission process has been the semiclassical treatment of electron motion. When the relativistic nature of the electron is taken into account, the connection of Eq.~(\ref{RadiationForm1}) with classical radiation theory~\cite{landau2013classical,schwinger2019classical} becomes transparent. The time-dependent electron energy before and after axion emission is given by $\varepsilon \equiv \sqrt{\boldsymbol{p}_c(\tau)^2 + m_e^2}$ and $\varepsilon' \equiv \varepsilon - \omega$, respectively, where the time dependence is suppressed for compactness.
The axion is emitted predominantly in the forward direction relative to the relativistically moving electron, which permits the following binomial expansion
\begin{equation}
\sqrt{\boldsymbol{p}_c^{\prime\,2}(\tau) + m_e^2} \approx \varepsilon' 
    + \omega\frac{\varepsilon}{\varepsilon'}  \left( 1 - \frac{k}{\omega} \boldsymbol{v}_c(\tau) \cdot \boldsymbol{n} \right) 
    - \frac{1}{2} \frac{m_\phi^2}{\varepsilon'},
\label{Energy2}
\end{equation}
whose validity is demonstrated at the end of this subsection [see Eq.~(\ref{vcn})].

With Eq.~(\ref{Energy2}), the classical action $S_c$ in the exponent can be approximated as
\begin{equation}
S_c
\approx \int_{t}^{t^\prime} \mathrm{d}\tau \left( -\omega\frac{\varepsilon}{\varepsilon'} + \frac{1}{2} \frac{m_\phi^2}{\varepsilon'} + k\frac{\varepsilon}{\varepsilon'} \boldsymbol{v}_c(\tau) \cdot \boldsymbol{n} \right).
\end{equation}
The prefactor can be calculated by substituting the expression of spinor basis [Eq.~(\ref{psiibasis})] into Eq.~(\ref{PrefactorC}), which yields:
\begin{widetext} 
\begin{equation}
\begin{aligned}
Q_c^*(t') Q_c(t) 
&= \frac{1}{2} \Bigl[ 
   (\boldsymbol{\zeta}_i \cdot \boldsymbol{\mathfrak{B}}(t') ) (\boldsymbol{\zeta}_f \cdot \boldsymbol{\mathfrak{B}}(t)) 
   + (\boldsymbol{\zeta}_i \cdot \boldsymbol{\mathfrak{B}}(t) ) (\boldsymbol{\zeta}_f \cdot \boldsymbol{\mathfrak{B}}(t')) 
   \Bigr. \\ 
&\quad + (\boldsymbol{\mathfrak{B}}(t') \cdot \boldsymbol{\mathfrak{B}}(t))  (1 - \boldsymbol{\zeta}_i \cdot \boldsymbol{\zeta}_f) 
   + i (\boldsymbol{\zeta}_i - \boldsymbol{\zeta}_f) \cdot (\boldsymbol{\mathfrak{B}}(t') \times \boldsymbol{\mathfrak{B}}(t))
   \Bigr],
\end{aligned}\label{fullEqQc}
\end{equation}
\end{widetext} 
where we use the abbreviation 
\begin{equation}
\boldsymbol{\mathfrak{B}}(t) \approx -\frac{1}{2} \frac{\omega}{\varepsilon'} \left( 1-\frac{m_e}{\varepsilon} \right) \boldsymbol{v}_c(t) + \frac{1}{2} \frac{\omega}{\varepsilon'} \boldsymbol{n},\label{EqB}
\end{equation}
which is accurate to order $O(\frac{m_e}{\varepsilon})$.

Gathering all the results above, we arrive at a general expression for the axion radiation probability, which closely resembles the classical radiation theory of electromagnetism, but with recoil effects and the nonvanishing axion mass taken into account:
\begin{equation}
\begin{aligned}
\frac{\mathrm{d}^3 P}{\mathrm{d}^3 k}
&= \frac{1}{16} \frac{g_{ea}^2}{\pi^3} \frac{1}{\omega}
\int \mathrm{d}t_0 \int \mathrm{d}\tau\,
Q_c^*\left(t_0 + \frac{\tau}{2}\right) Q_c\left(t_0 - \frac{\tau}{2}\right) \\
& \times \exp\left\{ i \int_{t_0 - \frac{\tau}{2}}^{t_0 + \frac{\tau}{2}} \mathrm{d}t \left[ -\omega\frac{\varepsilon}{\varepsilon'} + \frac{1}{2} \frac{m_\phi^2}{\varepsilon'} + k\frac{\varepsilon}{\varepsilon'} \boldsymbol{v}_c(t) \cdot \boldsymbol{n} \right] \right\},
\label{eqDP}
\end{aligned}
\end{equation}
where we have implemented the coordinate transformation $t_0 = \frac{t + t'}{2}$ and $\tau = t' - t$.
With the radiation formula in Eq.~(\ref{eqDP}),  we can show using integration by parts that irrespective of the details of the trajectory,
the $\boldsymbol{v}_c(t) \cdot \boldsymbol{n}$ term in the prefactor can be replaced as
\begin{equation}
\boldsymbol{v}_c(t) \cdot \boldsymbol{n} \to \frac{\omega}{k}\left( 1 - \frac{1}{2} \frac{m^2_\phi}{\omega \varepsilon} \right),\label{vcn}
\end{equation}
which justifies the expansion in Eq.~(\ref{Energy2}) and can be used to simplify the calculation below.
Equation~(\ref{vcn}) also implies that $\boldsymbol{n} \to \boldsymbol{v}_c(\tau)$, with the error being $O(\frac{m_e}{\varepsilon})$, ensuring the validity of Eq.~(\ref{Energy2}) [see also Eq.~(\ref{nExpansion}) and Appendix~\ref{Time Integration}].

%and validates the omission of the higher order corrections in Eq.~(\ref{Energy2}).

\subsection{Spin-Resolved Axion Emission Probability under the LCFA}
To obtain an explicit expression for the axion production rate, information about the corresponding trajectory is required. In general, a compact analytical result cannot be derived from Eq.~(\ref{eqDP}); however, one can proceed by performing a Taylor expansion in powers of $\tau$, which corresponds to the LCFA. The LCFA is valid when the classical nonlinear parameter is much greater than unity, i.e., $a_0 \gg 1$, allowing the laser field to be treated as quasi-static at a given instant $t_0$ [see Appendix~\ref{Time Integration}]. In the presence of the quasi-static laser field, the classical equation of motion takes the form:
\begin{equation}
\frac{\mathrm{d}\boldsymbol{p}(\tau)}{\mathrm{d}\tau} 
= -\left( \boldsymbol{E}_{\mathrm{eff}} + \frac{\boldsymbol{p}(\tau) \times \boldsymbol{B}_{\mathrm{eff}}}{\varepsilon(\tau)} \right),
\end{equation}
where $\boldsymbol{E}_{\mathrm{eff}}$ is the effective electric field responsible for longitudinal acceleration and $\boldsymbol{B}_{\mathrm{eff}}$ is the effective static magnetic field that generates transverse acceleration.
The connection between the effective electric field and the physical electric field $\boldsymbol{E}_{\mathrm{phy}}$ is given by
\begin{equation}
\boldsymbol{E}_{\mathrm{eff}} = g_e \left( \boldsymbol{E}_{\mathrm{phy}} \cdot \mathbf{T} \right) \mathbf{T},
\end{equation}
while the connection between the effective magnetic field and the physical magnetic field $\boldsymbol{B}_{\mathrm{phy}}$ is
\begin{equation}
\boldsymbol{B}_{\mathrm{eff}} = \frac{g_e}{v_0^2} \left[ \left( \boldsymbol{E}_{\mathrm{phy}} - \boldsymbol{E}_{\mathrm{eff}} + \boldsymbol{v}(t_0) \times \boldsymbol{B}_{\mathrm{phy}} \right) \times \boldsymbol{v}(t_0) \right].
\end{equation}
For a head-on collision between a laser pulse and a relativistic electron, we typically have $\frac{E_{\mathrm{eff}}}{B_{\mathrm{eff}}} < 1$. Based on this relation, an explicit calculation demonstrating that the contribution of longitudinal acceleration is relativistically suppressed is provided in Appendix~\ref{APlongitudinal}.
However, the Taylor expansion in powers of $\tau$ breaks down when $\frac{E_{\mathrm{eff}}}{B_{\mathrm{eff}}}  > \frac{\varepsilon_0}{m_e}$. This extreme case, corresponding to constant acceleration where the Unruh effect~\cite{unruh1976notes} becomes relevant, is beyond the scope of this paper.

As implied by Eq.~(\ref{eqDP}), changes in velocity are crucial for radiative emission. For relativistic particles, the effect of longitudinal acceleration on velocity change is relativistically suppressed, since the magnitude of the velocity cannot exceed the speed of light. Consequently, transverse acceleration dominates the radiation process for relativistic particles. Accordingly, we approximate the trajectory as that of an electron in an effective static magnetic field $B_{\mathrm{eff}}$, which generates the transverse acceleration at the given instant $t_0$, while neglecting the instantaneous longitudinal acceleration.
Explicit calculations can be carried out using the Frenet--Serret frame $\left(\mathbf{T}, \mathbf{N}, \mathbf{B}\right)$ at the instant $t_0$, where $\mathbf{T}$ is the unit vector in the direction of motion, $\mathbf{N}$ the unit vector in the direction of transverse acceleration, and $\mathbf{B} = \mathbf{T} \times \mathbf{N}$ the unit vector along the direction of the effective magnetic field. Using $\mathbf{T}$, $\mathbf{N}$, and $\mathbf{B}$ as basis vectors, the relevant classical trajectory can be approximated as
\begin{equation}
\boldsymbol{v}(t_0 + \Delta t) \approx v_0 \left( 1 - \frac{1}{2} \omega_{\mathrm{eff}}^2 \Delta t^2,\, \omega_{\mathrm{eff}} \Delta t,\, 0 \right),
\label{EqvExpansion}
\end{equation}
where $\varepsilon_0 \equiv \varepsilon(t_0)$ is the electron energy at the instant $t_0$, $v_0 \equiv |\boldsymbol{v}(t_0)|$ the magnitude of the velocity, and $\omega_{\mathrm{eff}} = \frac{B_{\mathrm{eff}}}{\varepsilon_0}$ the effective cyclotron frequency.
The contribution of the $\omega_{\mathrm{eff}} \Delta t$ term is of order $O\left(\frac{m_e}{\varepsilon_0}\right)$ (see Appendix\ref{Time Integration}), which means that contributions to the axion yield from higher-order terms in the expansion of Eq.~(\ref{EqvExpansion}) are relativistically suppressed.

In the Frenet-Serret frame, the axion emission direction can be specified by the polar angles as $\boldsymbol{n}=\left(\cos\alpha \cos \beta, \cos \alpha \sin \beta, \sin \alpha\right)$,
which can be expanded as
\begin{equation}
\begin{aligned} 
\boldsymbol{n}\approx \left(1-\frac{1}{2}\alpha^2-\frac{1}{2}\beta^2, \beta, \alpha\right).
\end{aligned} \label{nExpansion}
\end{equation}
The domains of the angular variables are $ \alpha\in (-\frac{\pi}{2},\frac{\pi}{2})$ and $\beta\in (-\pi,\pi)$, with the integration measure being $\cos\alpha \mathrm{d}\alpha \mathrm{d}\beta \approx \mathrm{d}\alpha \mathrm{d}\beta$.
The expansion is valid as the contributions from the $\alpha$ and $\beta$ terms are of the order $O(\frac{m_e}{\varepsilon_0})$ [see Appendix~\ref{Time Integration}].
Thus, the expansion in Eq.~(\ref{nExpansion}) is accurate to the order of $O(\frac{m^2_e}{\varepsilon^2_0})$.

\begin{figure*}
\centering
\includegraphics[width=1\textwidth]{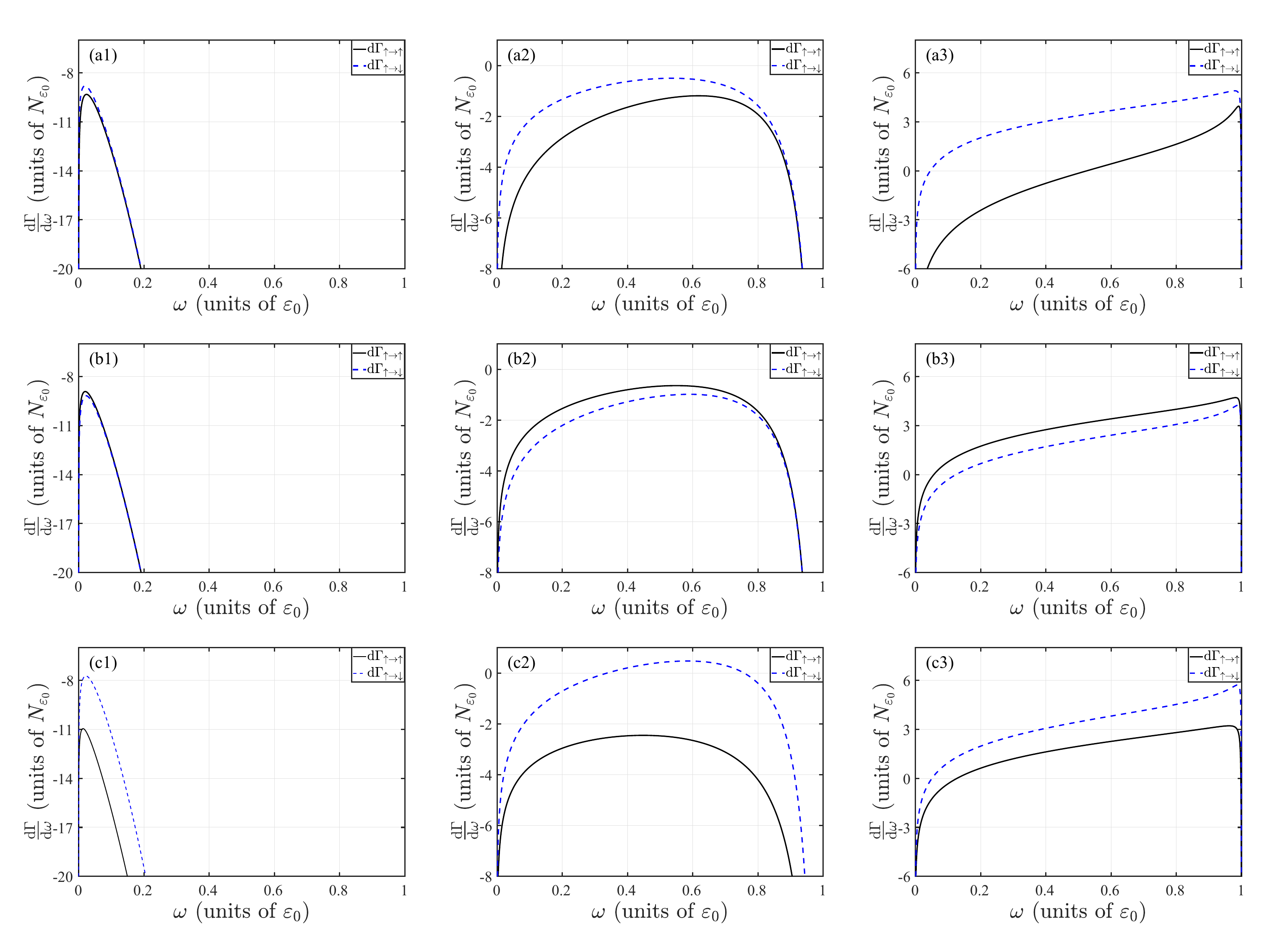}
\caption{The logarithm of the spin-resolved differential axion radiation rate $\frac{\mathrm{d} \Gamma}{\mathrm{d} \omega}$ (units of $N_{\varepsilon_0} = \frac{\sqrt{3} g_{ea}^2}{48 \pi^2} \frac{m_e^2}{\varepsilon_0^2}$) as a function of the emitted axion energy $\delta_e = \frac{\omega} { \varepsilon_0}$ for different electron spin channels. The nonlinear quantum parameters for the left, middle, and right columns are $\chi_e = 0.01$, $1$, and $100$, respectively. The initial spin polarizations for the top, middle, and bottom rows are $\boldsymbol{\zeta}_i = \mathbf{T}$, $\mathbf{N}$, and $\mathbf{B}$, respectively. The solid black line corresponds to the case where the final electron spin is parallel to the initial spin, while the dashed blue line corresponds to the antiparallel case. The initial electron energy is $1~\mathrm{GeV}$ and the axion mass is $\delta_m = 0$.
}
\label{FigdGamma}
\end{figure*}

Using the Taylor expansion of $\boldsymbol{v}$ [Eq.~(\ref{EqvExpansion})] and $\boldsymbol{n}$ [Eq.~(\ref{nExpansion})], the electron spin-resolved prefactor [Eq.~(\ref{fullEqQc})] reads as:
\begin{widetext} 
\begin{equation}
\begin{aligned}
&Q_c^*(t') Q_c(t) 
\approx \frac{1}{4} \frac{\omega^2}{\varepsilon_0^{\prime2}} 
\Biggl\{ 
   \biggl( -\frac{m_\phi^2}{\omega^2} \frac{\varepsilon_0^{\prime}}{\varepsilon_0} - \frac{1}{2} \omega_{\text{eff}}^2 \tau^2 \biggr) 
   (1 - \boldsymbol{\zeta}_i \cdot \boldsymbol{\zeta}_f) + i \frac{m_e}{\varepsilon_0} \omega_{\text{eff}} \tau (\boldsymbol{\zeta}_i - \boldsymbol{\zeta}_f) \cdot \mathbf{B} \\
&- \frac{1}{2} \omega_{\text{eff}}^2 \tau^2 (\boldsymbol{\zeta}_i \cdot \mathbf{B}) (\boldsymbol{\zeta}_f \cdot \mathbf{B})+ 2\frac{m_e^2}{\varepsilon_0^{2}} (\boldsymbol{\zeta}_i \cdot \mathbf{T}) (\boldsymbol{\zeta}_f \cdot \mathbf{T}) 
    + 2\alpha^2 (\boldsymbol{\zeta}_i \cdot \mathbf{B}) (\boldsymbol{\zeta}_f \cdot \mathbf{B}) + 2\beta^2 (\boldsymbol{\zeta}_i \cdot \mathbf{N}) (\boldsymbol{\zeta}_f \cdot \mathbf{N})
\Biggr\},
\label{EqPrefactor}
\end{aligned}
\end{equation}
\end{widetext} 
where Eq.~(\ref{vcn}) is used to simplify the calculation, and the prefactor is accurate to the order $O(\frac{m^2_e}{\varepsilon^2_0})$. 
To the same order of accuracy, the classical action $S_c$ reads as
\begin{equation}
S_c\approx -(a\tau^3+b\tau), \label{EqScA}
\end{equation}
where we use the abbreviations:
\begin{equation}
\left\{\begin{aligned}
a &=\frac{1}{24} \frac{\delta_e}{1-\delta_e} \varepsilon_0\omega_{\mathrm{eff}}^2, \\
b&=\frac{1}{2} \frac{\delta_e}{1-\delta_e} \frac{m_e^2}{\varepsilon_0} \Delta\left[1+\frac{1}{\Delta} \frac{\varepsilon_0^2}{m_e^2}\left(\alpha^2+\beta^2\right)\right]. \\
\end{aligned}
\right. 
\end{equation}
Here, $\delta_e = \frac{\omega}{\varepsilon_0}$ is the energy ratio between the emitted axion and the initial electron, $\delta_m =\frac{ m_\phi}{m_e}$ their mass ratio, and $\Delta = 1 + \delta_m^2 \frac{1 - \delta_e}{\delta_e^2}$ the axion mass correction factor.

With the prefactor Eq.~(\ref{EqPrefactor}) and the action Eq.~(\ref{EqScA}), the integration that needs to be evaluated can be summarized as
\begin{equation}
F = \int_{-\infty }^{\infty} \! \mathrm{d}\alpha\int_{-\infty }^{\infty} \! \mathrm{d}\beta \int_{-\infty }^{\infty} \! \mathrm{d}\tau \, f(\tau,\alpha,\beta) \, \exp \left[ -i \left(a\tau^3+b\tau \right) \right],\label{IntTable}
\end{equation}
where the integration range of the angles is extended to infinity, since the radiated axions are concentrated in the forward direction of the motion with $\alpha\sim \frac{m_e}{\varepsilon_0}$ and $\beta\sim \frac{m_e}{\varepsilon_0}$ [see Appendix~\ref{Time Integration}]. Thus extending the integration range of the angular variables to infinity does not affect the results.

The evaluated integrals are summarized in the integration table in Appendix~\ref{Time Integration}, with which the electron spin-resolved axion emission rate $\frac{\mathrm{d} \Gamma}{\mathrm{d} \omega}=\frac{\mathrm{d}^2 P}{\mathrm{d} t_0\mathrm{d} \omega} $ can be obtained as
\begin{widetext}
\begin{equation}
\begin{aligned}
\frac{\mathrm{d} \Gamma}{\mathrm{d} \omega}
&= \frac{\sqrt{3} g_{ea}^2}{48 \pi^2} \frac{m_e^2}{\varepsilon_0^2} \frac{\delta_e^2}{1-\delta_e}
\Biggl\{
(1 - \boldsymbol{\zeta}_i \cdot \boldsymbol{\zeta}_f) \Biggl[
\Delta K_{\frac{2}{3}}(z_q)
- \frac{1}{2} \frac{\delta_m^2 (1-\delta_e)}{\delta_e^2} \int_{z_q}^{+\infty} \! \mathrm{d}t \, K_{\frac{1}{3}}(t)
\Biggr]
\\
&\quad + \frac{1}{2} (\boldsymbol{\zeta}_i \cdot \mathbf{B}) (\boldsymbol{\zeta}_f \cdot \mathbf{B})
\Delta \left( K_{\frac{2}{3}}(z_q) - \int_{z_q}^{+\infty} \! \mathrm{d}t \, K_{\frac{1}{3}}(t) \right) + (\boldsymbol{\zeta}_i - \boldsymbol{\zeta}_f) \cdot \mathbf{B} \, \Delta^{1/2} \, K_{\frac{1}{3}}(z_q)
\\
&\quad + (\boldsymbol{\zeta}_i \cdot \mathbf{N}) (\boldsymbol{\zeta}_f \cdot \mathbf{N})
\Delta \left( \frac{3}{2} K_{\frac{2}{3}}(z_q) - \frac{1}{2} \int_{z_q}^{+\infty} \! \mathrm{d}t \, K_{\frac{1}{3}}(t) \right)  + (\boldsymbol{\zeta}_i \cdot \mathbf{T}) (\boldsymbol{\zeta}_f \cdot \mathbf{T})
\int_{z_q}^{+\infty} \! \mathrm{d}t \, K_{\frac{1}{3}}(t)
\Biggr\},
\end{aligned}\label{radiationProbability}
\end{equation}
\end{widetext}
where $z_q = \frac{2}{3\chi_e} \frac{\delta_e}{1-\delta_e} \Delta^{\frac{3}{2}}$ is the argument of the modified Bessel function, and $\chi_e=\frac{B_{\mathrm{eff}}\varepsilon_0}{m_e^3}$ the nonlinear quantum parameter.
According to the definition of the effective magnetic field strength, we have $\chi_e \approx \frac{g_e \sqrt{(F_{\mu\nu} p^\nu)^2}}{m_e^3}$. After averaging over the initial electron spin and summing over the final spin, the radiation rate in Eq.~(\ref{radiationProbability}) is consistent with the spin-unresolved results from Refs.~\cite{dillon2018alp,king2019axion}.
%After averaging over the electron spin, Eq.~(\ref{radiationProbability}) is consistent with Refs.~\cite{dillon2018alp,king2019axion}, which are obtained based on the Volkov wave function approach.

The implicit assumption behind the radiation rate in Eq.~(\ref{radiationProbability}) is that the variation of the laser field can be neglected during the radiation formation time, i.e., the LCFA applies. Since the formation time for axion emission is $\tau\sim\frac{m_e}{B_{\mathrm{eff}}}$ [see Appendix~\ref{Time Integration}], the LCFA is valid when the classical nonlinear parameter is much larger than unity, i.e., $a_0 \gg 1$.

\begin{figure*}
\centering
\includegraphics[width=1\textwidth]{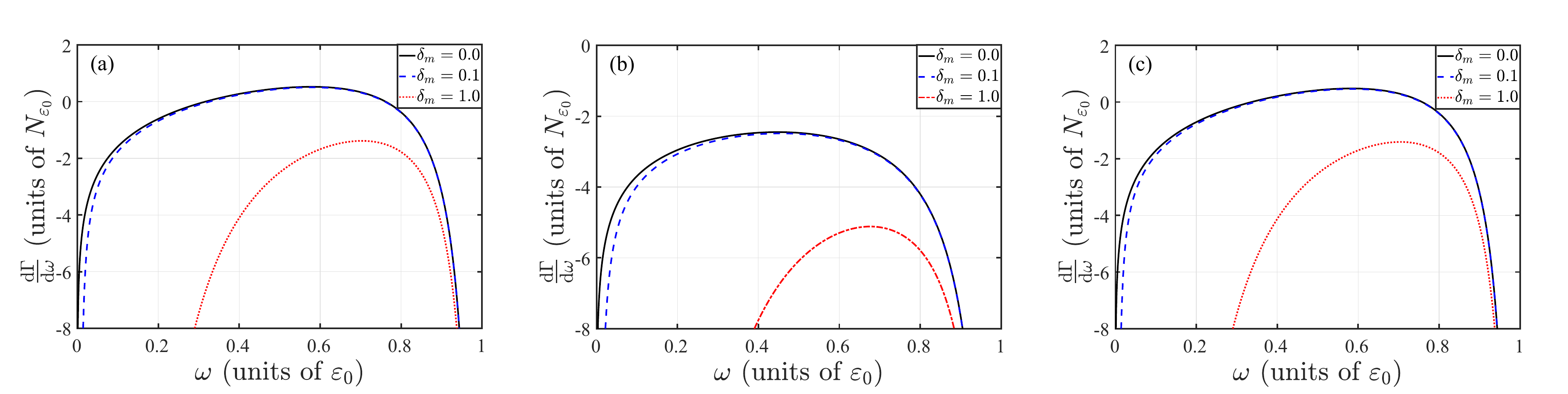}
\caption{
The logarithm of the spin-resolved differential axion radiation rate $\frac{\mathrm{d} \Gamma}{\mathrm{d} \omega}$ (units of $N_{\varepsilon_0} = \frac{\sqrt{3} g_{ea}^2}{48 \pi^2} \frac{m_e^2}{\varepsilon_0^2}$) as a function of the emitted axion energy $\delta_e = \frac{\omega }{ \varepsilon_0}$ for nonzero axion mass. The initial electron energy is $1~\mathrm{GeV}$, with initial spin polarization $\boldsymbol{\zeta}_i = \mathbf{B}$. The final electron spin direction is (a) summed over spin-up and spin-down, (b) spin-up, and (c) spin-down. The solid black line corresponds to $\delta_m = 0.0$, the blue dashed line to $\delta_m = 0.1$, and the red dotted line to $\delta_m = 1.0$. The nonlinear quantum parameter is $\chi_e = 1$.
}
\label{FigMa}
\end{figure*}

\section{Properties of the Radiation}\label{SectionRadiation}
With the derived axion production rate $\frac{\mathrm{d} \Gamma}{\mathrm{d} \omega}$, we explore the properties of nonlinear Compton-like scattering with the electron spin polarization resolved in this section.

\subsection{Differential Spectrum of Radiation}\label{SectionDRadiation}
Constraints from cosmology suggest that the axion mass lies in the range $10^{-11} < \frac{m_\phi}{m_e} < 10^{-6}$ \cite{kim1987light}. Thus, we first consider the properties of radiation in the limit where the axion is massless and then consider corrections due to the finite axion mass.

The axion emission rate is significant when the argument of the modified Bessel functions in Eq.~(\ref{radiationProbability}) is of order unity. This condition yields the characteristic energy of the emitted axion as
\begin{equation}
\omega_c = \frac{\chi_e}{\frac{2}{3} + \chi_e} \varepsilon_0,
\label{omegac}
\end{equation}
which is analogous to the characteristic energy of photon emission~\cite{berestetskii2012quantum}. Given this characteristic energy loss, the condition $\chi_e\ll \frac{\varepsilon_0}{m_e}$ must be satisfied to ensure that the electron motion remains relativistic ($\frac{m_e}{\varepsilon'_0}\ll1$) after axion emission.

The differential axion emission rate $\frac{\mathrm{d} \Gamma}{\mathrm{d} \omega}$ is dramatically different from that of photon emission. For a fixed value of $\chi_e$, the low-energy behavior of Eq.~(\ref{radiationProbability}) scales as $\omega^{\frac{4}{3}}$ and thus exhibits no infrared divergence even in the massless limit.
Figure~\ref{FigdGamma} displays the electron spin-resolved differential rate of axion emission for different values of the nonlinear quantum parameter. From left to right, the columns correspond to $\chi_e = 0.01$, $1$, and $100$, representing the transition from the weak- to the strong-field limit.
From top to bottom rows, the initial spin directions are parallel to the direction of motion $\mathbf{T}$, the direction of acceleration $\mathbf{N}$, and the direction of the effective magnetic field $\mathbf{B}$, respectively.
For all spin channels, the characteristic energy of the radiation is qualitatively well described by Eq.~(\ref{omegac}). With increasing nonlinear quantum parameters, both the characteristic radiation energy and the total yield increase.
Compared to photon emission, axion production displays a more pronounced sensitivity to the electron spin polarization [see also Eqs.~(\ref{weak_axion}) and (\ref{weak_photon})]. 
Regardless of the value of $\chi_e$, when the initial spin is parallel with the $\mathbf{T}$ or $\mathbf{B}$ directions, the dashed blue lines are higher than the solid black line, indicating that the spin-flip process is dominant. In contrast, when the initial spin is parallel with the $\mathbf{N}$ direction, the solid black lines are higher than the dashed blue lines, indicating that the spin-preserving process dominates. This spin dependence of axion emission thus contrasts sharply with photon emission, where the spin-preserving process is preferred.

The above analysis proceeds in the massless limit. To assess the effects of a nonzero axion mass, we present in Fig.~\ref{FigMa} the electron spin-resolved axion radiation rate $\frac{\mathrm{d} \Gamma}{\mathrm{d} \omega}$ for axion masses $\delta_m = 0.0$ (solid black line), $\delta_m = 0.1$ (blue dashed line), and $\delta_m = 1.0$ (red dotted line). 
Panel (a) shows the results with the final electron spin summed over both spin-up and spin-down states; panel (b) shows the spin-up channel; and panel (c) shows the spin-down channel. It is evident that the axion mass modulates the relative strengths of the spin channels, yet the dominant spin channel remains unchanged.
The primary effect of the axion mass correction is that increasing the axion mass exponentially suppresses the radiation rate [see also Fig.~\ref{FigTot}] due to the finite energy gap introduced by the nonvanishing axion mass, while the peak position remains relatively insensitive to this parameter. Meanwhile, the energy distribution becomes narrower as the axion mass increases.
The correction becomes significant when the axion mass correction factor $\Delta = 1 + \delta_m^2 \frac{1 - \delta_e}{\delta_e^2}$ deviates from unity, leading to the criterion:
\begin{equation}
\delta_m^2 > \frac{3}{2} \frac{\chi_e^2}{\frac{2}{3} + \chi_e}.
\label{omegac2}
\end{equation}
Thus, given the mass range constraints from cosmology, the correction due to the finite axion mass is important only in the weak-field regime, i.e., $\chi_e<\frac{m_\phi}{m_e}$.
The nonlinear quantum parameter is $\chi_e=1$ in Fig.~\ref{FigMa}, thus the dashed blue lines [$\delta_m=0.1$] only show minor suppression compared to the solid black lines [$\delta_m=0$], while the dotted dashed red line [$\delta_m=1$] shows significant suppression, consistent with the criterion Eq.~(\ref{omegac2}).

%Remarkably, the spin dependence reveals a pronounced asymmetry: while panels (a) and (c) exhibit similar radiation patterns with peak intensities reaching log(dΓ) ≈ 0.5 for massless axions, panel (b) shows significantly suppressed radiation rates with peak values around log(dΓ) ≈ -2.5, indicating that transitions to final spin-up states are substantially less probable than those to spin-down states or averaged configurations.

\subsection{Integrated Axion Spectrum}\label{SectionDRadiation}

\begin{figure*}
\centering
\includegraphics[width=1\textwidth]{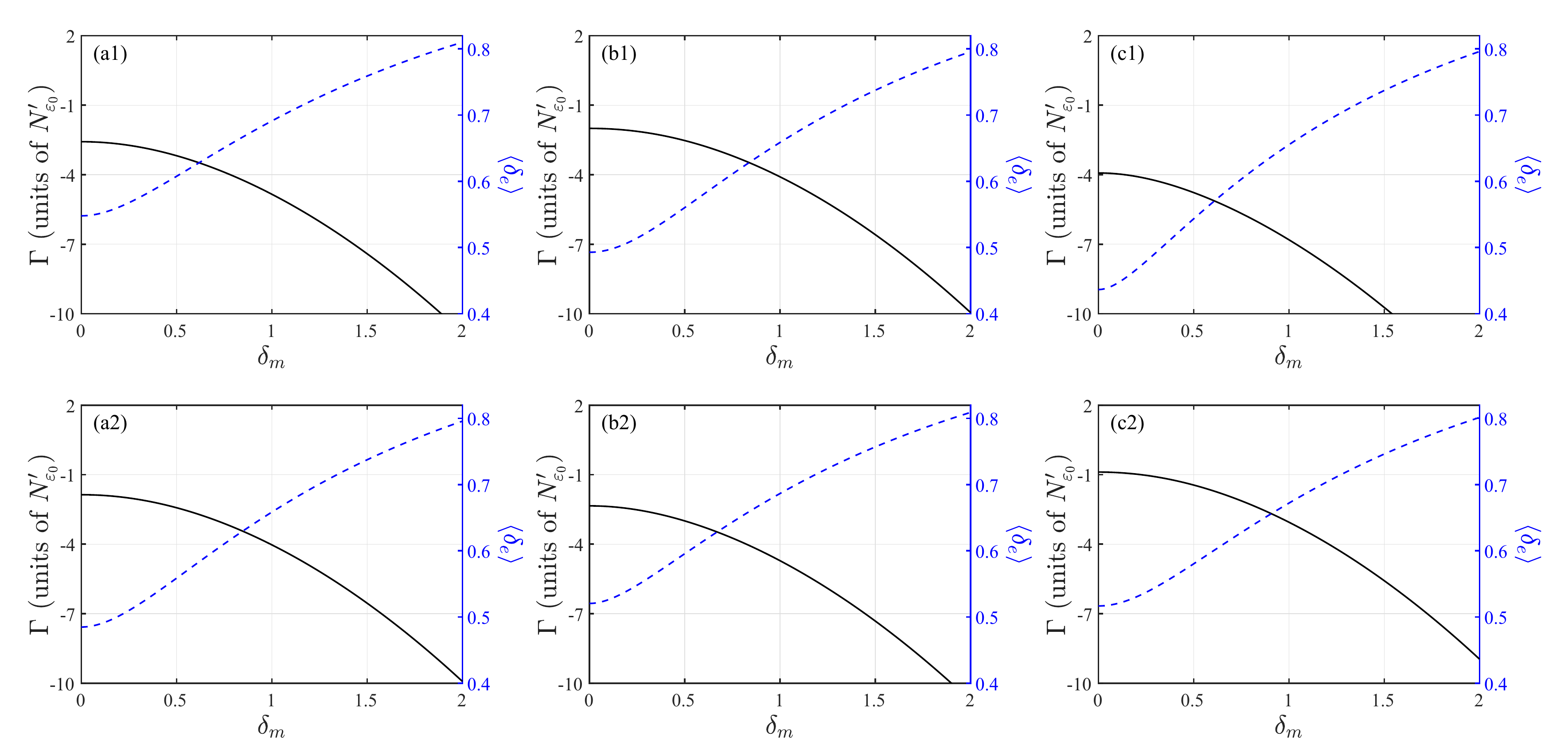}
\caption{
The effect of finite axion mass on the spin-resolved total radiation rate $\Gamma$ (solid black line) and the average emitted axion energy $\delta_e = \omega / \varepsilon_0$ (blue dashed line). The total rate $\Gamma$ is given in units of $N'_{\varepsilon_0} = \frac{\sqrt{3} g_{ea}^2}{48 \pi^2} \frac{m_e^2}{\varepsilon_0}$ and is plotted on a logarithmic scale. The initial spin polarization is $\boldsymbol{\zeta}_i = \mathbf{T}$ (left column), $\boldsymbol{\zeta}_i = \mathbf{N}$ (middle column), and $\boldsymbol{\zeta}_i = \mathbf{B}$ (right column). The final spin polarization is parallel (top row) or antiparallel (bottom row) to the initial spin. The nonlinear quantum parameter is $\chi_e = 1$, and the initial electron energy is $1~\mathrm{GeV}$.
}
\label{FigTot}
\end{figure*}

As shown in Fig.~\ref{FigMa}, a nonvanishing axion mass not only suppresses the axion production probability but also narrows the spectral distribution, leading to an increase in the average energy of the emitted axions.
Figure~\ref{FigTot} provides a detailed investigation of how a finite axion mass affects both the total radiation yield and the average emitted energy of axions for different electron spin configurations. In the figure, the solid black lines represent the logarithm of the integrated radiation rate $\Gamma$ (in units of $N'_{\varepsilon_0} = \frac{\sqrt{3} g_{ea}^2}{48 \pi^2} \frac{m_e^2}{\varepsilon_0}$), while the blue dashed lines show the average emitted axion energy (in units of $\varepsilon_0$).
The columns correspond to different initial electron spin polarizations: (a) $\boldsymbol{\zeta}_i = \mathbf{T}$, (b) $\boldsymbol{\zeta}_i = \mathbf{N}$, and (c) $\boldsymbol{\zeta}_i = \mathbf{B}$, while the rows distinguish final electron spins that are (top row) parallel and (bottom row) antiparallel to the initial spin orientation. 
In all cases, the black solid curves exhibit a consistent, monotonic decrease in radiation probability as the axion mass increases, indicating strong mass suppression of axion production. Conversely, the blue dashed curves show a systematic increase in the average emitted axion energy with increasing mass, regardless of the initial spin polarization direction or whether the final electron spin is parallel or antiparallel to the initial state.
The spin-flip process has a higher emission probability when the initial spin is parallel to $\mathbf{T}$ and $\mathbf{B}$ [panels (a2) and (c2)], whereas the spin-preserving process is favored when the initial spin is parallel to $\mathbf{N}$ [panel (b1)]. While the spin-flip process results in a slightly higher average emitted axion energy when the initial spin is parallel to $\mathbf{N}$ and $\mathbf{B}$ [panels (b2) and (c2)], the spin-preserving process leads to a higher average energy when the initial spin is parallel to $\mathbf{T}$ [panel (a1)].
The suppression due to the nonvanishing axion mass can be estimated from Eq.~(\ref{radiationProbability}), which gives
\begin{equation}
\begin{aligned}
\Gamma \sim \exp\left[ -\frac{\delta_m^2}{\chi_e^2} \left( \frac{2}{3} + \chi_e \right) \right],
\end{aligned}
\end{equation}
indicating that the mass suppression effect is important in the weak-field regime $\chi_e<\delta_m$.

\begin{figure*}
\centering
\includegraphics[width=1\textwidth]{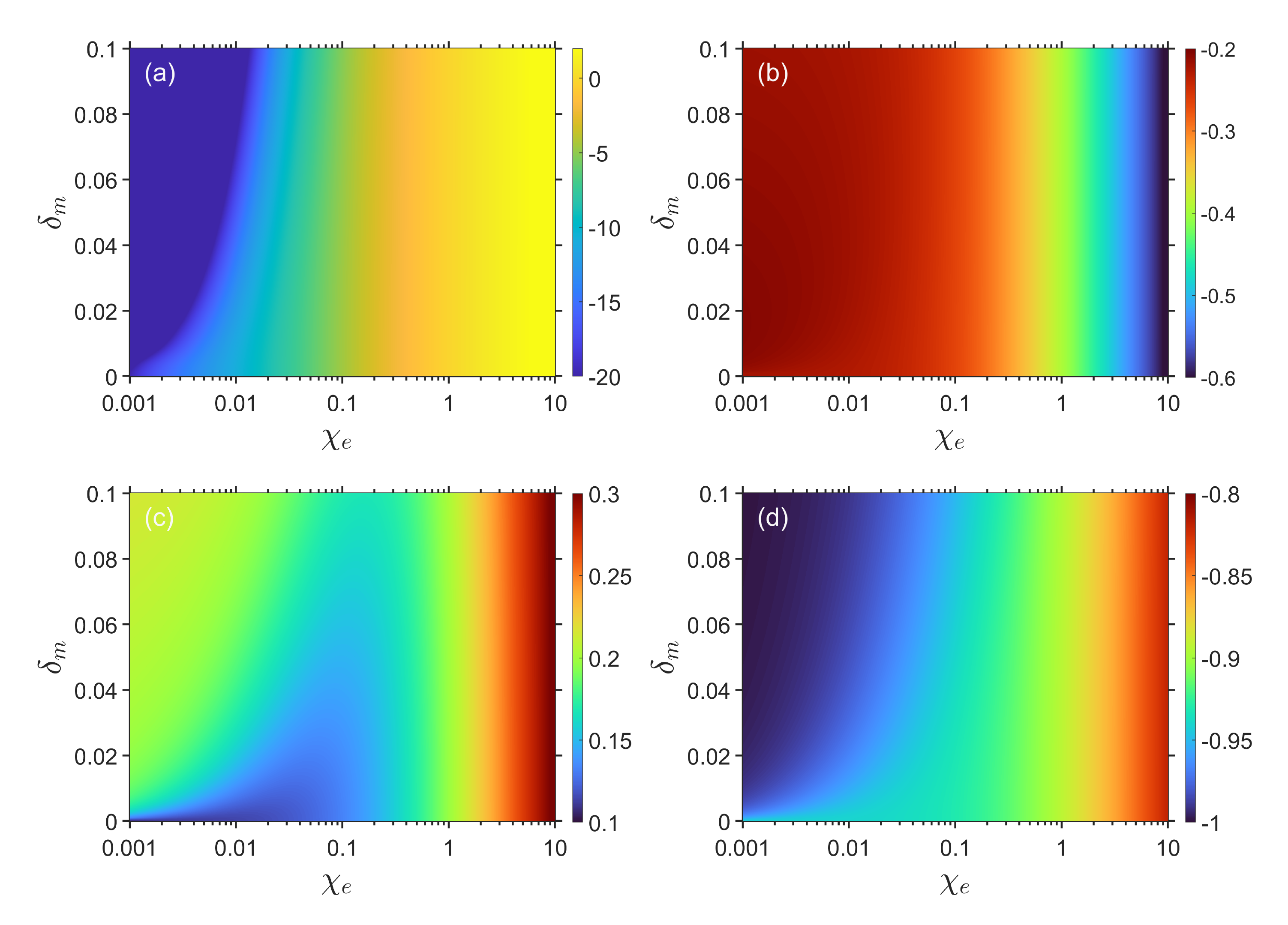}
\caption{
Dependence of axion production on the nonlinear quantum parameter $\chi_e$ and axion mass $\delta_m$. 
(a) The spin-summed axion production rate is shown in units of $N'_{\varepsilon_0} = \frac{\sqrt{3} g_{ea}^2}{48 \pi^2} \frac{m_e^2}{\varepsilon_0}$ and plotted on a logarithmic scale.
(b-d) The asymmetry between spin-preserved and spin-flip rates, $Asy=\frac{\Gamma_{\uparrow\to\uparrow} - \Gamma_{\uparrow\to\downarrow}}{\Gamma_{\uparrow\to\uparrow} + \Gamma_{\uparrow\to\downarrow}}$ is displayed for initial spin polarizations: (b) $\boldsymbol{\zeta}_i = \mathbf{T}$, (c) $\boldsymbol{\zeta}_i = \mathbf{N}$, and (d) $\boldsymbol{\zeta}_i = \mathbf{B}$. The initial electron energy is $1~\mathrm{GeV}$.
}
\label{FigTotMass}
\end{figure*}

Figure~\ref{FigTotMass} provides a comprehensive exploration of how the total axion production rates and spin asymmetries depend on both the nonlinear quantum parameter $\chi_e$ and the axion mass $\delta_m$ for various initial spin configurations. Panel~(a) shows the logarithm of the total axion production rate $\Gamma$, revealing a dramatic enhancement of more than 20 orders of magnitude as $\chi_e$ increases from $0.001$ to $10$. In contrast, the axion mass $\delta_m$ has a suppressive effect, which is most pronounced at low field strengths [see Eq.~(\ref{omegac2})].
Panels~(b)--(d) examine the spin asymmetry parameter 
\begin{equation}
Asy=\frac{\Gamma_{\uparrow\to\uparrow} - \Gamma_{\uparrow\to\downarrow}}{\Gamma_{\uparrow\to\uparrow} + \Gamma_{\uparrow\to\downarrow}},
\end{equation}
 which quantifies the degree of final spin polarization along the quantization axis defined by the initial spin direction $\boldsymbol{\zeta}_i$. 
%which quantifies the difference between spin-preserving and spin-flipping transitions 
The cases shown correspond to initial spin aligned with: (b) $\boldsymbol{\zeta}_i = \mathbf{T}$, (c) $\boldsymbol{\zeta}_i = \mathbf{N}$, and (d) $\boldsymbol{\zeta}_i = \mathbf{B}$, respectively.
Panel~(b) shows that the asymmetry decreases from approximately $-0.25$ to $-0.6$ as $\chi_e$ increases, indicating that spin-flipping processes become increasingly dominant in stronger fields when the initial spin polarization is parallel to the direction of electron motion. The axion mass plays only a minor role in this case.
Panel~(c) exhibits a more complex dependence on $\chi_e$. In the weak-field limit, when $\chi_e < \delta_m$, the asymmetry decreases as the field strength increases, while in the intense-field regime, the asymmetry increases with $\chi_e$. At fixed $\chi_e$, the asymmetry increases with $\delta_m$. The spin-preserving processes are always favored when the initial spin polarization is parallel to the direction of acceleration.
Panel~(d) demonstrates that the asymmetry values increase from approximately $-1.0$ to $-0.8$ as $\chi_e$ increases, indicating that although spin-flipping processes are always favored when the initial spin polarization is parallel to the direction of the effective magnetic field, their dominance diminishes slightly as the field strength increases.

It is possible to investigate the asymptotic behavior of axion emission analytically in the massless axion limit.
After integrating over the axion energy $\delta_e$ in Eq.~(\ref{radiationProbability}), the integrated radiation rate in the weak-field limit $\chi_e \ll 1$ is given by
\begin{equation}
\begin{aligned}
\Gamma \approx \frac{5\sqrt{3}\,g_{ea}^2}{128\pi} \frac{m_e^2}{\varepsilon_0} \chi_e^3 &\Biggl[
1 - \frac{2}{9} (\boldsymbol{\zeta}_i \cdot \mathbf{T}) (\boldsymbol{\zeta}_f \cdot \mathbf{T})
+ \frac{1}{9} (\boldsymbol{\zeta}_i \cdot \mathbf{N}) (\boldsymbol{\zeta}_f \cdot \mathbf{N}) \\
&- \frac{8}{9} (\boldsymbol{\zeta}_i \cdot \mathbf{B}) (\boldsymbol{\zeta}_f \cdot \mathbf{B})
+ \frac{8\sqrt{3}}{15} (\boldsymbol{\zeta}_i - \boldsymbol{\zeta}_f) \cdot \mathbf{B}
\Biggr],
\end{aligned}
\label{weak_axion}
\end{equation}
where the identity
\begin{equation}
\boldsymbol{\zeta}_i \cdot \boldsymbol{\zeta}_f =
(\boldsymbol{\zeta}_i \cdot \mathbf{T})(\boldsymbol{\zeta}_f \cdot \mathbf{T})
+ (\boldsymbol{\zeta}_i \cdot \mathbf{N})(\boldsymbol{\zeta}_f \cdot \mathbf{N})
+ (\boldsymbol{\zeta}_i \cdot \mathbf{B})(\boldsymbol{\zeta}_f \cdot \mathbf{B})
\end{equation}
is used.
The axion radiation exhibits significant differences compared to photon emission. For comparison, the probability for photon emission in the same regime is
\begin{equation}
\begin{aligned}
\Gamma_\gamma &\approx \frac{\sqrt{3}\,g_e^2}{24\pi} \frac{m_e^2}{\varepsilon_0}
\Biggl\{  \frac{5}{2}\chi_e \bigl(1 + \boldsymbol{\zeta}_i \cdot \boldsymbol{\zeta}_f\bigr)
- \frac{3}{4} \chi_e^2 \bigl( \mathbf{B} \cdot \boldsymbol{\zeta}_f + \mathbf{B} \cdot \boldsymbol{\zeta}_i \bigr) \\
&\quad + \frac{15}{8} \chi_e^3 \biggl[
1 + \frac{2}{9} (\boldsymbol{\zeta}_i \cdot \mathbf{T}) (\boldsymbol{\zeta}_f \cdot \mathbf{T})
- \frac{8\sqrt{3}}{15} \mathbf{B} \cdot \boldsymbol{\zeta}_f
\biggr] \Biggr\}.
\end{aligned}
\label{weak_photon}
\end{equation}
Thus, in the weak-field limit, the axion production rate is proportional to $\chi_e^3$, whereas the photon emission rate is proportional to $\chi_e$. While spin-preserving processes dominate in photon emission, axion emission exhibits a strong spin dependence: the spin-flip process is favored when the initial spin is along the $\mathbf{T}$ or $\mathbf{B}$ directions, whereas the spin-preserving process is favored when the initial spin is along the $\mathbf{N}$ direction.

The significant difference between axion radiation and photon emission persists in the strong-field limit $\chi_e \gg 1$, where the axion emission rate is given by
\begin{equation}
\begin{aligned}
\Gamma &\approx 0.1634\, \frac{g_{ea}^2}{4\pi} \frac{m_e^2}{\varepsilon_0} \chi_e^{\frac{2}{3}} 
\Biggl[ 
1 - (\boldsymbol{\zeta}_i \cdot \mathbf{T}) (\boldsymbol{\zeta}_f \cdot \mathbf{T}) \\
&\qquad + \frac{1}{2} (\boldsymbol{\zeta}_i \cdot \mathbf{N}) (\boldsymbol{\zeta}_f \cdot \mathbf{N}) - \frac{1}{2} (\boldsymbol{\zeta}_i \cdot \mathbf{B}) (\boldsymbol{\zeta}_f \cdot \mathbf{B})
\Biggr],
\end{aligned}
\label{strong_axion}
\end{equation}
while the photon emission rate is
\begin{equation}
\begin{aligned}
\Gamma_\gamma &\approx 0.6259\, \frac{g_e^2}{4\pi} \frac{m_e^2}{\varepsilon_0} \chi_e^{\frac{2}{3}} \biggl\{ 1 + \boldsymbol{\zeta}_i \cdot \boldsymbol{\zeta}_f \\
&\qquad + \frac{1}{6} \left[ 1 + (\boldsymbol{\zeta}_i \cdot \mathbf{T}) (\boldsymbol{\zeta}_f \cdot \mathbf{T}) \right] \biggr\}.
\end{aligned}
\label{strong_photon}
\end{equation}
In the strong-field limit, both axion and photon production rates scale as $\chi_e^{\frac{2}{3}}$. However, the spin dependence remains pronounced. Similar to the weak-field case, for axion emission, the spin-flip process is favored when the initial spin is aligned with either the $\mathbf{T}$ or $\mathbf{B}$ directions, whereas the spin-preserving process is favored when the initial spin is along the $\mathbf{N}$ direction. In contrast, for photon emission, the spin-preserving process is favored when the initial spin is along the $\mathbf{T}$ direction.

\subsection{Polarization Induced by Axion Emission}
Up to this point, our discussion has focused on the case where the initial electron spin is polarized. However, an initially unpolarized electron can acquire a nonzero polarization after an axion emission. 

\begin{figure}
\centering
\includegraphics[width=0.5\textwidth]{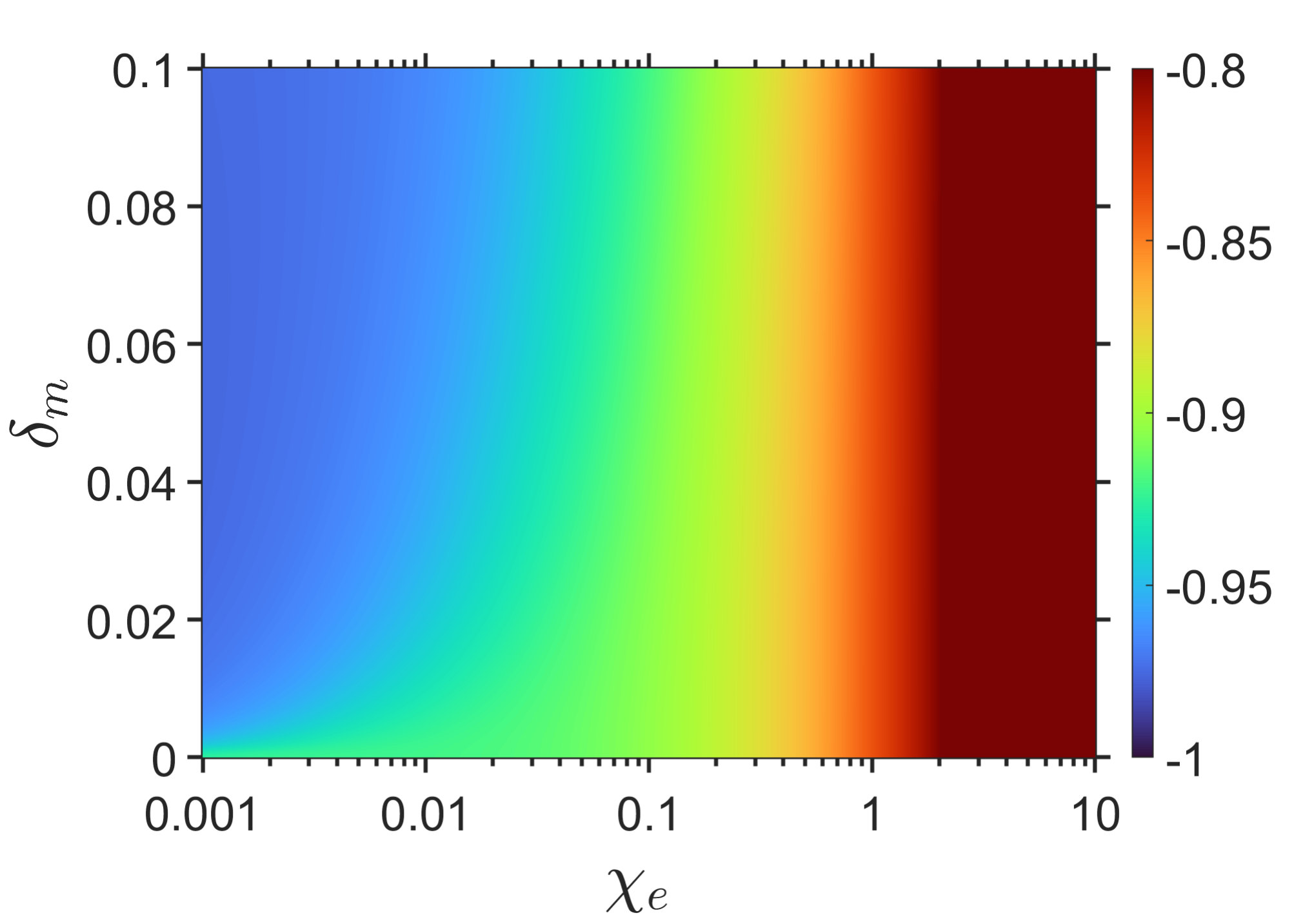}
\caption{
The net electron polarization $\left\langle \zeta_B \right\rangle = \frac{\Gamma_{\uparrow\to\uparrow} + \Gamma_{\downarrow\to\uparrow} - \Gamma_{\uparrow\to\downarrow} - \Gamma_{\downarrow\to\downarrow}}{\Gamma_{\uparrow\to\uparrow} + \Gamma_{\downarrow\to\uparrow} + \Gamma_{\uparrow\to\downarrow} + \Gamma_{\downarrow\to\downarrow}}$ induced by axion radiation as a function of the nonlinear quantum parameter $\chi_e$ and axion mass $\delta_m$. The initial electron is unpolarized with an energy of $1~\mathrm{GeV}$.
The spin quantization axis is taken along the $\mathbf{B}$ direction.
}
\label{NetF}
\end{figure}

Averaging over the initial spin directions in Eq.~(\ref{radiationProbability}), it is evident that the net final electron polarization persists only along the $\mathbf{B}$ direction, vanishing in the $\mathbf{T}$ and $\mathbf{N}$ directions. 
Figure~\ref{NetF} shows the final spin polarization as a function of the nonlinear quantum parameter $\chi_e$ and the axion mass $\delta_m$:
\begin{equation}
\left\langle \zeta_B \right\rangle = \frac{\Gamma_{\uparrow\to\uparrow} + \Gamma_{\downarrow\to\uparrow} - \Gamma_{\uparrow\to\downarrow} - \Gamma_{\downarrow\to\downarrow}}{\Gamma_{\uparrow\to\uparrow} + \Gamma_{\downarrow\to\uparrow} + \Gamma_{\uparrow\to\downarrow} + \Gamma_{\downarrow\to\downarrow}},
\end{equation}
where the spin quantization axis is taken along the $\mathbf{B}$ direction, and the energy of the initially unpolarized electron is $1~\mathrm{GeV}$.
The color scale in Fig. \ref{NetF} ranges from $-1.0$ to $-0.8$, indicating that axion emission consistently induces a negative net spin polarization along the $\mathbf{B}$ direction. 
In the weak-field regime ($\chi_e \ll 1$), the polarization approaches $-\frac{8\sqrt{3}}{15}$, consistent with Eq.~(\ref{weak_axion}). As the nonlinear quantum parameter increases to $\chi_e = 10$, the net polarization becomes less negative, reaching approximately $-0.8$. Furthermore, Eq.~(\ref{strong_axion}) implies that in the strong-field regime ($\chi_e \gg 1$), the polarization gradually tends to zero. This decay is slow, however, since the contribution from the omitted term $(\boldsymbol{\zeta}_i - \boldsymbol{\zeta}_f) \cdot \mathbf{B}$ in Eq.~(\ref{strong_axion}) is proportional to $\chi_e^{\frac{1}{3}}$. The net polarization also increases with greater axion mass, which becomes noticeable when $\delta_m > \chi_e$.

The qualitative features of Fig.~\ref{NetF} are similar to those of Fig.~\ref{FigTotMass}(d), indicating that the final electron spin tends to align anti-parallel to the effective magnetic field direction. The pronounced spin dependence of the axion radiation rate suggests the possibility of controlling the axion emission direction via the electron spin~\cite{JointSub}.

\section{Conclusion}\label{section_Conclusion}

The ALP production via nonlinear Compton-like scattering processes is investigated using the Baier-Katkov operator method with electron spin channels resolved. Our approach explicitly exploits the semiclassical nature of strong-field QED through the eikonal spinor wave function, neglecting higher-order terms in powers of $\frac{m_e}{\varepsilon_0}$. Several reasonable assumptions are made, summarized below along with their physical implications:
\begin{itemize}
\item The laser field is intense such that the classical nonlinear parameter $a_0 \gg 1$, which ensures that LCFA are valid.
\item The electron is relativistic before axion emission
($\frac{m_e} {\varepsilon_0} \ll 1$), 
as set by the initial conditions.
\item The electron motion remains relativistic after axion emission ($\frac{m_e}{\varepsilon'_0} \ll 1$),
 which requires that $\chi_e \ll \frac{\varepsilon_0}{m_e}$.
\item The emitted axions are also relativistic 
($\frac{m_\phi}{\omega} \ll 1$), 
which requires $\chi_e\gg\frac{m_\phi}{\varepsilon_0}$.
\item The electron dynamics are semiclassical in the sense that the quantized Landau energy level spacing in the effective magnetic field is negligible, which requires $\chi_e < \frac{\varepsilon_0^3} {m_e^3}$. This condition is trivially satisfied here, as the electron motion is relativistic both before and after axion emission. Since $\frac{\varepsilon_0}{m_e} \gg 1$, it follows that $\chi_e \ll \frac{\varepsilon_0}{m_e} < \frac{\varepsilon_0^3} {m_e^3}$.
\item The electron dynamics are also semiclassical in the sense that the electron can be treated as a point particle in the intense laser field-an implicit assumption of the semiclassical Monte Carlo model-which requires $\varepsilon_0 \lambda_L \gg 1$.
\item The effective electric field and the magnetic field should satisfy the relationship $\frac{E_{\mathrm{eff}}}{B_{\mathrm{eff}}}< \frac{\varepsilon_0}{m_e}$, which excludes situations of constant acceleration where the Unruh effect plays a role.
\end{itemize}
The above assumptions are readily achieved with typical intense strong-field QED parameters, for example, with an initial electron energy of $1~\mathrm{GeV}$ and a laser intensity of $10^{22}~\mathrm{W/cm}^2$ at a near-infrared wavelength.

The spin-resolved ALP production rate [Eq.~(\ref{radiationProbability})], derived under the LCF approximation, is a key result of our work, revealing dramatic differences in behavior compared to photon emission. The derived spin-resolved axion emission rates are also well-suited for incorporating axion dynamics into semiclassical Monte Carlo models of strong-field QED.
In our jointly submitted paper~\cite{JointSub}, we have demonstrated that electron spin can serve as a degree of freedom to control axion dynamics and may improve the sensitivity of laboratory-based axion searches, highlighting the potential of strong-field QED as a promising tool for exploring axion physics.
While in the weak-field limit ($\chi_e \ll 1$), the axion yield via the nonlinear Compton-like process is not significant [$\Gamma \propto \chi_e^3$], in the strong-field regime ($\chi_e \gg 1$), the yield is enhanced by several orders of magnitude [$\Gamma \propto \chi_e^{\frac{2}{3}}$]. Moreover, for $\chi_e > 1$, electron-positron pairs can be generated via the nonlinear Breit--Wheeler process; the resulting pairs can subsequently emit axions via the nonlinear Compton-like process. Thus, a large number of axions are expected to be produced in this regime, providing a promising experimental source for axion detection.

%In the QED cascade, a large number of photons and electron-positron pairs can be generated.
%The generated gamma photons can convert to axions in intense laser fields. Meanwhile, the charged particles can emit axions via nonlinear Compton-like scattering.
%As a consequence, a large number of axions are produced, which provide a promising experimental source for detection.

%To get the last line, use the fact that $\frac{m}{\varepsilon}$ is a relativisically small quantity, and the recoil effect $\frac{\omega}{\varepsilon}$ due to the emission of the axion is a small quatntity.

{\it Acknowledgement:}
P.-L. H. thanks the sponsorship from Yangyang Development Fund.
\vspace{10pt}

\appendix

\section{Integration Formula}\label{Time Integration}
This appendix deals with the integration in Eq.~(\ref{IntTable}) of the main text:
\begin{equation}
F = \int_{-\infty }^{\infty} \mathrm{d}\alpha \int_{-\infty }^{\infty}  \mathrm{d}\beta \int_{-\infty }^{\infty} \mathrm{d}\tau\, f(\tau,\alpha,\beta) \exp \left[ -i \left(a\tau^3+b\tau \right) \right].
\end{equation}

For the integration over the time variable, we use the integral representation of the modified Bessel function:
\begin{equation}
\begin{aligned}
&\int_{-\infty}^{+\infty} \! \mathrm{d}t \, \exp\left[-i(a t^3 + b t)\right]
= \frac{2}{\sqrt{3}} \sqrt{\frac{b}{3a}}\, K_{\frac{1}{3}}(z_q) ,
\\  
&\int_{-\infty}^{+\infty} \! \mathrm{d}t \, t \, \exp\left[-i(a t^3 + b t)\right]
= \frac{-2i}{\sqrt{3}} \frac{b}{3a}\, K_{\frac{2}{3}}(z_q) ,
\\  
&\int_{-\infty}^{+\infty} \! \mathrm{d}t \, t^2 \, \exp\left[-i(a t^3 + b t)\right] 
= \frac{-2}{\sqrt{3}} \left(\frac{b}{3a}\right)^{\frac{3}{2}}\, K_{\frac{1}{3}}(z_q),
\end{aligned}\label{A1}
\end{equation}
where the parameters are defined as $a =\frac{1}{24} \frac{\omega}{\varepsilon'_0} \varepsilon_0\omega_{\text{eff}}^2$, $\Delta 
= 1 + \frac{m_\phi^2}{m_e^2} \frac{\varepsilon \varepsilon'}{\omega^2}$, $b=\frac{1}{2} \frac{\omega}{\varepsilon'_0} \frac{m_e^2}{\varepsilon_0} \Delta\left[1+\frac{1}{\Delta} \frac{\varepsilon_0^2}{m_e^2}\left(\alpha^2+\beta^2\right)\right]$, and $z_q= \frac{2}{3} \frac{\omega}{\omega_{\text{eff}}} \frac{m_e^3}{\varepsilon^2_0 \varepsilon'_0} \Delta^{3/2}\left[ 1 + \frac{\varepsilon_0^2}{m_e^2 } \frac{\left(\alpha^2+\beta^2\right)}{\Delta} \right]^{3/2}$ as stated in the main text.

The remaining integration over the angular variables can be performed using polar coordinates, where $q = \left[ 1 + \frac{\varepsilon_0^2}{m_e^2 } \frac{\rho^2}{\Delta} \right]^{3/2}$ and $\rho = \sqrt{\alpha^2 + \beta^2}$, with the following integration formulas:
\begin{equation}
\begin{aligned}
& \int_1^{\infty}\!\mathrm{d}x\, x^{\frac{1}{3}} K_{\frac{2}{3}}(x \rho) =\frac{K_{\frac{1}{3}}(\rho)}{\rho}, \\
& \int_1^{\infty}\!\mathrm{d}x\, x^{\frac{2}{3}} K_{\frac{1}{3}}(x \rho) =\frac{K_{\frac{2}{3}}(\rho)}{\rho}, \\
& \int_1^{\infty}\!\mathrm{d}x\,  K_{\frac{1}{3}}(x \rho) =\frac{\int_\rho^{\infty} K_{\frac{1}{3}}(x) \, \mathrm{d} x}{\rho}.
\end{aligned}\label{A2}
\end{equation}

Gathering the results of Eqs.~(\ref{A1}) and~(\ref{A2}), we have the following table of integrals:
\begin{table}[H]
\centering
\caption{Table of integrals for Eq.~(\ref{IntTable}) for different prefactors $f(\tau,\alpha,\beta)$.}
\begin{tabular}{c|l}
\toprule
$f$ & \multicolumn{1}{c}{$F$} \\
\midrule
$1$ & $\displaystyle \frac{4\pi}{\sqrt{3}} \frac{\varepsilon'_0}{\varepsilon_0} \frac{1}{\omega} \int_{z_q}^{+\infty} \! \mathrm{d}t \, K_{\frac{1}{3}}(t)$ \\
$\omega_{\text{eff}}\tau$ & $\displaystyle -\frac{8\pi i}{\sqrt{3}} \frac{\varepsilon'_0}{\varepsilon_0} \frac{\Delta^{1/2}}{\omega} \frac{m_e}{\varepsilon_0} K_{\frac{1}{3}}(z_q)$ \\
$\omega_{\text{eff}}^2\tau^2$ & $\displaystyle -\frac{16\pi}{\sqrt{3}}\frac{\varepsilon'_0}{\varepsilon_0} \frac{\Delta}{\omega} \frac{m_e^2}{\varepsilon_0^2} K_{\frac{2}{3}}(z_q)$ \\
$\alpha$ & 0\\
$\beta$ & 0\\
$\alpha^2$ & $\displaystyle \frac{2\pi}{\sqrt{3}} \frac{\varepsilon'_0}{\varepsilon_0} \frac{\Delta}{\omega} \frac{m_e^2}{\varepsilon_0^2} \left( K_{\frac{2}{3}}(z_q) - \int_{z_q}^{+\infty} \! \mathrm{d}t \, K_{\frac{1}{3}}(t) \right)$ \\
$\beta^2$ & $\displaystyle \frac{2\pi}{\sqrt{3}} \frac{\varepsilon'_0}{\varepsilon_0} \frac{\Delta}{\omega} \frac{m_e^2}{\varepsilon_0^2} \left( K_{\frac{2}{3}}(z_q) - \int_{z_q}^{+\infty} \! \mathrm{d}t \, K_{\frac{1}{3}}(t) \right)$ \\
\bottomrule
\end{tabular}\label{TableInt}
\end{table}

From these results, the expansions in Eqs.~(\ref{EqvExpansion}) and (\ref{nExpansion}) for a relativistic particle are justified: the contributions from the time variable $\omega_{\mathrm{eff}} \tau$ and the angular variables $\alpha$ and $\beta$ are of order $O\left( \frac{m_e}{\varepsilon_0} \right)$. Thus, the omitted higher-order terms are relativistically suppressed. Physically, this implies that the formation time of axion radiation is $\frac{m_e}{B_{\mathrm{eff}}}$, and the emission is concentrated in the forward direction within an angle $\sim \frac{m_e}{\varepsilon_0}$.

%\section{photon emission probability}
%For reference, the electron spin-resolved photon emission probability is
%\begin{equation}
%\begin{aligned}
%& \frac{\mathrm{d}P_{\gamma}}{\mathrm{d}t_0 \mathrm{d}\omega}=\frac{   \sqrt{3}e^2}{24\pi^2} \frac{m^2}{ \varepsilon^2} \\
%& \left\{-\int_{z}^{\infty} \mathrm{d} t\,K_{1 / 3}(t) \left[1+\boldsymbol{\zeta}_{i} \cdot\boldsymbol{\zeta}_{f}+\frac{\delta^2}{1-\delta}\left(\boldsymbol{\zeta}_{i} \cdot \boldsymbol{v}\right)\left(\boldsymbol{\zeta}_{f}\cdot \boldsymbol{v}\right)\right]\right. \\
%& +K_{2 / 3}\left(z\right)\left[2\left(1+\boldsymbol{\zeta}_{i} \cdot\boldsymbol{\zeta}_{f}\right)+\frac{\delta^2}{1-\delta}\left(1+\left(\boldsymbol{\zeta}_{i} \cdot \boldsymbol{v}\right)\left(\boldsymbol{\zeta}_{f}\cdot \boldsymbol{v}\right)\right)\right] \\
%& \left.-K_{1 / 3}\left(z\right) \frac{\delta}{1-\delta}\left(\boldsymbol{e}_H\cdot \boldsymbol{\zeta}_{f}+(1-\delta) \boldsymbol{e}_H \cdot\boldsymbol{\zeta}_{i}\right)\right\}
%\end{aligned}
%\end{equation}

\section{Contribution from the Longitudinal Acceleration}\label{APlongitudinal}

The equation of motion for the classical trajectory in constant crossed fields reads
\begin{equation}
\left\{
\begin{aligned}
\frac{\mathrm{d}\boldsymbol{p}(t)}{\mathrm{d}t} 
&= -\left( \boldsymbol{E}_0 + \frac{\boldsymbol{p}(t) \times \boldsymbol{B}_0}{\varepsilon(t)} \right), 
\\  
\frac{\mathrm{d}\varepsilon(t)}{\mathrm{d}t} 
&= -\frac{\boldsymbol{p}(t) \cdot \boldsymbol{E}_0}{\varepsilon(t)}.
\end{aligned}
\right.
\end{equation}
Here, $\boldsymbol{E}_0$ is the effective electric field generating the longitudinal force, and $\boldsymbol{B}_0$ the effective magnetic field responsible for the transverse force. In the instantaneous Frenet-Serret frame, the equation of motion reads
\begin{equation}
\left\{
\begin{aligned}
\varepsilon &= \varepsilon_0 - E_0 v_0 t + \frac{1}{2} \frac{E_0^2}{\varepsilon_0} \frac{m_e^2}{\varepsilon_0^2} t^2, \\
v_x &= v_0 - \frac{1}{2} \frac{B_0^2}{\varepsilon_0^2} v_0 t^2 - \frac{m_e^2}{\varepsilon_0^2} \frac{E_0}{\varepsilon_0} t - \frac{3}{2} \frac{E_0^2}{\varepsilon_0^2} \frac{m_e^2}{\varepsilon_0^2} t^2, \\
v_y &= \frac{B_0}{\varepsilon_0} v_0 t - \frac{E_0 B_0}{2 \varepsilon_0^2} \left( \frac{3 m_e^2}{\varepsilon_0^2} - 1 \right) t^2, \\
v_z &= 0,
\end{aligned}
\right.
\end{equation}
where a Taylor expansion up to second order in $t$ is performed. According to Appendix~\ref{Time Integration}, each factor of $t$ in the prefactor corresponds to a factor of $\frac{m_e}{B_0}$ in the axion yield. 
Thus, except in the extreme case where $\frac{E_0}{B_0} > \frac{\varepsilon_0}{m_e}$ (corresponding to constant acceleration), the $t$ factor does not affect the order of relativistic suppression.
For a laser pulse, $\frac{E_0}{B_0} \sim 1$. Therefore, up to order $O(\frac{m_e^2}{\varepsilon_0^2})$, the trajectory simplifies to
\begin{equation}
\left\{
\begin{aligned}
\varepsilon &= \varepsilon_0 - E_0 v_0 t, \\
v_x &= v_0 - \frac{1}{2} \frac{B_0^2}{\varepsilon_0^2} v_0 t^2, \\
v_y &= \frac{B_0}{\varepsilon_0} v_0 t + \frac{E_0 B_0}{2 \varepsilon_0^2} t^2, \\
v_z &= 0.
\end{aligned}
\right.
\label{LCFtra}
\end{equation}

Using the trajectory from Eq.~(\ref{LCFtra}), the action $S_c$ takes the form
\begin{equation}
S_c
\approx -\int_{-\frac{\tau}{2}}^{\frac{\tau}{2}} \mathrm{d}t\, \left[ b'(t) + c'(t)t + a'(t)t^2 \right],
\end{equation}
where $a' = -\frac{k}{2} \frac{\varepsilon}{\varepsilon'} \boldsymbol{n} \cdot \ddot{\boldsymbol{v}}_0$, $b' = \frac{\omega \varepsilon}{\varepsilon'} \left( 1 - \frac{k}{\omega} \boldsymbol{n} \cdot \boldsymbol{v}_0 - \frac{1}{2} \frac{m_\phi^2}{\varepsilon \omega} \right)$, and $c' = -k \frac{\varepsilon}{\varepsilon'} \boldsymbol{n} \cdot \dot{\boldsymbol{v}}_0$.
As in the main text, the action $S_c$ should be accurate up to order $O(\frac{m_e^2}{\varepsilon_0^2})$.
First, we estimate
\begin{equation}
\begin{aligned}
\frac{\varepsilon}{\varepsilon'} &\approx \frac{\varepsilon_0}{\varepsilon_0'} \left( 1 + \frac{\omega v_0}{\varepsilon_0 \varepsilon_0'} t \right) \\
&\approx \frac{\varepsilon_0}{\varepsilon_0'} + O\left( \frac{m_e}{\varepsilon_0} \right),
\end{aligned}
\end{equation}
which will be frequently used below:
\begin{itemize}
    \item \textbf{For the $a'$ term:} We have $\boldsymbol{n} \cdot \ddot{\boldsymbol{v}}_0 \approx -\frac{B_0^2}{\varepsilon_0^2} + O(\frac{m_e^3}{\varepsilon_0^3})$, so $a' \approx \frac{k}{2} \frac{B_0^2}{\varepsilon_0 \varepsilon_0'}$ is accurate up to $O(\frac{m_e^2}{\varepsilon_0^2})$.
    \item \textbf{For the $b'$ term:} 
    Expanding over the angular variables gives $b' \approx \frac{1}{2} \frac{\omega}{\varepsilon'} \frac{m_e^2}{\varepsilon}
    \left[ 1 + \frac{m_\phi^2}{m_e^2} \frac{\varepsilon \varepsilon'}{\omega^2} + \frac{\varepsilon^2 (\alpha^2 + \beta^2)}{m_e^2} \right]$.
    Therefore, $ b' \approx \frac{1}{2} \frac{\omega}{\varepsilon_0'} \frac{m_e^2}{\varepsilon_0}
    \left[ 1 + \frac{m_\phi^2}{m_e^2} \frac{\varepsilon_0 \varepsilon_0'}{\omega^2} + \frac{\varepsilon_0^2 (\alpha^2 + \beta^2)}{m_e^2} \right]$,
    which is accurate up to $O(\frac{m_e^2}{\varepsilon_0^2})$.
    \item \textbf{For the $c'$ term:} 
    We have $c' \approx -k\beta \frac{B_0}{\varepsilon_0'} t$, which is accurate up to $O(\frac{m_e^2}{\varepsilon_0^2})$ by power counting. However, its contribution to $S_c$ vanishes upon integration over $t$.
\end{itemize}
Thus, we conclude that the contributions from longitudinal acceleration constitute higher-order relativistic corrections to the action $S_c$.

For the prefactor, we note that $\boldsymbol{\mathfrak{B}}(t)$ in Eq.~(\ref{EqB}) is accurate up to order $O(\frac{m_e}{\varepsilon_0})$. Using the trajectory from Eq.~(\ref{LCFtra}), we obtain
\begin{equation}
\boldsymbol{\mathfrak{B}}(t) \approx -\frac{1}{2} \frac{\omega}{\varepsilon_0'} \left( -\frac{m_e}{\varepsilon_0},\, \frac{B_0}{\varepsilon_0} v_0 t - \beta,\,-\alpha \right),
\end{equation}
which coincides with the result in the absence of longitudinal acceleration. It follows that the correction to the prefactor in Eq.~(\ref{EqPrefactor}) is also unaffected by longitudinal acceleration up to order $O(\frac{m_e^2}{\varepsilon_0^2})$.

In summary, both the prefactor and the exponent in the integration are unaltered by longitudinal acceleration up to order $O(\frac{m_e^2}{\varepsilon_0^2})$. This concludes our proof that the contribution from longitudinal acceleration to axion radiation is a higher-order relativistic correction, suppressed by the factor $O(\frac{m_e}{\varepsilon_0})$, and can thus be omitted as in the main text.

\bibliography{references}

\begin{thebibliography}{85}%
\makeatletter
\providecommand \@ifxundefined [1]{%
 \@ifx{#1\undefined}
}%
\providecommand \@ifnum [1]{%
 \ifnum #1\expandafter \@firstoftwo
 \else \expandafter \@secondoftwo
 \fi
}%
\providecommand \@ifx [1]{%
 \ifx #1\expandafter \@firstoftwo
 \else \expandafter \@secondoftwo
 \fi
}%
\providecommand \natexlab [1]{#1}%
\providecommand \enquote  [1]{``#1''}%
\providecommand \bibnamefont  [1]{#1}%
\providecommand \bibfnamefont [1]{#1}%
\providecommand \citenamefont [1]{#1}%
\providecommand \href@noop [0]{\@secondoftwo}%
\providecommand \href [0]{\begingroup \@sanitize@url \@href}%
\providecommand \@href[1]{\@@startlink{#1}\@@href}%
\providecommand \@@href[1]{\endgroup#1\@@endlink}%
\providecommand \@sanitize@url [0]{\catcode `\\12\catcode `\$12\catcode
  `\&12\catcode `\#12\catcode `\^12\catcode `\_12\catcode `\%12\relax}%
\providecommand \@@startlink[1]{}%
\providecommand \@@endlink[0]{}%
\providecommand \url  [0]{\begingroup\@sanitize@url \@url }%
\providecommand \@url [1]{\endgroup\@href {#1}{\urlprefix }}%
\providecommand \urlprefix  [0]{URL }%
\providecommand \Eprint [0]{\href }%
\providecommand \doibase [0]{https://doi.org/}%
\providecommand \selectlanguage [0]{\@gobble}%
\providecommand \bibinfo  [0]{\@secondoftwo}%
\providecommand \bibfield  [0]{\@secondoftwo}%
\providecommand \translation [1]{[#1]}%
\providecommand \BibitemOpen [0]{}%
\providecommand \bibitemStop [0]{}%
\providecommand \bibitemNoStop [0]{.\EOS\space}%
\providecommand \EOS [0]{\spacefactor3000\relax}%
\providecommand \BibitemShut  [1]{\csname bibitem#1\endcsname}%
\let\auto@bib@innerbib\@empty
%</preamble>
\bibitem [{\citenamefont {Polyakov}(1975)}]{POLYAKOV197582}%
  \BibitemOpen
  \bibfield  {author} {\bibinfo {author} {\bibfnamefont {A.}~\bibnamefont
  {Polyakov}},\ }\bibfield  {title} {\bibinfo {title} {Compact gauge fields and
  the infrared catastrophe},\ }\href
  {https://doi.org/https://doi.org/10.1016/0370-2693(75)90162-8} {\bibfield
  {journal} {\bibinfo  {journal} {Physics Letters B}\ }\textbf {\bibinfo
  {volume} {59}},\ \bibinfo {pages} {82} (\bibinfo {year} {1975})}\BibitemShut
  {NoStop}%
\bibitem [{\citenamefont {Jackiw}\ and\ \citenamefont
  {Rebbi}(1976)}]{PhysRevLett.37.172}%
  \BibitemOpen
  \bibfield  {author} {\bibinfo {author} {\bibfnamefont {R.}~\bibnamefont
  {Jackiw}}\ and\ \bibinfo {author} {\bibfnamefont {C.}~\bibnamefont {Rebbi}},\
  }\bibfield  {title} {\bibinfo {title} {Vacuum periodicity in a
  $\mathrm{Y}$ang-$\mathrm{M}$ills quantum theory},\ }\href
  {https://doi.org/10.1103/PhysRevLett.37.172} {\bibfield  {journal} {\bibinfo
  {journal} {Phys. Rev. Lett.}\ }\textbf {\bibinfo {volume} {37}},\ \bibinfo
  {pages} {172} (\bibinfo {year} {1976})}\BibitemShut {NoStop}%
\bibitem [{\citenamefont {Callan}\ \emph {et~al.}(1976)\citenamefont {Callan},
  \citenamefont {Dashen},\ and\ \citenamefont {Gross}}]{CALLAN1976334}%
  \BibitemOpen
  \bibfield  {author} {\bibinfo {author} {\bibfnamefont {C.}~\bibnamefont
  {Callan}}, \bibinfo {author} {\bibfnamefont {R.}~\bibnamefont {Dashen}},\
  and\ \bibinfo {author} {\bibfnamefont {D.}~\bibnamefont {Gross}},\ }\bibfield
   {title} {\bibinfo {title} {The structure of the gauge theory vacuum},\
  }\href {https://doi.org/https://doi.org/10.1016/0370-2693(76)90277-X}
  {\bibfield  {journal} {\bibinfo  {journal} {Physics Letters B}\ }\textbf
  {\bibinfo {volume} {63}},\ \bibinfo {pages} {334} (\bibinfo {year}
  {1976})}\BibitemShut {NoStop}%
\bibitem [{\citenamefont {Peccei}\ and\ \citenamefont
  {Quinn}(1977{\natexlab{a}})}]{PhysRevD.16.1791}%
  \BibitemOpen
  \bibfield  {author} {\bibinfo {author} {\bibfnamefont {R.~D.}\ \bibnamefont
  {Peccei}}\ and\ \bibinfo {author} {\bibfnamefont {H.~R.}\ \bibnamefont
  {Quinn}},\ }\bibfield  {title} {\bibinfo {title} {Constraints imposed by
  $\mathrm{CP}$ conservation in the presence of pseudoparticles},\ }\href
  {https://doi.org/10.1103/PhysRevD.16.1791} {\bibfield  {journal} {\bibinfo
  {journal} {Phys. Rev. D}\ }\textbf {\bibinfo {volume} {16}},\ \bibinfo
  {pages} {1791} (\bibinfo {year} {1977}{\natexlab{a}})}\BibitemShut {NoStop}%
\bibitem [{\citenamefont {Peccei}\ and\ \citenamefont
  {Quinn}(1977{\natexlab{b}})}]{PhysRevLett.38.1440}%
  \BibitemOpen
  \bibfield  {author} {\bibinfo {author} {\bibfnamefont {R.~D.}\ \bibnamefont
  {Peccei}}\ and\ \bibinfo {author} {\bibfnamefont {H.~R.}\ \bibnamefont
  {Quinn}},\ }\bibfield  {title} {\bibinfo {title} {$\mathrm{CP}$ conservation
  in the presence of pseudoparticles},\ }\href
  {https://doi.org/10.1103/PhysRevLett.38.1440} {\bibfield  {journal} {\bibinfo
   {journal} {Phys. Rev. Lett.}\ }\textbf {\bibinfo {volume} {38}},\ \bibinfo
  {pages} {1440} (\bibinfo {year} {1977}{\natexlab{b}})}\BibitemShut {NoStop}%
\bibitem [{\citenamefont {Wilczek}(1978)}]{PhysRevLett.40.279}%
  \BibitemOpen
  \bibfield  {author} {\bibinfo {author} {\bibfnamefont {F.}~\bibnamefont
  {Wilczek}},\ }\bibfield  {title} {\bibinfo {title} {Problem of strong
  $\mathrm{P}$ and $\mathrm{T}$ invariance in the presence of instantons},\
  }\href {https://doi.org/10.1103/PhysRevLett.40.279} {\bibfield  {journal}
  {\bibinfo  {journal} {Phys. Rev. Lett.}\ }\textbf {\bibinfo {volume} {40}},\
  \bibinfo {pages} {279} (\bibinfo {year} {1978})}\BibitemShut {NoStop}%
\bibitem [{\citenamefont {Weinberg}(1978)}]{PhysRevLett.40.223}%
  \BibitemOpen
  \bibfield  {author} {\bibinfo {author} {\bibfnamefont {S.}~\bibnamefont
  {Weinberg}},\ }\bibfield  {title} {\bibinfo {title} {A new light boson?},\
  }\href {https://doi.org/10.1103/PhysRevLett.40.223} {\bibfield  {journal}
  {\bibinfo  {journal} {Phys. Rev. Lett.}\ }\textbf {\bibinfo {volume} {40}},\
  \bibinfo {pages} {223} (\bibinfo {year} {1978})}\BibitemShut {NoStop}%
\bibitem [{\citenamefont {Kim}(1987)}]{kim1987light}%
  \BibitemOpen
  \bibfield  {author} {\bibinfo {author} {\bibfnamefont {J.~E.}\ \bibnamefont
  {Kim}},\ }\bibfield  {title} {\bibinfo {title} {Light pseudoscalars, particle
  physics and cosmology},\ }\href@noop {} {\bibfield  {journal} {\bibinfo
  {journal} {Physics Reports}\ }\textbf {\bibinfo {volume} {150}},\ \bibinfo
  {pages} {1} (\bibinfo {year} {1987})}\BibitemShut {NoStop}%
\bibitem [{\citenamefont {Kim}(1979)}]{PhysRevLett.43.103}%
  \BibitemOpen
  \bibfield  {author} {\bibinfo {author} {\bibfnamefont {J.~E.}\ \bibnamefont
  {Kim}},\ }\bibfield  {title} {\bibinfo {title} {Weak-interaction singlet and
  strong $\mathrm{CP}$ invariance},\ }\href
  {https://doi.org/10.1103/PhysRevLett.43.103} {\bibfield  {journal} {\bibinfo
  {journal} {Phys. Rev. Lett.}\ }\textbf {\bibinfo {volume} {43}},\ \bibinfo
  {pages} {103} (\bibinfo {year} {1979})}\BibitemShut {NoStop}%
\bibitem [{\citenamefont {Shifman}\ \emph {et~al.}(1980)\citenamefont
  {Shifman}, \citenamefont {Vainshtein},\ and\ \citenamefont
  {Zakharov}}]{SHIFMAN1980493}%
  \BibitemOpen
  \bibfield  {author} {\bibinfo {author} {\bibfnamefont {M.}~\bibnamefont
  {Shifman}}, \bibinfo {author} {\bibfnamefont {A.}~\bibnamefont
  {Vainshtein}},\ and\ \bibinfo {author} {\bibfnamefont {V.}~\bibnamefont
  {Zakharov}},\ }\bibfield  {title} {\bibinfo {title} {Can confinement ensure
  natural $\mathrm{CP}$ invariance of strong interactions?},\ }\href
  {https://doi.org/https://doi.org/10.1016/0550-3213(80)90209-6} {\bibfield
  {journal} {\bibinfo  {journal} {Nuclear Physics B}\ }\textbf {\bibinfo
  {volume} {166}},\ \bibinfo {pages} {493} (\bibinfo {year}
  {1980})}\BibitemShut {NoStop}%
\bibitem [{\citenamefont {Zhitnitsky}(1980)}]{Zhitnitsky:1980tq}%
  \BibitemOpen
  \bibfield  {author} {\bibinfo {author} {\bibfnamefont {A.~R.}\ \bibnamefont
  {Zhitnitsky}},\ }\bibfield  {title} {\bibinfo {title} {{On Possible
  Suppression of the Axion Hadron Interactions. (In Russian)}},\ }\href@noop {}
  {\bibfield  {journal} {\bibinfo  {journal} {Sov. J. Nucl. Phys.}\ }\textbf
  {\bibinfo {volume} {31}},\ \bibinfo {pages} {260} (\bibinfo {year}
  {1980})}\BibitemShut {NoStop}%
\bibitem [{\citenamefont {Dine}\ \emph {et~al.}(1981)\citenamefont {Dine},
  \citenamefont {Fischler},\ and\ \citenamefont {Srednicki}}]{dine1981simple}%
  \BibitemOpen
  \bibfield  {author} {\bibinfo {author} {\bibfnamefont {M.}~\bibnamefont
  {Dine}}, \bibinfo {author} {\bibfnamefont {W.}~\bibnamefont {Fischler}},\
  and\ \bibinfo {author} {\bibfnamefont {M.}~\bibnamefont {Srednicki}},\
  }\bibfield  {title} {\bibinfo {title} {A simple solution to the strong
  $\mathrm{CP}$ problem with a harmless axion},\ }\href@noop {} {\bibfield
  {journal} {\bibinfo  {journal} {Physics Letters B}\ }\textbf {\bibinfo
  {volume} {104}},\ \bibinfo {pages} {199} (\bibinfo {year}
  {1981})}\BibitemShut {NoStop}%
\bibitem [{\citenamefont {Peccei}\ \emph {et~al.}(1986)\citenamefont {Peccei},
  \citenamefont {Wu},\ and\ \citenamefont {Yanagida}}]{PECCEI1986435}%
  \BibitemOpen
  \bibfield  {author} {\bibinfo {author} {\bibfnamefont {R.}~\bibnamefont
  {Peccei}}, \bibinfo {author} {\bibfnamefont {T.~T.}\ \bibnamefont {Wu}},\
  and\ \bibinfo {author} {\bibfnamefont {T.}~\bibnamefont {Yanagida}},\
  }\bibfield  {title} {\bibinfo {title} {A viable axion model},\ }\href
  {https://doi.org/https://doi.org/10.1016/0370-2693(86)90284-4} {\bibfield
  {journal} {\bibinfo  {journal} {Physics Letters B}\ }\textbf {\bibinfo
  {volume} {172}},\ \bibinfo {pages} {435} (\bibinfo {year}
  {1986})}\BibitemShut {NoStop}%
\bibitem [{\citenamefont {Marsh}(2016)}]{marsh2016axion}%
  \BibitemOpen
  \bibfield  {author} {\bibinfo {author} {\bibfnamefont {D.~J.}\ \bibnamefont
  {Marsh}},\ }\bibfield  {title} {\bibinfo {title} {Axion cosmology},\
  }\href@noop {} {\bibfield  {journal} {\bibinfo  {journal} {Physics Reports}\
  }\textbf {\bibinfo {volume} {643}},\ \bibinfo {pages} {1} (\bibinfo {year}
  {2016})}\BibitemShut {NoStop}%
\bibitem [{\citenamefont {Witten}(1984)}]{witten1984some}%
  \BibitemOpen
  \bibfield  {author} {\bibinfo {author} {\bibfnamefont {E.}~\bibnamefont
  {Witten}},\ }\bibfield  {title} {\bibinfo {title} {Some properties of
  $\mathrm{O}$(32) superstrings},\ }\href@noop {} {\bibfield  {journal}
  {\bibinfo  {journal} {Physics Letters B}\ }\textbf {\bibinfo {volume}
  {149}},\ \bibinfo {pages} {351} (\bibinfo {year} {1984})}\BibitemShut
  {NoStop}%
\bibitem [{\citenamefont {Arvanitaki}\ \emph {et~al.}(2010)\citenamefont
  {Arvanitaki}, \citenamefont {Dimopoulos}, \citenamefont {Dubovsky},
  \citenamefont {Kaloper},\ and\ \citenamefont
  {March-Russell}}]{arvanitaki2010string}%
  \BibitemOpen
  \bibfield  {author} {\bibinfo {author} {\bibfnamefont {A.}~\bibnamefont
  {Arvanitaki}}, \bibinfo {author} {\bibfnamefont {S.}~\bibnamefont
  {Dimopoulos}}, \bibinfo {author} {\bibfnamefont {S.}~\bibnamefont
  {Dubovsky}}, \bibinfo {author} {\bibfnamefont {N.}~\bibnamefont {Kaloper}},\
  and\ \bibinfo {author} {\bibfnamefont {J.}~\bibnamefont {March-Russell}},\
  }\bibfield  {title} {\bibinfo {title} {String axiverse},\ }\href@noop {}
  {\bibfield  {journal} {\bibinfo  {journal} {Phys. Rev. D}\ }\textbf {\bibinfo
  {volume} {81}},\ \bibinfo {pages} {123530} (\bibinfo {year}
  {2010})}\BibitemShut {NoStop}%
\bibitem [{\citenamefont {Wu}\ \emph {et~al.}(2016)\citenamefont {Wu},
  \citenamefont {Salehi}, \citenamefont {Koirala}, \citenamefont {Moon},
  \citenamefont {Oh},\ and\ \citenamefont {Armitage}}]{wu2016quantized}%
  \BibitemOpen
  \bibfield  {author} {\bibinfo {author} {\bibfnamefont {L.}~\bibnamefont
  {Wu}}, \bibinfo {author} {\bibfnamefont {M.}~\bibnamefont {Salehi}}, \bibinfo
  {author} {\bibfnamefont {N.}~\bibnamefont {Koirala}}, \bibinfo {author}
  {\bibfnamefont {J.}~\bibnamefont {Moon}}, \bibinfo {author} {\bibfnamefont
  {S.}~\bibnamefont {Oh}},\ and\ \bibinfo {author} {\bibfnamefont
  {N.}~\bibnamefont {Armitage}},\ }\bibfield  {title} {\bibinfo {title}
  {Quantized $\mathrm{F}$araday and $\mathrm{K}$err rotation and axion
  electrodynamics of a 3$\mathrm{D}$ topological insulator},\ }\href@noop {}
  {\bibfield  {journal} {\bibinfo  {journal} {Science}\ }\textbf {\bibinfo
  {volume} {354}},\ \bibinfo {pages} {1124} (\bibinfo {year}
  {2016})}\BibitemShut {NoStop}%
\bibitem [{\citenamefont {Okada}\ \emph {et~al.}(2016)\citenamefont {Okada},
  \citenamefont {Takahashi}, \citenamefont {Mogi}, \citenamefont {Yoshimi},
  \citenamefont {Tsukazaki}, \citenamefont {Takahashi}, \citenamefont {Ogawa},
  \citenamefont {Kawasaki},\ and\ \citenamefont {Tokura}}]{okada2016terahertz}%
  \BibitemOpen
  \bibfield  {author} {\bibinfo {author} {\bibfnamefont {K.~N.}\ \bibnamefont
  {Okada}}, \bibinfo {author} {\bibfnamefont {Y.}~\bibnamefont {Takahashi}},
  \bibinfo {author} {\bibfnamefont {M.}~\bibnamefont {Mogi}}, \bibinfo {author}
  {\bibfnamefont {R.}~\bibnamefont {Yoshimi}}, \bibinfo {author} {\bibfnamefont
  {A.}~\bibnamefont {Tsukazaki}}, \bibinfo {author} {\bibfnamefont {K.~S.}\
  \bibnamefont {Takahashi}}, \bibinfo {author} {\bibfnamefont {N.}~\bibnamefont
  {Ogawa}}, \bibinfo {author} {\bibfnamefont {M.}~\bibnamefont {Kawasaki}},\
  and\ \bibinfo {author} {\bibfnamefont {Y.}~\bibnamefont {Tokura}},\
  }\bibfield  {title} {\bibinfo {title} {Terahertz spectroscopy on
  $\mathrm{Faraday}$ and $\mathrm{Kerr}$ rotations in a quantum anomalous
  $\mathrm{H}$all state},\ }\href@noop {} {\bibfield  {journal} {\bibinfo
  {journal} {Nature communications}\ }\textbf {\bibinfo {volume} {7}},\
  \bibinfo {pages} {12245} (\bibinfo {year} {2016})}\BibitemShut {NoStop}%
\bibitem [{\citenamefont {Zhang}\ \emph {et~al.}(2019)\citenamefont {Zhang},
  \citenamefont {Shi}, \citenamefont {Zhu}, \citenamefont {Xing}, \citenamefont
  {Zhang},\ and\ \citenamefont {Wang}}]{zhang2019topological}%
  \BibitemOpen
  \bibfield  {author} {\bibinfo {author} {\bibfnamefont {D.}~\bibnamefont
  {Zhang}}, \bibinfo {author} {\bibfnamefont {M.}~\bibnamefont {Shi}}, \bibinfo
  {author} {\bibfnamefont {T.}~\bibnamefont {Zhu}}, \bibinfo {author}
  {\bibfnamefont {D.}~\bibnamefont {Xing}}, \bibinfo {author} {\bibfnamefont
  {H.}~\bibnamefont {Zhang}},\ and\ \bibinfo {author} {\bibfnamefont
  {J.}~\bibnamefont {Wang}},\ }\bibfield  {title} {\bibinfo {title}
  {Topological axion states in the magnetic insulator
  $\mathrm{MnBi}_2\mathrm{Te}_4$ with the quantized magnetoelectric effect},\
  }\href@noop {} {\bibfield  {journal} {\bibinfo  {journal} {Phys. Rev. Lett.}\
  }\textbf {\bibinfo {volume} {122}},\ \bibinfo {pages} {206401} (\bibinfo
  {year} {2019})}\BibitemShut {NoStop}%
\bibitem [{\citenamefont {Liu}\ \emph {et~al.}(2020)\citenamefont {Liu},
  \citenamefont {Wang}, \citenamefont {Li}, \citenamefont {Wu}, \citenamefont
  {Li}, \citenamefont {Li}, \citenamefont {He}, \citenamefont {Xu},
  \citenamefont {Zhang},\ and\ \citenamefont {Wang}}]{liu2020robust}%
  \BibitemOpen
  \bibfield  {author} {\bibinfo {author} {\bibfnamefont {C.}~\bibnamefont
  {Liu}}, \bibinfo {author} {\bibfnamefont {Y.}~\bibnamefont {Wang}}, \bibinfo
  {author} {\bibfnamefont {H.}~\bibnamefont {Li}}, \bibinfo {author}
  {\bibfnamefont {Y.}~\bibnamefont {Wu}}, \bibinfo {author} {\bibfnamefont
  {Y.}~\bibnamefont {Li}}, \bibinfo {author} {\bibfnamefont {J.}~\bibnamefont
  {Li}}, \bibinfo {author} {\bibfnamefont {K.}~\bibnamefont {He}}, \bibinfo
  {author} {\bibfnamefont {Y.}~\bibnamefont {Xu}}, \bibinfo {author}
  {\bibfnamefont {J.}~\bibnamefont {Zhang}},\ and\ \bibinfo {author}
  {\bibfnamefont {Y.}~\bibnamefont {Wang}},\ }\bibfield  {title} {\bibinfo
  {title} {Robust axion insulator and chern insulator phases in a
  two-dimensional antiferromagnetic topological insulator},\ }\href@noop {}
  {\bibfield  {journal} {\bibinfo  {journal} {Nature materials}\ }\textbf
  {\bibinfo {volume} {19}},\ \bibinfo {pages} {522} (\bibinfo {year}
  {2020})}\BibitemShut {NoStop}%
\bibitem [{\citenamefont {Wilczek}(1987)}]{PhysRevLett.58.1799}%
  \BibitemOpen
  \bibfield  {author} {\bibinfo {author} {\bibfnamefont {F.}~\bibnamefont
  {Wilczek}},\ }\bibfield  {title} {\bibinfo {title} {Two applications of axion
  electrodynamics},\ }\href {https://doi.org/10.1103/PhysRevLett.58.1799}
  {\bibfield  {journal} {\bibinfo  {journal} {Phys. Rev. Lett.}\ }\textbf
  {\bibinfo {volume} {58}},\ \bibinfo {pages} {1799} (\bibinfo {year}
  {1987})}\BibitemShut {NoStop}%
\bibitem [{\citenamefont {Gooth}\ \emph {et~al.}(2019)\citenamefont {Gooth},
  \citenamefont {Bradlyn}, \citenamefont {Honnali}, \citenamefont {Schindler},
  \citenamefont {Kumar}, \citenamefont {Noky}, \citenamefont {Qi},
  \citenamefont {Shekhar}, \citenamefont {Sun}, \citenamefont {Wang} \emph
  {et~al.}}]{gooth2019axionic}%
  \BibitemOpen
  \bibfield  {author} {\bibinfo {author} {\bibfnamefont {J.}~\bibnamefont
  {Gooth}}, \bibinfo {author} {\bibfnamefont {B.}~\bibnamefont {Bradlyn}},
  \bibinfo {author} {\bibfnamefont {S.}~\bibnamefont {Honnali}}, \bibinfo
  {author} {\bibfnamefont {C.}~\bibnamefont {Schindler}}, \bibinfo {author}
  {\bibfnamefont {N.}~\bibnamefont {Kumar}}, \bibinfo {author} {\bibfnamefont
  {J.}~\bibnamefont {Noky}}, \bibinfo {author} {\bibfnamefont {Y.}~\bibnamefont
  {Qi}}, \bibinfo {author} {\bibfnamefont {C.}~\bibnamefont {Shekhar}},
  \bibinfo {author} {\bibfnamefont {Y.}~\bibnamefont {Sun}}, \bibinfo {author}
  {\bibfnamefont {Z.}~\bibnamefont {Wang}}, \emph {et~al.},\ }\bibfield
  {title} {\bibinfo {title} {Axionic charge-density wave in the $\mathrm{W}$eyl
  semimetal $(\mathrm{TaSe}_4)_2\mathrm{I}$},\ }\href@noop {} {\bibfield
  {journal} {\bibinfo  {journal} {Nature}\ }\textbf {\bibinfo {volume} {575}},\
  \bibinfo {pages} {315} (\bibinfo {year} {2019})}\BibitemShut {NoStop}%
\bibitem [{\citenamefont {Irastorza}\ and\ \citenamefont
  {Redondo}(2018)}]{IRASTORZA201889}%
  \BibitemOpen
  \bibfield  {author} {\bibinfo {author} {\bibfnamefont {I.~G.}\ \bibnamefont
  {Irastorza}}\ and\ \bibinfo {author} {\bibfnamefont {J.}~\bibnamefont
  {Redondo}},\ }\bibfield  {title} {\bibinfo {title} {New experimental
  approaches in the search for axion-like particles},\ }\href
  {https://doi.org/https://doi.org/10.1016/j.ppnp.2018.05.003} {\bibfield
  {journal} {\bibinfo  {journal} {Progress in Particle and Nuclear Physics}\
  }\textbf {\bibinfo {volume} {102}},\ \bibinfo {pages} {89} (\bibinfo {year}
  {2018})}\BibitemShut {NoStop}%
\bibitem [{\citenamefont {Di~Luzio}\ \emph {et~al.}(2020)\citenamefont
  {Di~Luzio}, \citenamefont {Giannotti}, \citenamefont {Nardi},\ and\
  \citenamefont {Visinelli}}]{di2020landscape}%
  \BibitemOpen
  \bibfield  {author} {\bibinfo {author} {\bibfnamefont {L.}~\bibnamefont
  {Di~Luzio}}, \bibinfo {author} {\bibfnamefont {M.}~\bibnamefont {Giannotti}},
  \bibinfo {author} {\bibfnamefont {E.}~\bibnamefont {Nardi}},\ and\ \bibinfo
  {author} {\bibfnamefont {L.}~\bibnamefont {Visinelli}},\ }\bibfield  {title}
  {\bibinfo {title} {The landscape of $\mathrm{QCD}$ axion models},\
  }\href@noop {} {\bibfield  {journal} {\bibinfo  {journal} {Physics Reports}\
  }\textbf {\bibinfo {volume} {870}},\ \bibinfo {pages} {1} (\bibinfo {year}
  {2020})}\BibitemShut {NoStop}%
\bibitem [{\citenamefont {Sikivie}(2021)}]{RevModPhys.93.015004}%
  \BibitemOpen
  \bibfield  {author} {\bibinfo {author} {\bibfnamefont {P.}~\bibnamefont
  {Sikivie}},\ }\bibfield  {title} {\bibinfo {title} {Invisible axion search
  methods},\ }\href {https://doi.org/10.1103/RevModPhys.93.015004} {\bibfield
  {journal} {\bibinfo  {journal} {Rev. Mod. Phys.}\ }\textbf {\bibinfo {volume}
  {93}},\ \bibinfo {pages} {015004} (\bibinfo {year} {2021})}\BibitemShut
  {NoStop}%
\bibitem [{\citenamefont {Carenza}\ \emph {et~al.}(2025)\citenamefont
  {Carenza}, \citenamefont {Giannotti}, \citenamefont {Isern}, \citenamefont
  {Mirizzi},\ and\ \citenamefont {Straniero}}]{carenza2025axion}%
  \BibitemOpen
  \bibfield  {author} {\bibinfo {author} {\bibfnamefont {P.}~\bibnamefont
  {Carenza}}, \bibinfo {author} {\bibfnamefont {M.}~\bibnamefont {Giannotti}},
  \bibinfo {author} {\bibfnamefont {J.}~\bibnamefont {Isern}}, \bibinfo
  {author} {\bibfnamefont {A.}~\bibnamefont {Mirizzi}},\ and\ \bibinfo {author}
  {\bibfnamefont {O.}~\bibnamefont {Straniero}},\ }\bibfield  {title} {\bibinfo
  {title} {Axion astrophysics},\ }\href@noop {} {\bibfield  {journal} {\bibinfo
   {journal} {Physics Reports}\ }\textbf {\bibinfo {volume} {1117}},\ \bibinfo
  {pages} {1} (\bibinfo {year} {2025})}\BibitemShut {NoStop}%
\bibitem [{\citenamefont {Pugnat}\ \emph {et~al.}(2014)\citenamefont {Pugnat},
  \citenamefont {Ballou}, \citenamefont {Schott}, \citenamefont {Husek},
  \citenamefont {Sulc}, \citenamefont {Deferne}, \citenamefont {Duvillaret},
  \citenamefont {Finger}, \citenamefont {Finger}, \citenamefont {Flekova} \emph
  {et~al.}}]{pugnat2014search}%
  \BibitemOpen
  \bibfield  {author} {\bibinfo {author} {\bibfnamefont {P.}~\bibnamefont
  {Pugnat}}, \bibinfo {author} {\bibfnamefont {R.}~\bibnamefont {Ballou}},
  \bibinfo {author} {\bibfnamefont {M.}~\bibnamefont {Schott}}, \bibinfo
  {author} {\bibfnamefont {T.}~\bibnamefont {Husek}}, \bibinfo {author}
  {\bibfnamefont {M.}~\bibnamefont {Sulc}}, \bibinfo {author} {\bibfnamefont
  {G.}~\bibnamefont {Deferne}}, \bibinfo {author} {\bibfnamefont
  {L.}~\bibnamefont {Duvillaret}}, \bibinfo {author} {\bibfnamefont
  {M.}~\bibnamefont {Finger}}, \bibinfo {author} {\bibfnamefont
  {M.}~\bibnamefont {Finger}}, \bibinfo {author} {\bibfnamefont
  {L.}~\bibnamefont {Flekova}}, \emph {et~al.},\ }\bibfield  {title} {\bibinfo
  {title} {Search for weakly interacting sub-ev particles with the
  $\mathrm{OSQAR}$ laser-based experiment: results and perspectives},\
  }\href@noop {} {\bibfield  {journal} {\bibinfo  {journal} {The European
  Physical Journal C}\ }\textbf {\bibinfo {volume} {74}},\ \bibinfo {pages} {1}
  (\bibinfo {year} {2014})}\BibitemShut {NoStop}%
\bibitem [{\citenamefont {Ballou}\ \emph {et~al.}(2015)\citenamefont {Ballou},
  \citenamefont {Deferne}, \citenamefont {Finger~Jr}, \citenamefont {Finger},
  \citenamefont {Flekova}, \citenamefont {Hosek}, \citenamefont {Kunc},
  \citenamefont {Macuchova}, \citenamefont {Meissner}, \citenamefont {Pugnat}
  \emph {et~al.}}]{ballou2015new}%
  \BibitemOpen
  \bibfield  {author} {\bibinfo {author} {\bibfnamefont {R.}~\bibnamefont
  {Ballou}}, \bibinfo {author} {\bibfnamefont {G.}~\bibnamefont {Deferne}},
  \bibinfo {author} {\bibfnamefont {M.}~\bibnamefont {Finger~Jr}}, \bibinfo
  {author} {\bibfnamefont {M.}~\bibnamefont {Finger}}, \bibinfo {author}
  {\bibfnamefont {L.}~\bibnamefont {Flekova}}, \bibinfo {author} {\bibfnamefont
  {J.}~\bibnamefont {Hosek}}, \bibinfo {author} {\bibfnamefont
  {S.}~\bibnamefont {Kunc}}, \bibinfo {author} {\bibfnamefont {K.}~\bibnamefont
  {Macuchova}}, \bibinfo {author} {\bibfnamefont {K.}~\bibnamefont {Meissner}},
  \bibinfo {author} {\bibfnamefont {P.}~\bibnamefont {Pugnat}}, \emph
  {et~al.},\ }\bibfield  {title} {\bibinfo {title} {New exclusion limits on
  scalar and pseudoscalar axionlike particles from light shining through a
  wall},\ }\href@noop {} {\bibfield  {journal} {\bibinfo  {journal} {Phys. Rev.
  D}\ }\textbf {\bibinfo {volume} {92}},\ \bibinfo {pages} {092002} (\bibinfo
  {year} {2015})}\BibitemShut {NoStop}%
\bibitem [{\citenamefont {Della~Valle}\ \emph {et~al.}(2016)\citenamefont
  {Della~Valle}, \citenamefont {Ejlli}, \citenamefont {Gastaldi}, \citenamefont
  {Messineo}, \citenamefont {Milotti}, \citenamefont {Pengo}, \citenamefont
  {Ruoso},\ and\ \citenamefont {Zavattini}}]{della2016pvlas}%
  \BibitemOpen
  \bibfield  {author} {\bibinfo {author} {\bibfnamefont {F.}~\bibnamefont
  {Della~Valle}}, \bibinfo {author} {\bibfnamefont {A.}~\bibnamefont {Ejlli}},
  \bibinfo {author} {\bibfnamefont {U.}~\bibnamefont {Gastaldi}}, \bibinfo
  {author} {\bibfnamefont {G.}~\bibnamefont {Messineo}}, \bibinfo {author}
  {\bibfnamefont {E.}~\bibnamefont {Milotti}}, \bibinfo {author} {\bibfnamefont
  {R.}~\bibnamefont {Pengo}}, \bibinfo {author} {\bibfnamefont
  {G.}~\bibnamefont {Ruoso}},\ and\ \bibinfo {author} {\bibfnamefont
  {G.}~\bibnamefont {Zavattini}},\ }\bibfield  {title} {\bibinfo {title} {The
  pvlas experiment: measuring vacuum magnetic birefringence and dichroism with
  a birefringent $\mathrm{Fabry}$--$\mathrm{Perot}$ cavity},\ }\href@noop {}
  {\bibfield  {journal} {\bibinfo  {journal} {The European Physical Journal C}\
  }\textbf {\bibinfo {volume} {76}},\ \bibinfo {pages} {1} (\bibinfo {year}
  {2016})}\BibitemShut {NoStop}%
\bibitem [{\citenamefont {Konaka}\ \emph {et~al.}(1986)\citenamefont {Konaka},
  \citenamefont {Imai}, \citenamefont {Kobayashi}, \citenamefont {Masaike},
  \citenamefont {Miyake}, \citenamefont {Nakamura}, \citenamefont {Nagamine},
  \citenamefont {Sasao}, \citenamefont {Enomoto}, \citenamefont {Fukushima}
  \emph {et~al.}}]{konaka1986search}%
  \BibitemOpen
  \bibfield  {author} {\bibinfo {author} {\bibfnamefont {A.}~\bibnamefont
  {Konaka}}, \bibinfo {author} {\bibfnamefont {K.}~\bibnamefont {Imai}},
  \bibinfo {author} {\bibfnamefont {H.}~\bibnamefont {Kobayashi}}, \bibinfo
  {author} {\bibfnamefont {A.}~\bibnamefont {Masaike}}, \bibinfo {author}
  {\bibfnamefont {K.}~\bibnamefont {Miyake}}, \bibinfo {author} {\bibfnamefont
  {T.}~\bibnamefont {Nakamura}}, \bibinfo {author} {\bibfnamefont
  {N.}~\bibnamefont {Nagamine}}, \bibinfo {author} {\bibfnamefont
  {N.}~\bibnamefont {Sasao}}, \bibinfo {author} {\bibfnamefont
  {A.}~\bibnamefont {Enomoto}}, \bibinfo {author} {\bibfnamefont
  {Y.}~\bibnamefont {Fukushima}}, \emph {et~al.},\ }\bibfield  {title}
  {\bibinfo {title} {Search for neutral particles in electron-beam-dump
  experiment},\ }\href@noop {} {\bibfield  {journal} {\bibinfo  {journal}
  {Phys. Rev. Lett.}\ }\textbf {\bibinfo {volume} {57}},\ \bibinfo {pages}
  {659} (\bibinfo {year} {1986})}\BibitemShut {NoStop}%
\bibitem [{\citenamefont {Davier}\ and\ \citenamefont
  {Ngoc}(1989)}]{davier1989unambiguous}%
  \BibitemOpen
  \bibfield  {author} {\bibinfo {author} {\bibfnamefont {M.}~\bibnamefont
  {Davier}}\ and\ \bibinfo {author} {\bibfnamefont {H.~N.}\ \bibnamefont
  {Ngoc}},\ }\bibfield  {title} {\bibinfo {title} {An unambiguous search for a
  light higgs boson},\ }\href@noop {} {\bibfield  {journal} {\bibinfo
  {journal} {Physics Letters B}\ }\textbf {\bibinfo {volume} {229}},\ \bibinfo
  {pages} {150} (\bibinfo {year} {1989})}\BibitemShut {NoStop}%
\bibitem [{\citenamefont {Bross}\ \emph {et~al.}(1991)\citenamefont {Bross},
  \citenamefont {Crisler}, \citenamefont {Pordes}, \citenamefont {Volk},
  \citenamefont {Errede},\ and\ \citenamefont {Wrbanek}}]{bross1991search}%
  \BibitemOpen
  \bibfield  {author} {\bibinfo {author} {\bibfnamefont {A.}~\bibnamefont
  {Bross}}, \bibinfo {author} {\bibfnamefont {M.}~\bibnamefont {Crisler}},
  \bibinfo {author} {\bibfnamefont {S.}~\bibnamefont {Pordes}}, \bibinfo
  {author} {\bibfnamefont {J.}~\bibnamefont {Volk}}, \bibinfo {author}
  {\bibfnamefont {S.}~\bibnamefont {Errede}},\ and\ \bibinfo {author}
  {\bibfnamefont {J.}~\bibnamefont {Wrbanek}},\ }\bibfield  {title} {\bibinfo
  {title} {Search for short-lived particles produced in an electron beam
  dump},\ }\href@noop {} {\bibfield  {journal} {\bibinfo  {journal} {Phys. Rev.
  Lett.}\ }\textbf {\bibinfo {volume} {67}},\ \bibinfo {pages} {2942} (\bibinfo
  {year} {1991})}\BibitemShut {NoStop}%
\bibitem [{\citenamefont {Abrahamyan}\ \emph {et~al.}(2011)\citenamefont
  {Abrahamyan}, \citenamefont {Ahmed}, \citenamefont {Allada}, \citenamefont
  {Anez}, \citenamefont {Averett}, \citenamefont {Barbieri}, \citenamefont
  {Bartlett}, \citenamefont {Beacham}, \citenamefont {Bono}, \citenamefont
  {Boyce} \emph {et~al.}}]{abrahamyan2011search}%
  \BibitemOpen
  \bibfield  {author} {\bibinfo {author} {\bibfnamefont {S.}~\bibnamefont
  {Abrahamyan}}, \bibinfo {author} {\bibfnamefont {Z.}~\bibnamefont {Ahmed}},
  \bibinfo {author} {\bibfnamefont {K.}~\bibnamefont {Allada}}, \bibinfo
  {author} {\bibfnamefont {D.}~\bibnamefont {Anez}}, \bibinfo {author}
  {\bibfnamefont {T.}~\bibnamefont {Averett}}, \bibinfo {author} {\bibfnamefont
  {A.}~\bibnamefont {Barbieri}}, \bibinfo {author} {\bibfnamefont
  {K.}~\bibnamefont {Bartlett}}, \bibinfo {author} {\bibfnamefont
  {J.}~\bibnamefont {Beacham}}, \bibinfo {author} {\bibfnamefont
  {J.}~\bibnamefont {Bono}}, \bibinfo {author} {\bibfnamefont {J.}~\bibnamefont
  {Boyce}}, \emph {et~al.},\ }\bibfield  {title} {\bibinfo {title} {Search for
  a new gauge boson in electron-nucleus fixed-target scattering by the
  $\mathrm{APEX}$ experiment},\ }\href@noop {} {\bibfield  {journal} {\bibinfo
  {journal} {Phys. Rev. Lett.}\ }\textbf {\bibinfo {volume} {107}},\ \bibinfo
  {pages} {191804} (\bibinfo {year} {2011})}\BibitemShut {NoStop}%
\bibitem [{\citenamefont {Mendon{\c{c}}a}(2007)}]{mendoncca2007axion}%
  \BibitemOpen
  \bibfield  {author} {\bibinfo {author} {\bibfnamefont {J.}~\bibnamefont
  {Mendon{\c{c}}a}},\ }\bibfield  {title} {\bibinfo {title} {Axion excitation
  by intense laser fields},\ }\href@noop {} {\bibfield  {journal} {\bibinfo
  {journal} {Europhysics Letters}\ }\textbf {\bibinfo {volume} {79}},\ \bibinfo
  {pages} {21001} (\bibinfo {year} {2007})}\BibitemShut {NoStop}%
\bibitem [{\citenamefont {Gies}(2009)}]{gies2009strong}%
  \BibitemOpen
  \bibfield  {author} {\bibinfo {author} {\bibfnamefont {H.}~\bibnamefont
  {Gies}},\ }\bibfield  {title} {\bibinfo {title} {Strong laser fields as a
  probe for fundamental physics},\ }\href@noop {} {\bibfield  {journal}
  {\bibinfo  {journal} {The European Physical Journal D}\ }\textbf {\bibinfo
  {volume} {55}},\ \bibinfo {pages} {311} (\bibinfo {year} {2009})}\BibitemShut
  {NoStop}%
\bibitem [{\citenamefont {D{\"o}brich}\ and\ \citenamefont
  {Gies}(2010)}]{dobrich2010axion}%
  \BibitemOpen
  \bibfield  {author} {\bibinfo {author} {\bibfnamefont {B.}~\bibnamefont
  {D{\"o}brich}}\ and\ \bibinfo {author} {\bibfnamefont {H.}~\bibnamefont
  {Gies}},\ }\bibfield  {title} {\bibinfo {title} {Axion-like-particle search
  with high-intensity lasers},\ }\href@noop {} {\bibfield  {journal} {\bibinfo
  {journal} {Journal of High Energy Physics}\ }\textbf {\bibinfo {volume}
  {2010}},\ \bibinfo {pages} {1} (\bibinfo {year} {2010})}\BibitemShut
  {NoStop}%
\bibitem [{\citenamefont {Dillon}\ and\ \citenamefont
  {King}(2018)}]{dillon2018alp}%
  \BibitemOpen
  \bibfield  {author} {\bibinfo {author} {\bibfnamefont {B.~M.}\ \bibnamefont
  {Dillon}}\ and\ \bibinfo {author} {\bibfnamefont {B.}~\bibnamefont {King}},\
  }\bibfield  {title} {\bibinfo {title} {$\mathrm{ALP}$ production through
  non-linear $\mathrm{C}$ompton scattering in intense fields},\ }\href@noop {}
  {\bibfield  {journal} {\bibinfo  {journal} {The European Physical Journal C}\
  }\textbf {\bibinfo {volume} {78}},\ \bibinfo {pages} {1} (\bibinfo {year}
  {2018})}\BibitemShut {NoStop}%
\bibitem [{\citenamefont {King}\ \emph {et~al.}(2019)\citenamefont {King},
  \citenamefont {Dillon}, \citenamefont {Beyer},\ and\ \citenamefont
  {Gregori}}]{king2019axion}%
  \BibitemOpen
  \bibfield  {author} {\bibinfo {author} {\bibfnamefont {B.}~\bibnamefont
  {King}}, \bibinfo {author} {\bibfnamefont {B.}~\bibnamefont {Dillon}},
  \bibinfo {author} {\bibfnamefont {K.}~\bibnamefont {Beyer}},\ and\ \bibinfo
  {author} {\bibfnamefont {G.}~\bibnamefont {Gregori}},\ }\bibfield  {title}
  {\bibinfo {title} {Axion-like-particle decay in strong electromagnetic
  backgrounds},\ }\href@noop {} {\bibfield  {journal} {\bibinfo  {journal}
  {Journal of High Energy Physics}\ }\textbf {\bibinfo {volume} {2019}},\
  \bibinfo {pages} {1} (\bibinfo {year} {2019})}\BibitemShut {NoStop}%
\bibitem [{\citenamefont {Ma}\ and\ \citenamefont
  {Li}(2025)}]{PhysRevD.111.055001}%
  \BibitemOpen
  \bibfield  {author} {\bibinfo {author} {\bibfnamefont {K.}~\bibnamefont
  {Ma}}\ and\ \bibinfo {author} {\bibfnamefont {T.}~\bibnamefont {Li}},\
  }\bibfield  {title} {\bibinfo {title} {Laser induced compton scattering to
  dark photon or axionlike particle},\ }\href
  {https://doi.org/10.1103/PhysRevD.111.055001} {\bibfield  {journal} {\bibinfo
   {journal} {Phys. Rev. D}\ }\textbf {\bibinfo {volume} {111}},\ \bibinfo
  {pages} {055001} (\bibinfo {year} {2025})}\BibitemShut {NoStop}%
\bibitem [{\citenamefont {Wolkow}(1935)}]{wolkow1935klasse}%
  \BibitemOpen
  \bibfield  {author} {\bibinfo {author} {\bibfnamefont {D.~M.}\ \bibnamefont
  {Wolkow}},\ }\bibfield  {title} {\bibinfo {title} {{\"U}ber eine klasse von
  l{\"o}sungen der diracschen gleichung},\ }\href@noop {} {\bibfield  {journal}
  {\bibinfo  {journal} {Zeitschrift f{\"u}r Physik}\ }\textbf {\bibinfo
  {volume} {94}},\ \bibinfo {pages} {250} (\bibinfo {year} {1935})}\BibitemShut
  {NoStop}%
\bibitem [{\citenamefont {King}(2018)}]{King:2018qbq}%
  \BibitemOpen
  \bibfield  {author} {\bibinfo {author} {\bibfnamefont {B.}~\bibnamefont
  {King}},\ }\bibfield  {title} {\bibinfo {title} {{Electron-seeded
  $\mathrm{ALP}$ production and $\mathrm{ALP}$ decay in an oscillating
  electromagnetic field}},\ }\href
  {https://doi.org/10.1016/j.physletb.2018.06.016} {\bibfield  {journal}
  {\bibinfo  {journal} {Phys. Lett. B}\ }\textbf {\bibinfo {volume} {782}},\
  \bibinfo {pages} {737} (\bibinfo {year} {2018})},\ \Eprint
  {https://arxiv.org/abs/1802.07507} {arXiv:1802.07507 [hep-ph]} \BibitemShut
  {NoStop}%
\bibitem [{\citenamefont {Dillon}\ and\ \citenamefont
  {King}(2019)}]{PhysRevD.99.035048}%
  \BibitemOpen
  \bibfield  {author} {\bibinfo {author} {\bibfnamefont {B.~M.}\ \bibnamefont
  {Dillon}}\ and\ \bibinfo {author} {\bibfnamefont {B.}~\bibnamefont {King}},\
  }\bibfield  {title} {\bibinfo {title} {Light scalars: Coherent nonlinear
  thomson scattering and detection},\ }\href
  {https://doi.org/10.1103/PhysRevD.99.035048} {\bibfield  {journal} {\bibinfo
  {journal} {Phys. Rev. D}\ }\textbf {\bibinfo {volume} {99}},\ \bibinfo
  {pages} {035048} (\bibinfo {year} {2019})}\BibitemShut {NoStop}%
\bibitem [{\citenamefont {Bai}\ \emph {et~al.}(2022)\citenamefont {Bai},
  \citenamefont {Blackburn}, \citenamefont {Borysov}, \citenamefont {Davidi},
  \citenamefont {Hartin}, \citenamefont {Heinemann}, \citenamefont {Ma},
  \citenamefont {Perez}, \citenamefont {Santra}, \citenamefont {Soreq},\ and\
  \citenamefont {Hod}}]{PhysRevD.106.115034}%
  \BibitemOpen
  \bibfield  {author} {\bibinfo {author} {\bibfnamefont {Z.}~\bibnamefont
  {Bai}}, \bibinfo {author} {\bibfnamefont {T.}~\bibnamefont {Blackburn}},
  \bibinfo {author} {\bibfnamefont {O.}~\bibnamefont {Borysov}}, \bibinfo
  {author} {\bibfnamefont {O.}~\bibnamefont {Davidi}}, \bibinfo {author}
  {\bibfnamefont {A.}~\bibnamefont {Hartin}}, \bibinfo {author} {\bibfnamefont
  {B.}~\bibnamefont {Heinemann}}, \bibinfo {author} {\bibfnamefont
  {T.}~\bibnamefont {Ma}}, \bibinfo {author} {\bibfnamefont {G.}~\bibnamefont
  {Perez}}, \bibinfo {author} {\bibfnamefont {A.}~\bibnamefont {Santra}},
  \bibinfo {author} {\bibfnamefont {Y.}~\bibnamefont {Soreq}},\ and\ \bibinfo
  {author} {\bibfnamefont {N.~T.}\ \bibnamefont {Hod}},\ }\bibfield  {title}
  {\bibinfo {title} {New physics searches with an optical dump at
  $\mathrm{LUXE}$},\ }\href {https://doi.org/10.1103/PhysRevD.106.115034}
  {\bibfield  {journal} {\bibinfo  {journal} {Phys. Rev. D}\ }\textbf {\bibinfo
  {volume} {106}},\ \bibinfo {pages} {115034} (\bibinfo {year}
  {2022})}\BibitemShut {NoStop}%
\bibitem [{\citenamefont {Huang}\ \emph {et~al.}(2022)\citenamefont {Huang},
  \citenamefont {Shen}, \citenamefont {Bu}, \citenamefont {Zhang},
  \citenamefont {Ji},\ and\ \citenamefont {Zhai}}]{huang2022axion}%
  \BibitemOpen
  \bibfield  {author} {\bibinfo {author} {\bibfnamefont {S.}~\bibnamefont
  {Huang}}, \bibinfo {author} {\bibfnamefont {B.}~\bibnamefont {Shen}},
  \bibinfo {author} {\bibfnamefont {Z.}~\bibnamefont {Bu}}, \bibinfo {author}
  {\bibfnamefont {X.}~\bibnamefont {Zhang}}, \bibinfo {author} {\bibfnamefont
  {L.}~\bibnamefont {Ji}},\ and\ \bibinfo {author} {\bibfnamefont
  {S.}~\bibnamefont {Zhai}},\ }\bibfield  {title} {\bibinfo {title} {Axion-like
  particle generation in laser-plasma interaction},\ }\href@noop {} {\bibfield
  {journal} {\bibinfo  {journal} {Physica Scripta}\ }\textbf {\bibinfo {volume}
  {97}},\ \bibinfo {pages} {105303} (\bibinfo {year} {2022})}\BibitemShut
  {NoStop}%
\bibitem [{\citenamefont {An}\ \emph {et~al.}(2024)\citenamefont {An},
  \citenamefont {Chen}, \citenamefont {Liu}, \citenamefont {Sheng},\ and\
  \citenamefont {Zhang}}]{an2024modeling}%
  \BibitemOpen
  \bibfield  {author} {\bibinfo {author} {\bibfnamefont {X.}~\bibnamefont
  {An}}, \bibinfo {author} {\bibfnamefont {M.}~\bibnamefont {Chen}}, \bibinfo
  {author} {\bibfnamefont {J.}~\bibnamefont {Liu}}, \bibinfo {author}
  {\bibfnamefont {Z.}~\bibnamefont {Sheng}},\ and\ \bibinfo {author}
  {\bibfnamefont {J.}~\bibnamefont {Zhang}},\ }\bibfield  {title} {\bibinfo
  {title} {Modeling of axion and electromagnetic fields interaction in
  particle-in-cell simulations},\ }\href@noop {} {\bibfield  {journal}
  {\bibinfo  {journal} {Matter and Radiation at Extremes}\ }\textbf {\bibinfo
  {volume} {9}} (\bibinfo {year} {2024})}\BibitemShut {NoStop}%
\bibitem [{\citenamefont {An}\ \emph {et~al.}(2025)\citenamefont {An},
  \citenamefont {Chen}, \citenamefont {Liu}, \citenamefont {Bai}, \citenamefont
  {Ji}, \citenamefont {Sheng},\ and\ \citenamefont {Zhang}}]{an2025situ}%
  \BibitemOpen
  \bibfield  {author} {\bibinfo {author} {\bibfnamefont {X.}~\bibnamefont
  {An}}, \bibinfo {author} {\bibfnamefont {M.}~\bibnamefont {Chen}}, \bibinfo
  {author} {\bibfnamefont {J.}~\bibnamefont {Liu}}, \bibinfo {author}
  {\bibfnamefont {Z.}~\bibnamefont {Bai}}, \bibinfo {author} {\bibfnamefont
  {L.}~\bibnamefont {Ji}}, \bibinfo {author} {\bibfnamefont {Z.}~\bibnamefont
  {Sheng}},\ and\ \bibinfo {author} {\bibfnamefont {J.}~\bibnamefont {Zhang}},\
  }\bibfield  {title} {\bibinfo {title} {In situ axion generation and detection
  in laser-driven wakefields},\ }\href@noop {} {\bibfield  {journal} {\bibinfo
  {journal} {arXiv preprint arXiv:2504.12500}\ } (\bibinfo {year}
  {2025})}\BibitemShut {NoStop}%
\bibitem [{\citenamefont {Yoon}\ \emph {et~al.}(2021)\citenamefont {Yoon},
  \citenamefont {Kim}, \citenamefont {Choi}, \citenamefont {Sung},
  \citenamefont {Lee}, \citenamefont {Lee},\ and\ \citenamefont
  {Nam}}]{Yoon:21}%
  \BibitemOpen
  \bibfield  {author} {\bibinfo {author} {\bibfnamefont {J.~W.}\ \bibnamefont
  {Yoon}}, \bibinfo {author} {\bibfnamefont {Y.~G.}\ \bibnamefont {Kim}},
  \bibinfo {author} {\bibfnamefont {I.~W.}\ \bibnamefont {Choi}}, \bibinfo
  {author} {\bibfnamefont {J.~H.}\ \bibnamefont {Sung}}, \bibinfo {author}
  {\bibfnamefont {H.~W.}\ \bibnamefont {Lee}}, \bibinfo {author} {\bibfnamefont
  {S.~K.}\ \bibnamefont {Lee}},\ and\ \bibinfo {author} {\bibfnamefont {C.~H.}\
  \bibnamefont {Nam}},\ }\bibfield  {title} {\bibinfo {title} {Realization of
  laser intensity over $10^{23}$ $\mathrm{W}$/cm$^2$},\ }\href
  {https://doi.org/10.1364/OPTICA.420520} {\bibfield  {journal} {\bibinfo
  {journal} {Optica}\ }\textbf {\bibinfo {volume} {8}},\ \bibinfo {pages} {630}
  (\bibinfo {year} {2021})}\BibitemShut {NoStop}%
\bibitem [{\citenamefont {Baier}\ and\ \citenamefont
  {Katkov}(2005)}]{baier2005concept}%
  \BibitemOpen
  \bibfield  {author} {\bibinfo {author} {\bibfnamefont {V.~N.}\ \bibnamefont
  {Baier}}\ and\ \bibinfo {author} {\bibfnamefont {V.~M.}\ \bibnamefont
  {Katkov}},\ }\bibfield  {title} {\bibinfo {title} {Concept of formation
  length in radiation theory},\ }\href@noop {} {\bibfield  {journal} {\bibinfo
  {journal} {Physics Reports}\ }\textbf {\bibinfo {volume} {409}},\ \bibinfo
  {pages} {261} (\bibinfo {year} {2005})}\BibitemShut {NoStop}%
\bibitem [{\citenamefont {Ritus}(1985)}]{ritus1985quantum}%
  \BibitemOpen
  \bibfield  {author} {\bibinfo {author} {\bibfnamefont {V.}~\bibnamefont
  {Ritus}},\ }\bibfield  {title} {\bibinfo {title} {Quantum effects of the
  interaction of elementary particles with an intense electromagnetic field},\
  }\href@noop {} {\bibfield  {journal} {\bibinfo  {journal} {J. Sov. Laser
  Res.}\ }\textbf {\bibinfo {volume} {6}} (\bibinfo {year} {1985})}\BibitemShut
  {NoStop}%
\bibitem [{\citenamefont {Di~Piazza}\ \emph {et~al.}(2010)\citenamefont
  {Di~Piazza}, \citenamefont {Hatsagortsyan},\ and\ \citenamefont
  {Keitel}}]{di2010quantum}%
  \BibitemOpen
  \bibfield  {author} {\bibinfo {author} {\bibfnamefont {A.}~\bibnamefont
  {Di~Piazza}}, \bibinfo {author} {\bibfnamefont {K.}~\bibnamefont
  {Hatsagortsyan}},\ and\ \bibinfo {author} {\bibfnamefont {C.~H.}\
  \bibnamefont {Keitel}},\ }\bibfield  {title} {\bibinfo {title} {Quantum
  radiation reaction effects in multiphoton $\mathrm{C}$ompton scattering},\
  }\href@noop {} {\bibfield  {journal} {\bibinfo  {journal} {Phys. Rev. Lett.}\
  }\textbf {\bibinfo {volume} {105}},\ \bibinfo {pages} {220403} (\bibinfo
  {year} {2010})}\BibitemShut {NoStop}%
\bibitem [{\citenamefont {Elkina}\ \emph {et~al.}(2011)\citenamefont {Elkina},
  \citenamefont {Fedotov}, \citenamefont {Kostyukov}, \citenamefont {Legkov},
  \citenamefont {Narozhny}, \citenamefont {Nerush},\ and\ \citenamefont
  {Ruhl}}]{elkina2011qed}%
  \BibitemOpen
  \bibfield  {author} {\bibinfo {author} {\bibfnamefont {N.}~\bibnamefont
  {Elkina}}, \bibinfo {author} {\bibfnamefont {A.}~\bibnamefont {Fedotov}},
  \bibinfo {author} {\bibfnamefont {I.~Y.}\ \bibnamefont {Kostyukov}}, \bibinfo
  {author} {\bibfnamefont {M.}~\bibnamefont {Legkov}}, \bibinfo {author}
  {\bibfnamefont {N.}~\bibnamefont {Narozhny}}, \bibinfo {author}
  {\bibfnamefont {E.}~\bibnamefont {Nerush}},\ and\ \bibinfo {author}
  {\bibfnamefont {H.}~\bibnamefont {Ruhl}},\ }\bibfield  {title} {\bibinfo
  {title} {$\mathrm{QED}$ cascades induced by circularly polarized laser
  fields},\ }\href@noop {} {\bibfield  {journal} {\bibinfo  {journal} {Phys.
  Rev. ST Accel. Beams}\ }\textbf {\bibinfo {volume} {14}},\ \bibinfo {pages}
  {054401} (\bibinfo {year} {2011})}\BibitemShut {NoStop}%
\bibitem [{\citenamefont {Ridgers}\ \emph {et~al.}(2014)\citenamefont
  {Ridgers}, \citenamefont {Kirk}, \citenamefont {Duclous}, \citenamefont
  {Blackburn}, \citenamefont {Brady}, \citenamefont {Bennett}, \citenamefont
  {Arber},\ and\ \citenamefont {Bell}}]{ridgers2014modelling}%
  \BibitemOpen
  \bibfield  {author} {\bibinfo {author} {\bibfnamefont {C.~P.}\ \bibnamefont
  {Ridgers}}, \bibinfo {author} {\bibfnamefont {J.~G.}\ \bibnamefont {Kirk}},
  \bibinfo {author} {\bibfnamefont {R.}~\bibnamefont {Duclous}}, \bibinfo
  {author} {\bibfnamefont {T.~G.}\ \bibnamefont {Blackburn}}, \bibinfo {author}
  {\bibfnamefont {C.~S.}\ \bibnamefont {Brady}}, \bibinfo {author}
  {\bibfnamefont {K.}~\bibnamefont {Bennett}}, \bibinfo {author} {\bibfnamefont
  {T.~D.}\ \bibnamefont {Arber}},\ and\ \bibinfo {author} {\bibfnamefont
  {A.~R.}\ \bibnamefont {Bell}},\ }\bibfield  {title} {\bibinfo {title}
  {Modelling gamma-ray photon emission and pair production in high-intensity
  laser--matter interactions},\ }\href@noop {} {\bibfield  {journal} {\bibinfo
  {journal} {Journal of Computational Physics}\ }\textbf {\bibinfo {volume}
  {260}},\ \bibinfo {pages} {273} (\bibinfo {year} {2014})}\BibitemShut
  {NoStop}%
\bibitem [{\citenamefont {Green}\ and\ \citenamefont
  {Harvey}(2015)}]{green2015simla}%
  \BibitemOpen
  \bibfield  {author} {\bibinfo {author} {\bibfnamefont {D.}~\bibnamefont
  {Green}}\ and\ \bibinfo {author} {\bibfnamefont {C.}~\bibnamefont {Harvey}},\
  }\bibfield  {title} {\bibinfo {title} {Simla: Simulating particle dynamics in
  intense laser and other electromagnetic fields via classical and quantum
  electrodynamics},\ }\href@noop {} {\bibfield  {journal} {\bibinfo  {journal}
  {Computer Physics Communications}\ }\textbf {\bibinfo {volume} {192}},\
  \bibinfo {pages} {313} (\bibinfo {year} {2015})}\BibitemShut {NoStop}%
\bibitem [{\citenamefont {Gonoskov}\ \emph {et~al.}(2015)\citenamefont
  {Gonoskov}, \citenamefont {Bastrakov}, \citenamefont {Efimenko},
  \citenamefont {Ilderton}, \citenamefont {Marklund}, \citenamefont {Meyerov},
  \citenamefont {Muraviev}, \citenamefont {Sergeev}, \citenamefont {Surmin},\
  and\ \citenamefont {Wallin}}]{gonoskov2015extended}%
  \BibitemOpen
  \bibfield  {author} {\bibinfo {author} {\bibfnamefont {A.}~\bibnamefont
  {Gonoskov}}, \bibinfo {author} {\bibfnamefont {S.}~\bibnamefont {Bastrakov}},
  \bibinfo {author} {\bibfnamefont {E.}~\bibnamefont {Efimenko}}, \bibinfo
  {author} {\bibfnamefont {A.}~\bibnamefont {Ilderton}}, \bibinfo {author}
  {\bibfnamefont {M.}~\bibnamefont {Marklund}}, \bibinfo {author}
  {\bibfnamefont {I.}~\bibnamefont {Meyerov}}, \bibinfo {author} {\bibfnamefont
  {A.}~\bibnamefont {Muraviev}}, \bibinfo {author} {\bibfnamefont
  {A.}~\bibnamefont {Sergeev}}, \bibinfo {author} {\bibfnamefont
  {I.}~\bibnamefont {Surmin}},\ and\ \bibinfo {author} {\bibfnamefont
  {E.}~\bibnamefont {Wallin}},\ }\bibfield  {title} {\bibinfo {title} {Extended
  particle-in-cell schemes for physics in ultrastrong laser fields: Review and
  developments},\ }\href@noop {} {\bibfield  {journal} {\bibinfo  {journal}
  {Phys. Rev. E}\ }\textbf {\bibinfo {volume} {92}},\ \bibinfo {pages} {023305}
  (\bibinfo {year} {2015})}\BibitemShut {NoStop}%
\bibitem [{\citenamefont {Zhuang}\ \emph {et~al.}(2023)\citenamefont {Zhuang},
  \citenamefont {Chen}, \citenamefont {Li}, \citenamefont {Hatsagortsyan},\
  and\ \citenamefont {Keitel}}]{zhuang2023laser}%
  \BibitemOpen
  \bibfield  {author} {\bibinfo {author} {\bibfnamefont {K.-H.}\ \bibnamefont
  {Zhuang}}, \bibinfo {author} {\bibfnamefont {Y.-Y.}\ \bibnamefont {Chen}},
  \bibinfo {author} {\bibfnamefont {Y.-F.}\ \bibnamefont {Li}}, \bibinfo
  {author} {\bibfnamefont {K.~Z.}\ \bibnamefont {Hatsagortsyan}},\ and\
  \bibinfo {author} {\bibfnamefont {C.~H.}\ \bibnamefont {Keitel}},\ }\bibfield
   {title} {\bibinfo {title} {Laser-driven lepton polarization in the quantum
  radiation-dominated reflection regime},\ }\href@noop {} {\bibfield  {journal}
  {\bibinfo  {journal} {Phys. Rev. D}\ }\textbf {\bibinfo {volume} {108}},\
  \bibinfo {pages} {033001} (\bibinfo {year} {2023})}\BibitemShut {NoStop}%
\bibitem [{\citenamefont {Schwinger}(1954)}]{schwinger1954quantum}%
  \BibitemOpen
  \bibfield  {author} {\bibinfo {author} {\bibfnamefont {J.}~\bibnamefont
  {Schwinger}},\ }\bibfield  {title} {\bibinfo {title} {The quantum correction
  in the radiation by energetic accelerated electrons},\ }\href@noop {}
  {\bibfield  {journal} {\bibinfo  {journal} {Proceedings of the National
  Academy of Sciences}\ }\textbf {\bibinfo {volume} {40}},\ \bibinfo {pages}
  {132} (\bibinfo {year} {1954})}\BibitemShut {NoStop}%
\bibitem [{\citenamefont {Baier}\ and\ \citenamefont
  {Katkov}(1968{\natexlab{a}})}]{baier1968processes}%
  \BibitemOpen
  \bibfield  {author} {\bibinfo {author} {\bibfnamefont {V.}~\bibnamefont
  {Baier}}\ and\ \bibinfo {author} {\bibfnamefont {V.}~\bibnamefont {Katkov}},\
  }\bibfield  {title} {\bibinfo {title} {Processes involved in the motion of
  high energy particles in a magnetic field},\ }\href@noop {} {\bibfield
  {journal} {\bibinfo  {journal} {Sov. Phys. JETP}\ }\textbf {\bibinfo {volume}
  {26}},\ \bibinfo {pages} {854} (\bibinfo {year}
  {1968}{\natexlab{a}})}\BibitemShut {NoStop}%
\bibitem [{\citenamefont {Baier}\ and\ \citenamefont
  {Katkov}(1968{\natexlab{b}})}]{baier1968quasiclassical}%
  \BibitemOpen
  \bibfield  {author} {\bibinfo {author} {\bibfnamefont {V.}~\bibnamefont
  {Baier}}\ and\ \bibinfo {author} {\bibfnamefont {V.}~\bibnamefont {Katkov}},\
  }\bibfield  {title} {\bibinfo {title} {Quasiclassical theory of
  bremsstrahlung by relativistic particles},\ }\href@noop {} {\bibfield
  {journal} {\bibinfo  {journal} {Sov. Phys. JETP}\ }\textbf {\bibinfo {volume}
  {26}},\ \bibinfo {pages} {854} (\bibinfo {year}
  {1968}{\natexlab{b}})}\BibitemShut {NoStop}%
\bibitem [{\citenamefont {Berestetskii}\ \emph {et~al.}(2012)\citenamefont
  {Berestetskii}, \citenamefont {Pitaevskii},\ and\ \citenamefont
  {Lifshitz}}]{berestetskii2012quantum}%
  \BibitemOpen
  \bibfield  {author} {\bibinfo {author} {\bibfnamefont {V.~B.}\ \bibnamefont
  {Berestetskii}}, \bibinfo {author} {\bibfnamefont {L.~P.}\ \bibnamefont
  {Pitaevskii}},\ and\ \bibinfo {author} {\bibfnamefont {E.~M.}\ \bibnamefont
  {Lifshitz}},\ }\href@noop {} {\emph {\bibinfo {title} {Quantum
  Electrodynamics: Volume 4}}},\ Vol.~\bibinfo {volume} {4}\ (\bibinfo
  {publisher} {Elsevier},\ \bibinfo {year} {2012})\BibitemShut {NoStop}%
\bibitem [{\citenamefont {Baier}\ \emph {et~al.}(1998)\citenamefont {Baier},
  \citenamefont {Katkov},\ and\ \citenamefont
  {Strakhovenko}}]{baier1998electromagnetic}%
  \BibitemOpen
  \bibfield  {author} {\bibinfo {author} {\bibfnamefont {V.~N.}\ \bibnamefont
  {Baier}}, \bibinfo {author} {\bibfnamefont {V.~M.}\ \bibnamefont {Katkov}},\
  and\ \bibinfo {author} {\bibfnamefont {V.~M.}\ \bibnamefont {Strakhovenko}},\
  }\bibfield  {title} {\bibinfo {title} {Electromagnetic processes at high
  energies in oriented single crystals},\ }\href@noop {} {\bibfield  {journal}
  {\bibinfo  {journal} {World Scientific}\ } (\bibinfo {year}
  {1998})}\BibitemShut {NoStop}%
\bibitem [{\citenamefont {Baier}\ and\ \citenamefont
  {Katkov}(2009)}]{baier2009recent}%
  \BibitemOpen
  \bibfield  {author} {\bibinfo {author} {\bibfnamefont {V.}~\bibnamefont
  {Baier}}\ and\ \bibinfo {author} {\bibfnamefont {V.}~\bibnamefont {Katkov}},\
  }\bibfield  {title} {\bibinfo {title} {Recent development of quasiclassical
  operator method},\ }in\ \href@noop {} {\emph {\bibinfo {booktitle} {Journal
  of Physics: Conference Series}}},\ Vol.\ \bibinfo {volume} {198}\ (\bibinfo
  {organization} {IOP Publishing},\ \bibinfo {year} {2009})\ p.\ \bibinfo
  {pages} {012003}\BibitemShut {NoStop}%
\bibitem [{\citenamefont {Bogdanov}\ \emph {et~al.}(2019)\citenamefont
  {Bogdanov}, \citenamefont {Kazinski},\ and\ \citenamefont
  {Lazarenko}}]{bogdanov2019semiclassical}%
  \BibitemOpen
  \bibfield  {author} {\bibinfo {author} {\bibfnamefont {O.~V.}\ \bibnamefont
  {Bogdanov}}, \bibinfo {author} {\bibfnamefont {P.}~\bibnamefont {Kazinski}},\
  and\ \bibinfo {author} {\bibfnamefont {G.~Y.}\ \bibnamefont {Lazarenko}},\
  }\bibfield  {title} {\bibinfo {title} {Semiclassical probability of radiation
  of twisted photons in the ultrarelativistic limit},\ }\href@noop {}
  {\bibfield  {journal} {\bibinfo  {journal} {Phys. Rev. D}\ }\textbf {\bibinfo
  {volume} {99}},\ \bibinfo {pages} {116016} (\bibinfo {year}
  {2019})}\BibitemShut {NoStop}%
\bibitem [{\citenamefont {Chen}\ \emph {et~al.}(2022)\citenamefont {Chen},
  \citenamefont {Hatsagortsyan}, \citenamefont {Keitel},\ and\ \citenamefont
  {Shaisultanov}}]{chen2022electron}%
  \BibitemOpen
  \bibfield  {author} {\bibinfo {author} {\bibfnamefont {Y.-Y.}\ \bibnamefont
  {Chen}}, \bibinfo {author} {\bibfnamefont {K.~Z.}\ \bibnamefont
  {Hatsagortsyan}}, \bibinfo {author} {\bibfnamefont {C.~H.}\ \bibnamefont
  {Keitel}},\ and\ \bibinfo {author} {\bibfnamefont {R.}~\bibnamefont
  {Shaisultanov}},\ }\bibfield  {title} {\bibinfo {title} {Electron spin-and
  photon polarization-resolved probabilities of strong-field $\mathrm{QED}$
  processes},\ }\href@noop {} {\bibfield  {journal} {\bibinfo  {journal} {Phys.
  Rev. D}\ }\textbf {\bibinfo {volume} {105}},\ \bibinfo {pages} {116013}
  (\bibinfo {year} {2022})}\BibitemShut {NoStop}%
\bibitem [{\citenamefont {Li}\ \emph {et~al.}(2023)\citenamefont {Li},
  \citenamefont {Chen}, \citenamefont {Hatsagortsyan}, \citenamefont
  {Di~Piazza}, \citenamefont {Tamburini},\ and\ \citenamefont
  {Keitel}}]{li2023strong}%
  \BibitemOpen
  \bibfield  {author} {\bibinfo {author} {\bibfnamefont {Y.-F.}\ \bibnamefont
  {Li}}, \bibinfo {author} {\bibfnamefont {Y.-Y.}\ \bibnamefont {Chen}},
  \bibinfo {author} {\bibfnamefont {K.~Z.}\ \bibnamefont {Hatsagortsyan}},
  \bibinfo {author} {\bibfnamefont {A.}~\bibnamefont {Di~Piazza}}, \bibinfo
  {author} {\bibfnamefont {M.}~\bibnamefont {Tamburini}},\ and\ \bibinfo
  {author} {\bibfnamefont {C.}~\bibnamefont {Keitel}},\ }\bibfield  {title}
  {\bibinfo {title} {Strong signature of one-loop self-energy in polarization
  resolved nonlinear compton scattering},\ }\href@noop {} {\bibfield  {journal}
  {\bibinfo  {journal} {Phys. Rev. D}\ }\textbf {\bibinfo {volume} {107}},\
  \bibinfo {pages} {116020} (\bibinfo {year} {2023})}\BibitemShut {NoStop}%
\bibitem [{\citenamefont {Bargmann}\ \emph {et~al.}(1959)\citenamefont
  {Bargmann}, \citenamefont {Michel},\ and\ \citenamefont
  {Telegdi}}]{bargmann1959precession}%
  \BibitemOpen
  \bibfield  {author} {\bibinfo {author} {\bibfnamefont {V.}~\bibnamefont
  {Bargmann}}, \bibinfo {author} {\bibfnamefont {L.}~\bibnamefont {Michel}},\
  and\ \bibinfo {author} {\bibfnamefont {V.}~\bibnamefont {Telegdi}},\
  }\bibfield  {title} {\bibinfo {title} {Precession of the polarization of
  particles moving in a homogeneous electromagnetic field},\ }\href@noop {}
  {\bibfield  {journal} {\bibinfo  {journal} {Phys. Rev. Lett.}\ }\textbf
  {\bibinfo {volume} {2}},\ \bibinfo {pages} {435} (\bibinfo {year}
  {1959})}\BibitemShut {NoStop}%
\bibitem [{\citenamefont {Li}\ \emph {et~al.}(2020{\natexlab{a}})\citenamefont
  {Li}, \citenamefont {Shaisultanov}, \citenamefont {Chen}, \citenamefont
  {Wan}, \citenamefont {Hatsagortsyan}, \citenamefont {Keitel},\ and\
  \citenamefont {Li}}]{li2020polarized}%
  \BibitemOpen
  \bibfield  {author} {\bibinfo {author} {\bibfnamefont {Y.-F.}\ \bibnamefont
  {Li}}, \bibinfo {author} {\bibfnamefont {R.}~\bibnamefont {Shaisultanov}},
  \bibinfo {author} {\bibfnamefont {Y.-Y.}\ \bibnamefont {Chen}}, \bibinfo
  {author} {\bibfnamefont {F.}~\bibnamefont {Wan}}, \bibinfo {author}
  {\bibfnamefont {K.~Z.}\ \bibnamefont {Hatsagortsyan}}, \bibinfo {author}
  {\bibfnamefont {C.~H.}\ \bibnamefont {Keitel}},\ and\ \bibinfo {author}
  {\bibfnamefont {J.-X.}\ \bibnamefont {Li}},\ }\bibfield  {title} {\bibinfo
  {title} {Polarized ultrashort brilliant multi-gev $\gamma$ rays via
  single-shot laser-electron interaction},\ }\href@noop {} {\bibfield
  {journal} {\bibinfo  {journal} {Phys. Rev. Lett.}\ }\textbf {\bibinfo
  {volume} {124}},\ \bibinfo {pages} {014801} (\bibinfo {year}
  {2020}{\natexlab{a}})}\BibitemShut {NoStop}%
\bibitem [{\citenamefont {Dai}\ \emph {et~al.}(2022)\citenamefont {Dai},
  \citenamefont {Shen}, \citenamefont {Li}, \citenamefont {Shaisultanov},
  \citenamefont {Hatsagortsyan}, \citenamefont {Keitel},\ and\ \citenamefont
  {Chen}}]{dai2022photon}%
  \BibitemOpen
  \bibfield  {author} {\bibinfo {author} {\bibfnamefont {Y.-N.}\ \bibnamefont
  {Dai}}, \bibinfo {author} {\bibfnamefont {B.-F.}\ \bibnamefont {Shen}},
  \bibinfo {author} {\bibfnamefont {J.-X.}\ \bibnamefont {Li}}, \bibinfo
  {author} {\bibfnamefont {R.}~\bibnamefont {Shaisultanov}}, \bibinfo {author}
  {\bibfnamefont {K.~Z.}\ \bibnamefont {Hatsagortsyan}}, \bibinfo {author}
  {\bibfnamefont {C.~H.}\ \bibnamefont {Keitel}},\ and\ \bibinfo {author}
  {\bibfnamefont {Y.-Y.}\ \bibnamefont {Chen}},\ }\bibfield  {title} {\bibinfo
  {title} {Photon polarization effects in polarized electron--positron pair
  production in a strong laser field},\ }\href@noop {} {\bibfield  {journal}
  {\bibinfo  {journal} {Matter and Radiation at Extremes}\ }\textbf {\bibinfo
  {volume} {7}} (\bibinfo {year} {2022})}\BibitemShut {NoStop}%
\bibitem [{\citenamefont {Dai}\ \emph {et~al.}(2024)\citenamefont {Dai},
  \citenamefont {Hatsagortsyan}, \citenamefont {Keitel},\ and\ \citenamefont
  {Chen}}]{PhysRevD.110.012008}%
  \BibitemOpen
  \bibfield  {author} {\bibinfo {author} {\bibfnamefont {Y.-N.}\ \bibnamefont
  {Dai}}, \bibinfo {author} {\bibfnamefont {K.~Z.}\ \bibnamefont
  {Hatsagortsyan}}, \bibinfo {author} {\bibfnamefont {C.~H.}\ \bibnamefont
  {Keitel}},\ and\ \bibinfo {author} {\bibfnamefont {Y.-Y.}\ \bibnamefont
  {Chen}},\ }\bibfield  {title} {\bibinfo {title} {Fermionic signal of vacuum
  polarization in strong laser fields},\ }\href
  {https://doi.org/10.1103/PhysRevD.110.012008} {\bibfield  {journal} {\bibinfo
   {journal} {Phys. Rev. D}\ }\textbf {\bibinfo {volume} {110}},\ \bibinfo
  {pages} {012008} (\bibinfo {year} {2024})}\BibitemShut {NoStop}%
\bibitem [{\citenamefont {Li}\ \emph {et~al.}(2019)\citenamefont {Li},
  \citenamefont {Shaisultanov}, \citenamefont {Hatsagortsyan}, \citenamefont
  {Wan}, \citenamefont {Keitel},\ and\ \citenamefont
  {Li}}]{li2019ultrarelativistic}%
  \BibitemOpen
  \bibfield  {author} {\bibinfo {author} {\bibfnamefont {Y.-F.}\ \bibnamefont
  {Li}}, \bibinfo {author} {\bibfnamefont {R.}~\bibnamefont {Shaisultanov}},
  \bibinfo {author} {\bibfnamefont {K.~Z.}\ \bibnamefont {Hatsagortsyan}},
  \bibinfo {author} {\bibfnamefont {F.}~\bibnamefont {Wan}}, \bibinfo {author}
  {\bibfnamefont {C.~H.}\ \bibnamefont {Keitel}},\ and\ \bibinfo {author}
  {\bibfnamefont {J.-X.}\ \bibnamefont {Li}},\ }\bibfield  {title} {\bibinfo
  {title} {Ultrarelativistic electron-beam polarization in single-shot
  interaction with an ultraintense laser pulse},\ }\href@noop {} {\bibfield
  {journal} {\bibinfo  {journal} {Phys. Rev. Lett.}\ }\textbf {\bibinfo
  {volume} {122}},\ \bibinfo {pages} {154801} (\bibinfo {year}
  {2019})}\BibitemShut {NoStop}%
\bibitem [{\citenamefont {Li}\ \emph {et~al.}(2022)\citenamefont {Li},
  \citenamefont {Chen}, \citenamefont {Hatsagortsyan},\ and\ \citenamefont
  {Keitel}}]{li2022helicity}%
  \BibitemOpen
  \bibfield  {author} {\bibinfo {author} {\bibfnamefont {Y.-F.}\ \bibnamefont
  {Li}}, \bibinfo {author} {\bibfnamefont {Y.-Y.}\ \bibnamefont {Chen}},
  \bibinfo {author} {\bibfnamefont {K.~Z.}\ \bibnamefont {Hatsagortsyan}},\
  and\ \bibinfo {author} {\bibfnamefont {C.~H.}\ \bibnamefont {Keitel}},\
  }\bibfield  {title} {\bibinfo {title} {Helicity transfer in strong laser
  fields via the electron anomalous magnetic moment},\ }\href@noop {}
  {\bibfield  {journal} {\bibinfo  {journal} {Phys. Rev. Lett.}\ }\textbf
  {\bibinfo {volume} {128}},\ \bibinfo {pages} {174801} (\bibinfo {year}
  {2022})}\BibitemShut {NoStop}%
\bibitem [{\citenamefont {Chen}\ \emph {et~al.}(2019)\citenamefont {Chen},
  \citenamefont {He}, \citenamefont {Shaisultanov}, \citenamefont
  {Hatsagortsyan},\ and\ \citenamefont {Keitel}}]{chen2019polarized}%
  \BibitemOpen
  \bibfield  {author} {\bibinfo {author} {\bibfnamefont {Y.-Y.}\ \bibnamefont
  {Chen}}, \bibinfo {author} {\bibfnamefont {P.-L.}\ \bibnamefont {He}},
  \bibinfo {author} {\bibfnamefont {R.}~\bibnamefont {Shaisultanov}}, \bibinfo
  {author} {\bibfnamefont {K.~Z.}\ \bibnamefont {Hatsagortsyan}},\ and\
  \bibinfo {author} {\bibfnamefont {C.~H.}\ \bibnamefont {Keitel}},\ }\bibfield
   {title} {\bibinfo {title} {Polarized positron beams via intense two-color
  laser pulses},\ }\href@noop {} {\bibfield  {journal} {\bibinfo  {journal}
  {Phys. Rev. Lett.}\ }\textbf {\bibinfo {volume} {123}},\ \bibinfo {pages}
  {174801} (\bibinfo {year} {2019})}\BibitemShut {NoStop}%
\bibitem [{\citenamefont {Li}\ \emph {et~al.}(2020{\natexlab{b}})\citenamefont
  {Li}, \citenamefont {Chen}, \citenamefont {Wang},\ and\ \citenamefont
  {Hu}}]{li2020production}%
  \BibitemOpen
  \bibfield  {author} {\bibinfo {author} {\bibfnamefont {Y.-F.}\ \bibnamefont
  {Li}}, \bibinfo {author} {\bibfnamefont {Y.-Y.}\ \bibnamefont {Chen}},
  \bibinfo {author} {\bibfnamefont {W.-M.}\ \bibnamefont {Wang}},\ and\
  \bibinfo {author} {\bibfnamefont {H.-S.}\ \bibnamefont {Hu}},\ }\bibfield
  {title} {\bibinfo {title} {Production of highly polarized positron beams via
  helicity transfer from polarized electrons in a strong laser field},\
  }\href@noop {} {\bibfield  {journal} {\bibinfo  {journal} {Phys. Rev. Lett.}\
  }\textbf {\bibinfo {volume} {125}},\ \bibinfo {pages} {044802} (\bibinfo
  {year} {2020}{\natexlab{b}})}\BibitemShut {NoStop}%
\bibitem [{\citenamefont {Gong}\ \emph {et~al.}(2021)\citenamefont {Gong},
  \citenamefont {Hatsagortsyan},\ and\ \citenamefont
  {Keitel}}]{gong2021retrieving}%
  \BibitemOpen
  \bibfield  {author} {\bibinfo {author} {\bibfnamefont {Z.}~\bibnamefont
  {Gong}}, \bibinfo {author} {\bibfnamefont {K.~Z.}\ \bibnamefont
  {Hatsagortsyan}},\ and\ \bibinfo {author} {\bibfnamefont {C.~H.}\
  \bibnamefont {Keitel}},\ }\bibfield  {title} {\bibinfo {title} {Retrieving
  transient magnetic fields of ultrarelativistic laser plasma via ejected
  electron polarization},\ }\href@noop {} {\bibfield  {journal} {\bibinfo
  {journal} {Phys. Rev. Lett.}\ }\textbf {\bibinfo {volume} {127}},\ \bibinfo
  {pages} {165002} (\bibinfo {year} {2021})}\BibitemShut {NoStop}%
\bibitem [{\citenamefont {Song}\ \emph {et~al.}(2022)\citenamefont {Song},
  \citenamefont {Wang},\ and\ \citenamefont {Li}}]{PhysRevLett.129.035001}%
  \BibitemOpen
  \bibfield  {author} {\bibinfo {author} {\bibfnamefont {H.-H.}\ \bibnamefont
  {Song}}, \bibinfo {author} {\bibfnamefont {W.-M.}\ \bibnamefont {Wang}},\
  and\ \bibinfo {author} {\bibfnamefont {Y.-T.}\ \bibnamefont {Li}},\
  }\bibfield  {title} {\bibinfo {title} {Dense polarized positrons from
  laser-irradiated foil targets in the $\mathrm{QED}$ regime},\ }\href
  {https://doi.org/10.1103/PhysRevLett.129.035001} {\bibfield  {journal}
  {\bibinfo  {journal} {Phys. Rev. Lett.}\ }\textbf {\bibinfo {volume} {129}},\
  \bibinfo {pages} {035001} (\bibinfo {year} {2022})}\BibitemShut {NoStop}%
\bibitem [{\citenamefont {Gong}\ \emph
  {et~al.}(2023{\natexlab{a}})\citenamefont {Gong}, \citenamefont
  {Hatsagortsyan},\ and\ \citenamefont {Keitel}}]{gong2023electron}%
  \BibitemOpen
  \bibfield  {author} {\bibinfo {author} {\bibfnamefont {Z.}~\bibnamefont
  {Gong}}, \bibinfo {author} {\bibfnamefont {K.~Z.}\ \bibnamefont
  {Hatsagortsyan}},\ and\ \bibinfo {author} {\bibfnamefont {C.~H.}\
  \bibnamefont {Keitel}},\ }\bibfield  {title} {\bibinfo {title} {Electron
  polarization in ultrarelativistic plasma current filamentation
  instabilities},\ }\href@noop {} {\bibfield  {journal} {\bibinfo  {journal}
  {Phys. Rev. Lett.}\ }\textbf {\bibinfo {volume} {130}},\ \bibinfo {pages}
  {015101} (\bibinfo {year} {2023}{\natexlab{a}})}\BibitemShut {NoStop}%
\bibitem [{\citenamefont {Gong}\ \emph
  {et~al.}(2023{\natexlab{b}})\citenamefont {Gong}, \citenamefont {Shen},
  \citenamefont {Hatsagortsyan},\ and\ \citenamefont
  {Keitel}}]{PhysRevLett.131.225101}%
  \BibitemOpen
  \bibfield  {author} {\bibinfo {author} {\bibfnamefont {Z.}~\bibnamefont
  {Gong}}, \bibinfo {author} {\bibfnamefont {X.}~\bibnamefont {Shen}}, \bibinfo
  {author} {\bibfnamefont {K.~Z.}\ \bibnamefont {Hatsagortsyan}},\ and\
  \bibinfo {author} {\bibfnamefont {C.~H.}\ \bibnamefont {Keitel}},\ }\bibfield
   {title} {\bibinfo {title} {Electron slingshot acceleration in relativistic
  preturbulent shocks explored via emitted photon polarization},\ }\href
  {https://doi.org/10.1103/PhysRevLett.131.225101} {\bibfield  {journal}
  {\bibinfo  {journal} {Phys. Rev. Lett.}\ }\textbf {\bibinfo {volume} {131}},\
  \bibinfo {pages} {225101} (\bibinfo {year} {2023}{\natexlab{b}})}\BibitemShut
  {NoStop}%
\bibitem [{\citenamefont {Xue}\ \emph {et~al.}(2023)\citenamefont {Xue},
  \citenamefont {Sun}, \citenamefont {Wei}, \citenamefont {Li}, \citenamefont
  {Zhao}, \citenamefont {Wan}, \citenamefont {Lv}, \citenamefont {Zhao},
  \citenamefont {Xu},\ and\ \citenamefont {Li}}]{PhysRevLett.131.175101}%
  \BibitemOpen
  \bibfield  {author} {\bibinfo {author} {\bibfnamefont {K.}~\bibnamefont
  {Xue}}, \bibinfo {author} {\bibfnamefont {T.}~\bibnamefont {Sun}}, \bibinfo
  {author} {\bibfnamefont {K.-J.}\ \bibnamefont {Wei}}, \bibinfo {author}
  {\bibfnamefont {Z.-P.}\ \bibnamefont {Li}}, \bibinfo {author} {\bibfnamefont
  {Q.}~\bibnamefont {Zhao}}, \bibinfo {author} {\bibfnamefont {F.}~\bibnamefont
  {Wan}}, \bibinfo {author} {\bibfnamefont {C.}~\bibnamefont {Lv}}, \bibinfo
  {author} {\bibfnamefont {Y.-T.}\ \bibnamefont {Zhao}}, \bibinfo {author}
  {\bibfnamefont {Z.-F.}\ \bibnamefont {Xu}},\ and\ \bibinfo {author}
  {\bibfnamefont {J.-X.}\ \bibnamefont {Li}},\ }\bibfield  {title} {\bibinfo
  {title} {Generation of high-density high-polarization positrons via
  single-shot strong laser-foil interaction},\ }\href
  {https://doi.org/10.1103/PhysRevLett.131.175101} {\bibfield  {journal}
  {\bibinfo  {journal} {Phys. Rev. Lett.}\ }\textbf {\bibinfo {volume} {131}},\
  \bibinfo {pages} {175101} (\bibinfo {year} {2023})}\BibitemShut {NoStop}%
\bibitem [{\citenamefont {Chen}\ \emph {et~al.}()\citenamefont {Chen},
  \citenamefont {Liu}, \citenamefont {Shen}, \citenamefont {He},\ and\
  \citenamefont {Chen}}]{JointSub}%
  \BibitemOpen
  \bibfield  {author} {\bibinfo {author} {\bibfnamefont {J.-D.}\ \bibnamefont
  {Chen}}, \bibinfo {author} {\bibfnamefont {H.}~\bibnamefont {Liu}}, \bibinfo
  {author} {\bibfnamefont {B.}~\bibnamefont {Shen}}, \bibinfo {author}
  {\bibfnamefont {P.-L.}\ \bibnamefont {He}},\ and\ \bibinfo {author}
  {\bibfnamefont {Y.-Y.}\ \bibnamefont {Chen}},\ }\bibfield  {title} {\bibinfo
  {title} {Spin-dependent axion generation with controllable emission angles in
  strong laser fields},\ }\href@noop {} {\bibfield  {journal} {\bibinfo
  {journal} {arxiv}\ }}\bibinfo {note} {Submitted}\BibitemShut {NoStop}%
\bibitem [{\citenamefont {Furry}(1951)}]{PhysRev.81.115}%
  \BibitemOpen
  \bibfield  {author} {\bibinfo {author} {\bibfnamefont {W.~H.}\ \bibnamefont
  {Furry}},\ }\bibfield  {title} {\bibinfo {title} {On bound states and
  scattering in positron theory},\ }\href
  {https://doi.org/10.1103/PhysRev.81.115} {\bibfield  {journal} {\bibinfo
  {journal} {Phys. Rev.}\ }\textbf {\bibinfo {volume} {81}},\ \bibinfo {pages}
  {115} (\bibinfo {year} {1951})}\BibitemShut {NoStop}%
\bibitem [{\citenamefont {Di~Piazza}\ \emph {et~al.}(2012)\citenamefont
  {Di~Piazza}, \citenamefont {M{\"u}ller}, \citenamefont {Hatsagortsyan},\ and\
  \citenamefont {Keitel}}]{di2012extremely}%
  \BibitemOpen
  \bibfield  {author} {\bibinfo {author} {\bibfnamefont {A.}~\bibnamefont
  {Di~Piazza}}, \bibinfo {author} {\bibfnamefont {C.}~\bibnamefont
  {M{\"u}ller}}, \bibinfo {author} {\bibfnamefont {K.}~\bibnamefont
  {Hatsagortsyan}},\ and\ \bibinfo {author} {\bibfnamefont {C.~H.}\
  \bibnamefont {Keitel}},\ }\bibfield  {title} {\bibinfo {title} {Extremely
  high-intensity laser interactions with fundamental quantum systems},\
  }\href@noop {} {\bibfield  {journal} {\bibinfo  {journal} {Rev. Mod. Phys}\
  }\textbf {\bibinfo {volume} {84}},\ \bibinfo {pages} {1177} (\bibinfo {year}
  {2012})}\BibitemShut {NoStop}%
\bibitem [{\citenamefont {Fedotov}\ \emph {et~al.}(2023)\citenamefont
  {Fedotov}, \citenamefont {Ilderton}, \citenamefont {Karbstein}, \citenamefont
  {King}, \citenamefont {Seipt}, \citenamefont {Taya},\ and\ \citenamefont
  {Torgrimsson}}]{fedotov2023advances}%
  \BibitemOpen
  \bibfield  {author} {\bibinfo {author} {\bibfnamefont {A.}~\bibnamefont
  {Fedotov}}, \bibinfo {author} {\bibfnamefont {A.}~\bibnamefont {Ilderton}},
  \bibinfo {author} {\bibfnamefont {F.}~\bibnamefont {Karbstein}}, \bibinfo
  {author} {\bibfnamefont {B.}~\bibnamefont {King}}, \bibinfo {author}
  {\bibfnamefont {D.}~\bibnamefont {Seipt}}, \bibinfo {author} {\bibfnamefont
  {H.}~\bibnamefont {Taya}},\ and\ \bibinfo {author} {\bibfnamefont
  {G.}~\bibnamefont {Torgrimsson}},\ }\bibfield  {title} {\bibinfo {title}
  {Advances in qed with intense background fields},\ }\href@noop {} {\bibfield
  {journal} {\bibinfo  {journal} {Physics Reports}\ }\textbf {\bibinfo {volume}
  {1010}},\ \bibinfo {pages} {1} (\bibinfo {year} {2023})}\BibitemShut
  {NoStop}%
\bibitem [{\citenamefont {Esarey}\ \emph {et~al.}(2009)\citenamefont {Esarey},
  \citenamefont {Schroeder},\ and\ \citenamefont
  {Leemans}}]{esarey2009physics}%
  \BibitemOpen
  \bibfield  {author} {\bibinfo {author} {\bibfnamefont {E.}~\bibnamefont
  {Esarey}}, \bibinfo {author} {\bibfnamefont {C.~B.}\ \bibnamefont
  {Schroeder}},\ and\ \bibinfo {author} {\bibfnamefont {W.~P.}\ \bibnamefont
  {Leemans}},\ }\bibfield  {title} {\bibinfo {title} {Physics of laser-driven
  plasma-based electron accelerators},\ }\href@noop {} {\bibfield  {journal}
  {\bibinfo  {journal} {Rev. Mod. Phys.}\ }\textbf {\bibinfo {volume} {81}},\
  \bibinfo {pages} {1229} (\bibinfo {year} {2009})}\BibitemShut {NoStop}%
\bibitem [{\citenamefont {Landau}(2013)}]{landau2013classical}%
  \BibitemOpen
  \bibfield  {author} {\bibinfo {author} {\bibfnamefont {L.~D.}\ \bibnamefont
  {Landau}},\ }\href@noop {} {\emph {\bibinfo {title} {The classical theory of
  fields}}},\ Vol.~\bibinfo {volume} {2}\ (\bibinfo  {publisher} {Elsevier},\
  \bibinfo {year} {2013})\BibitemShut {NoStop}%
\bibitem [{\citenamefont {Schwinger}\ \emph {et~al.}(2019)\citenamefont
  {Schwinger}, \citenamefont {DeRaad~Jr}, \citenamefont {Milton},\ and\
  \citenamefont {Tsai}}]{schwinger2019classical}%
  \BibitemOpen
  \bibfield  {author} {\bibinfo {author} {\bibfnamefont {J.}~\bibnamefont
  {Schwinger}}, \bibinfo {author} {\bibfnamefont {L.~L.}\ \bibnamefont
  {DeRaad~Jr}}, \bibinfo {author} {\bibfnamefont {K.}~\bibnamefont {Milton}},\
  and\ \bibinfo {author} {\bibfnamefont {W.-Y.}\ \bibnamefont {Tsai}},\
  }\href@noop {} {\emph {\bibinfo {title} {Classical electrodynamics}}}\
  (\bibinfo  {publisher} {CRC Press},\ \bibinfo {year} {2019})\BibitemShut
  {NoStop}%
\bibitem [{\citenamefont {Unruh}(1976)}]{unruh1976notes}%
  \BibitemOpen
  \bibfield  {author} {\bibinfo {author} {\bibfnamefont {W.~G.}\ \bibnamefont
  {Unruh}},\ }\bibfield  {title} {\bibinfo {title} {Notes on black-hole
  evaporation},\ }\href@noop {} {\bibfield  {journal} {\bibinfo  {journal}
  {Phys. Rev. D}\ }\textbf {\bibinfo {volume} {14}},\ \bibinfo {pages} {870}
  (\bibinfo {year} {1976})}\BibitemShut {NoStop}%
\end{thebibliography}%

\end{document}